\documentclass[twocolumn,reprint,amsmath,superscriptaddress,amssymb,aps]{revtex4-2}
\pdfoutput=1
\usepackage[utf8]{inputenc}
\usepackage{amsmath}
\usepackage{amsfonts}
\usepackage{amssymb}
\usepackage{mathtools}
\usepackage{graphicx}
\usepackage{dcolumn}
\usepackage{comment}
\usepackage{xcolor}
\usepackage{graphicx}
\usepackage{dsfont}
\usepackage{bm}
\usepackage{booktabs}
\usepackage{siunitx}
\usepackage{multirow}
\usepackage{longtable}
\usepackage{extarrows}
\usepackage[hypertexnames=false]{hyperref}

\newcommand{\RN}[1]{%
  \textup{\uppercase\expandafter{\romannumeral#1}}%
}

\newcommand{\imth}{\hspace{1pt}\mathrm{i}\hspace{1pt}}
\newcommand{\expval}[1]{\langle{#1}\rangle}

\renewcommand{\figurename}{Fig.}
\renewcommand{\tablename}{Table}

\makeatletter
\def\maketitle{
\@author@finish
\title@column\titleblock@produce
\suppressfloats[t]}
\makeatother

\begin{document}
\title{An efficient material search for room temperature topological magnons}
\author{Mohammed J. Karaki}
\email{karaki.4@osu.edu}
\affiliation{Department of Physics, The Ohio State University, Columbus, OH 43210, USA}
\author{Xu Yang}
\affiliation{Department of Physics, The Ohio State University, Columbus, OH 43210, USA}
\author{Archibald J. Williams}
\affiliation{Department of Chemistry and Biochemistry, The Ohio State University, Columbus, OH 43210, USA }
\author{Mohamed Nawwar}
\affiliation{Department of Materials Science and Engineering, The Ohio State University, Columbus, OH 43210, USA}
\author{Vicky Doan-Nguyen}
\affiliation{Department of Materials Science and Engineering, The Ohio State University, Columbus, OH 43210, USA}
\author{Joshua E. Goldberger}
\affiliation{Department of Chemistry and Biochemistry, The Ohio State University, Columbus, OH 43210, USA }
\author{Yuan-Ming Lu}
\email{lu.1435@osu.edu}
\affiliation{Department of Physics, The Ohio State University, Columbus, OH 43210, USA}
\maketitle

\textbf{Topologically protected magnon surface states are highly desirable as an ideal platform to engineer low-dissipation spintronics devices. However, theoretical prediction of topological magnons in strongly correlated materials proves to be challenging because the ab initio density functional theory calculations fail to reliably predict magnetic interactions in correlated materials. Here we present a symmetry-based approach, which predicts topological magnons in magnetically ordered crystals, upon applying external perturbations such as magnetic/electric fields and/or mechanical strains. We apply this approach to carry out an efficient search for magnetic materials in the Bilbao Crystallographic Server, where among 198 compounds with an over 300 K transition temperature, we identify 12 magnetic insulators that support room temperature topological magnons. They feature Weyl magnons and magnon axion insulators, which host protected surface or hinge magnon modes, offering a route to realize energy-efficient devices based on protected surface magnons.}\\

\begin{figure*}[t!]
  \centering
  \includegraphics[width=0.95\linewidth]{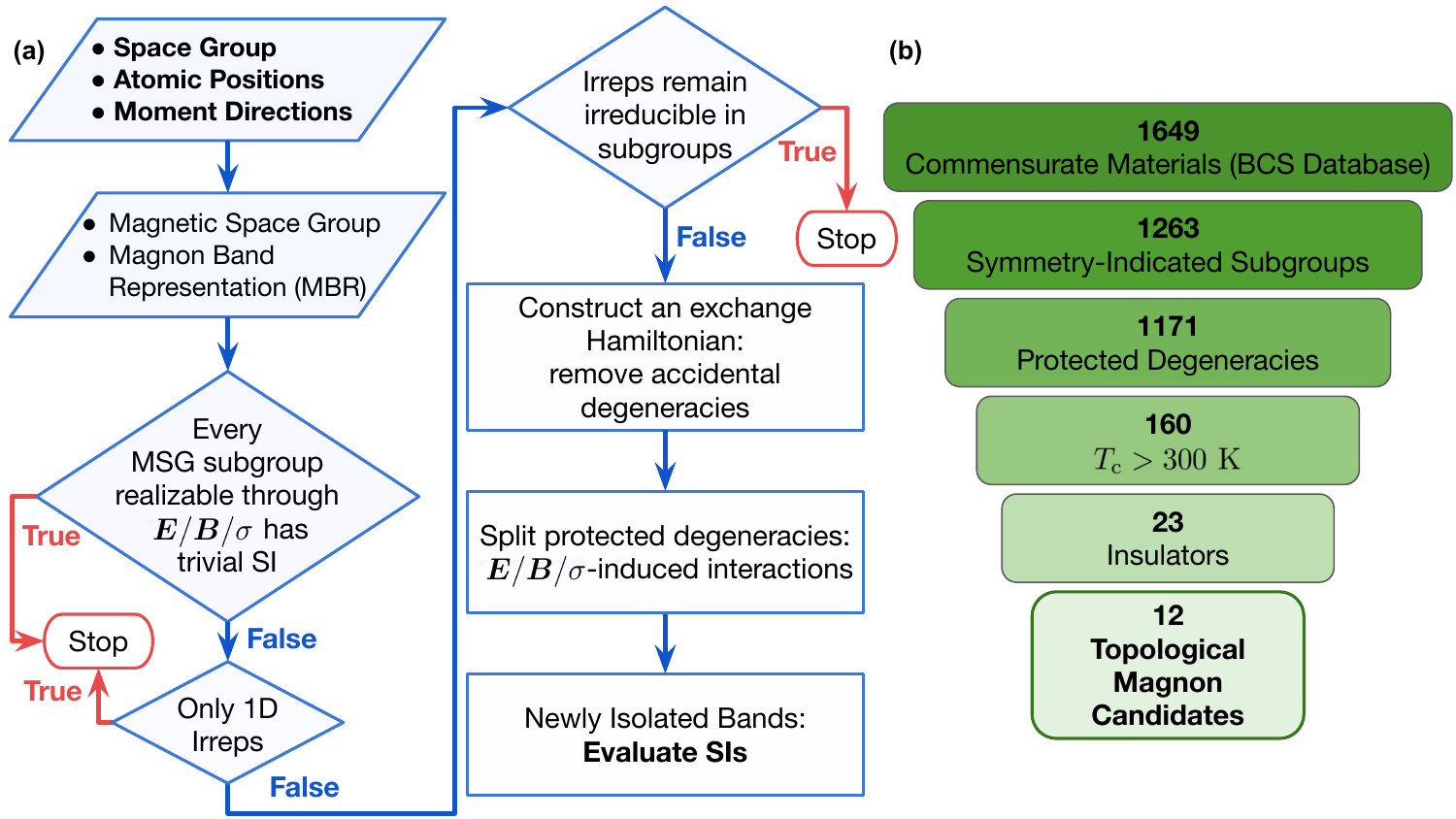}
  \caption{\textbf{Flowchart of the search process, and a summary of our search results from the Bilbao Crystallographic Server (BCS) magnetic materials database.} (a) Flowchart diagram highlighting the steps for systematic diagnosis of topological magnons in a generic magnetic material. (b) A summary of the search outcome. Explicit spin-wave calculations are performed in the last step to the 23 room-temperature magnetic insulators which pass all the group-theoretical filters.\label{fig:flowchart}}
\end{figure*}


Since the discovery of topological insulators and associated $Z_2$ topological invariants~\cite{Kane2005,Bernevig2006,Markus2007}, much progress has been made to reveal topological properties encoded in the electronic band structures of weakly correlated quantum materials~\cite{Hasan2010,Qi2011,Armitage2018,Chiu2016}. Not only has a full classification of topological bands been achieved, recent efforts have led to a complete catalogue of topological electronic materials based on \textit{ab initio} calculations of their band structures~\cite{Bradlyn2017,Tang2019,Vergniory2019,Zhang2019,Xu2020b}. One natural question arises: can the huge success of topological band theory be extended to predict the topology of strongly correlated materials?

In materials with long-range magnetic orders, the spin wave excitations can exhibit nontrivial topology and robust surface states, known as topological magnons (see Ref.~\cite{McClarty2022} and references therein for a review of the literature). Due to protected surface states insensitive to system geometry and defects, topological magnons provide a new route towards robust and low-dissipation magnon-based circuitry in magnon spintronics~\cite{Chumak2015,Wang2018}, making high quality materials hosting topological magnons very desirable. However, in contrast to efficiently searched and completely catalogued topological electronic materials, a systematic search for topological magnon materials has not been achieved so far~\cite{McClarty2022,Corticelli2022,Corticelli2022a}. In particular, two main challenges lie ahead of us.

First of all, unlike the electronic band topology which has been thoroughly understood and classified, much less is known about the magnon band topology. In electronic bands, after the 10-fold way classification of strong topological invariants protected by global symmetries~\cite{Chiu2016}, a complete classification of band topology in any space group has been achieved by the theoretical developments of symmetry indicators~\cite{Po2017,Haruki2022} and topological quantum chemistry~\cite{Bradlyn2017,Elcoro2021}. In contrast, the spin wave bands are described by a quadratic boson Hamiltonian, known to be very different from a fermion system~\cite{Colpa1978}. In spite of recent progress which established a map connecting the bosonic band topology to fermionic ones~\cite{Lu2018,Xu2020a,McClarty2022}, a complete classification of magnon band topology is still not available~\cite{McClarty2022,Corticelli2022,Corticelli2022a}.

Secondly, strong correlations of magnetic materials make it a challenging task to systematically predict the topology of magnon bands in materials. In electronic materials, with the help of \textit{ab initio} density functional theory calculations of band structures in existing materials, a catalogue of topological electronic materials has been achieved recently~\cite{Bradlyn2017,Tang2019,Vergniory2019,Zhang2019,Xu2020b}. In contrast, due to the lack of reliable \textit{ab initio} calculations, theoretically determining the microscopic spin model of a magnetic material, and hence its magnon bands, has been a longstanding problem in strongly correlated systems. Typically we rely on experimental inputs, for example by fitting inelastic neutron scattering data, to determine the exchange interactions on a material-by-material basis. This severely hinders any large-scale search for materials hosting topological magnons.

In this work, we show a systematic material search for topological magnons can be carried out efficiently, using a symmetry-based approach summarized in Fig.~1. We performed such a semi-automated search among all insulating materials with a room temperature ($T_c>300$~K) magnetic order in the Bilbao Crystallographic Server (BCS), and identified 12 candidate materials that can host room temperature topological magnons. In the main text, we outline our strategy, the search algorithm and outcome. We highlight two examples among the search results, and commenting on the synthesis of candidate materials. We list the detailed search process and data in the supplemental materials.

\subsection*{Theory framework}

\begin{table*}[t!]
  \centering
  \begin{tabular}{ccccc}
    \toprule
    Material ($T_{\mathrm{c}}[\si{\kelvin}]$) & MSG & Perturbation & Reduced MSG & Predicted Topology\\
    \midrule
    \multirow{1}{*}{NdFeO$_3$ ($760$)} & \multirow{6}{*}{$Pn'ma'~(62.448)$} & \multirow{3}{*}{$\bm{B}\parallel[001]$} & \multirow{3}{*}{$P2_1'/c'~(14.79)$} & \multirow{3}{*}{Weyl magnons\textsuperscript{\RN{1}}} \\
    \multirow{1}{*}{CeFeO$_3$ ($720$)} & & \multirow{3}{*}{uniaxial strain in $a-b$ plane} & \multirow{3}{*}{$P2_1'/c'~(14.79)$} & \multirow{3}{*}{Weyl magnons\textsuperscript{\RN{1}}} \\
    \multirow{1}{*}{NaOsO$_3$ ($410$)} & & \multirow{3}{*}{uniaxial strain in $b-c$ plane} & \multirow{3}{*}{$P2_1'/c'~(14.79)$} & \multirow{3}{*}{Weyl magnons\textsuperscript{\RN{1}}} \\
    \multirow{1}{*}{TbFeO$_3$ ($681$)} & & \multirow{3}{*}{uniaxial strain in $a-c$ plane} & \multirow{3}{*}{$P2_1/m~(11.50)$  }& \multirow{3}{*}{Trivial} \\
    \multirow{1}{*}{SmFeO$_3$ ($670$)} &  & &  &  \\
    \multirow{1}{*}{LaCrO$_3$ ($310$)} &  & &  &  \\

    \midrule

    \multirow{1}{*}{$\alpha$-Fe$_2$O$_3$ ($955$)} & \multirow{1}{*}{$C2'/c'~(15.89)$} & $\bm{B}\parallel[010]$ & $P\bar{1}~(2.4)$ & Weyl magnons\textsuperscript{\RN{3}}\\
    \midrule

    \multirow{1}{*}{SrRu$_2$O$_6$ ($565$)} & \multirow{1}{*}{$P_c\bar{3}1m~(162.78)$} & $\bm{B}\parallel[001]$ and generic strain & $P\bar{1}~(2.4)$ & Weyl/Magnon axion insulator\textsuperscript{\RN{3}} \\

    \midrule

    \multirow{1}{*}{CoF$_3$ ($460$)} & \multirow{1}{*}{$R\bar{3}c~(167.103)$} & $\bm{B}\parallel[001]$ and strain $\parallel [110]$& $P\bar{1}~(2.4)$ & Weyl magnons\textsuperscript{\RN{3}} \\

    \midrule
    \multirow{2}{*}{FeF$_3$ ($394$)} & \multirow{2}{*}{$C2'/c'~(15.89)$} & $\bm{B}\parallel[010]$ & $P\bar{1}~(2.4)$ & Weyl magnons\textsuperscript{\RN{2}}\\
    &  & $\bm{B}\parallel[010]$ and generic strain & $P\bar{1}~(2.4)$ & Magnon axion insulator\textsuperscript{\RN{2}}\\
    %

    \midrule

    \multirow{1}{*}{LaSrFeO$_4$ ($380$)} & \multirow{1}{*}{$P_C4_2/nnm~(134.481)$} & $\bm{E}\parallel[110]$ and strain $\perp[110]$ & $C_a2(5.17)$ & Weyl magnons \textsuperscript{\RN{3}}\\

    \midrule
    \multirow{2}{*}{MnTe ($323$)} & \multirow{2}{*}{$Cmcm~(63.457)$} & strain $\perp[100]$ & $C2/m~(12.58)$ & Weyl/Magnon axion insulator\textsuperscript{\RN{2}}\\
    &  & strain $\perp[010]$ and $\bm{B}\parallel[100]$ & $P\bar{1}~(2.4)$ & Weyl magnons\textsuperscript{\RN{2}} \\
    \bottomrule
  \end{tabular}
  \caption{\textbf{Room-temperature magnetic insulator candidates that host topological magnons.}\label{tab:results}}
\end{table*}

As shown previously~\cite{Colpa1978,Lu2018,Xu2020a}, the linear spin wave (LSW) problem of diagonalizing a quadratic Hamiltonian of bosons can be exactly mapped to the diagonalization of a fermionic Bogoliubov-de Gennes Hamiltonian, with the same symmetry group and band topology. This motivated us to apply the theory of symmetry indicators and topological quantum chemistry developed in electronic bands, to magnon excitations described by the LSW theory. In the theory of topological quantum chemistry~\cite{Bradlyn2017,Elcoro2021}, a generic band structure always arises from localized orbitals located at symmetry-respecting locations, known as the Wyckoff positions of a space group. The band representations in momentum space is strongly constrained by the possible Wyckoff positions and orbital symmetries. In particular, the band representations of a trivial atomic insulator can be enumerated for a given space group. If a band representation does not fit into any atomic insulator or their composites, the associated band structure must be topological, and these nontrivial band representations at high symmetry momenta are known as symmetry indicators (SIs) for topological bands~\cite{Fu2007,Kruthoff2017,Po2017,Haruki2022}. In the case of electronic bands in (weakly correlated) solid state materials, \textit{ab initio} calculations based on the density functional theory is used to compute the SIs at high symmetry momenta. If one set of connected bands is separated from other bands at the high symmetry momenta, its topology can be implied by its SIs, which enabled a complete search and catalogue for topological electronic materials~\cite{Bradlyn2017,Tang2019,Vergniory2019,Zhang2019,Xu2020b}.

In our problem of spin waves in magnetic insulators, the localized spin moments of ordered magnetic ions play the role of localized electronic orbitals in the theory of topological quantum chemistry. To be specific, for a magnetic ion whose spin $\vec S$ aligns along $z$ axis with $\expval{\vec S}=S\hat z$ ($S$ is the spin mangetic moment), its LSW variable $b=(S^x+\imth S^y)/\sqrt{2S}$ (known as Holstein-Primakoff boson) corresponds to the electron annihilation operator for the localized orbital in the electronic problem, and its symmetry transformations under the site symmetry group of the magnetic ion determine the orbital symmetry at its Wyckoff site. This allows us to constrain and classify all possible band representations of magnons in a magnetic space group (MSG), by mapping to its electronic counterparts discussed in Ref.~\cite{Elcoro2021,Haruki2022,Peng2021}.

This, unfortunately, is not enough for a large-sacle material search for topological magnons, mostly obstructed by the lack of microscopic spin models for a given magnetic material. In weakly-correlated electronic materials, \textit{ab initio} calculations are readily available to reliably predict the energy and SIs of electronic bands. On the contrary, although magnon band representations are constrained by the space group and magnetic order as mentioned above, the connectivity and SIs of magnon bands are usually not uniquely fixed by symmetry, and generally depend on the microscopic spin-spin interactions in the material. How to overcome this difficulty and systematically search for topological magnons?

Our strategy is to start from materials with symmetry-enforced magnon band degeneracy. More precisely, using the theory of topological quantum chemistry, we search for the MSGs and Wyckoff positions that host at least two connected bands in every possible band representation. Next, among the search results, we further identify those MSGs and Wyckoff positions where the protected degeneracy are lifted (i.e. the band crossings are gapped out) by external perturbations, including electric field $\vec E$, magnetic (Zeeman) field $\vec B$ and/or mechanical strains $\sigma$, which break the MSG down to a subgroup with separated magnon bands. The SIs of these separated bands will be used to diagnose the magnon band topology. Importantly, these search results are robust predictions from symmetry and representation theory, and independent of the microscopic spin models. Finally, we ask how the SIs of the separated bands depend on the external perturbations, and therefore determine the topological magnons induced by external perturbations. Our predictions are then consolidated by concrete spin models, which incorporate dominant nearest-neighbor Heisenberg exchange interactions and other symmetry-allowed terms.

\begin{figure*}[t!]
  \centering

  \includegraphics[width=0.95\linewidth]{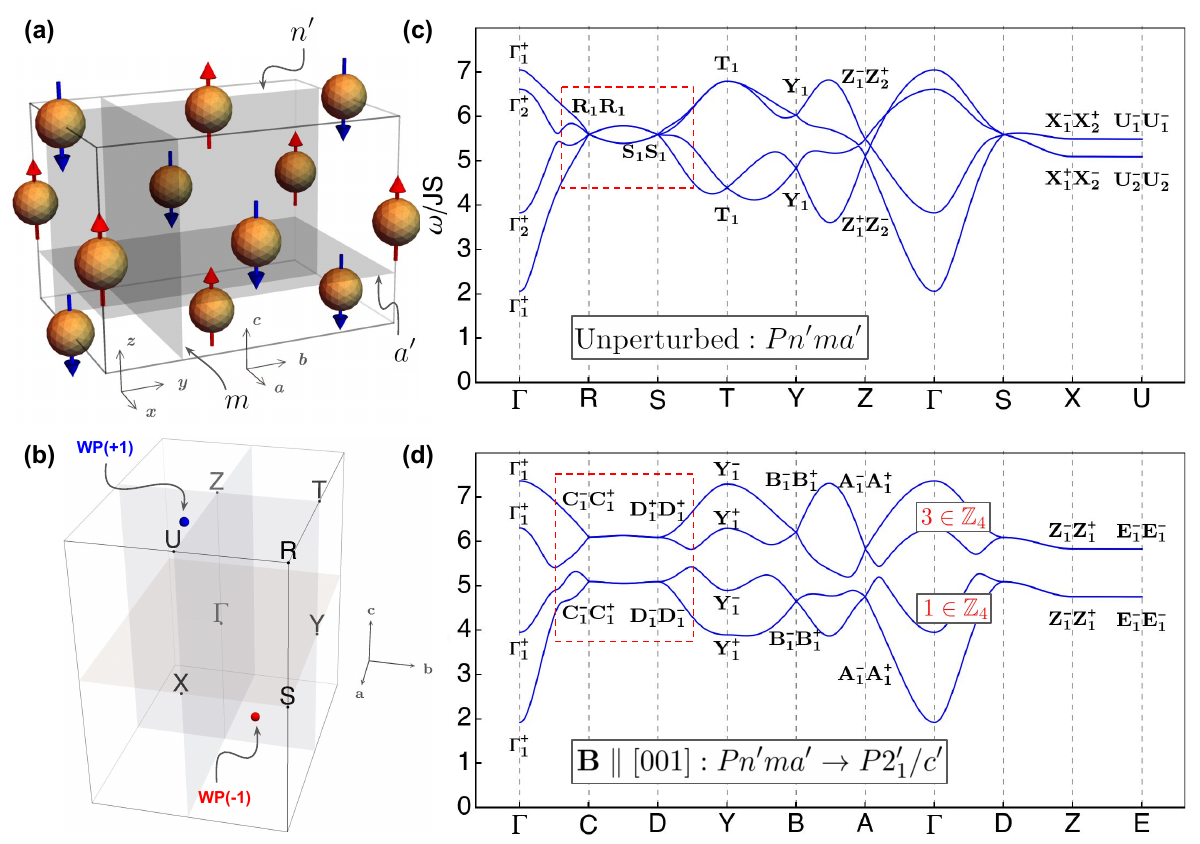}
  \caption{\textbf{Weyl magnons induced by a magnetic field or uniaxial strain in TbFeO$_3$.} \textbf{(a)}-\textbf{(b)} Magnetic unit cell and Brillouin zone of TbFeO$_3$, showing only the ordered Fe$^{3+}$ moments. \textbf{(c)} Magnon spectrum in the absence of external perturbations, showing two 4-fold degeneracies at momenta $R$ and $S$, protected by the MSG symmetry. The labels at each high-symmetry momentum indicate the irreducible representation (irrep) of each state, and the $\pm$ superscripts indicate whether the state is even or odd under inversion, respectively. \textbf{(d)} Magnon spectrum upon the application of a magnetic field, reducing the MSG and splitting the four connected bands into two pairs of bands. Each pair of bands acquires a non-trivial $\nu=1,3\in\mathbb{Z}_4$ inversion SI, corresponding to Weyl magnons shown in \textbf{(b)} and Fermi arcs on the material surface.\label{fig:tbfeo3}}
\end{figure*}

\subsection*{Search algorithm}
\begin{figure*}[t!]
  \centering
  \includegraphics[width=0.95\linewidth]{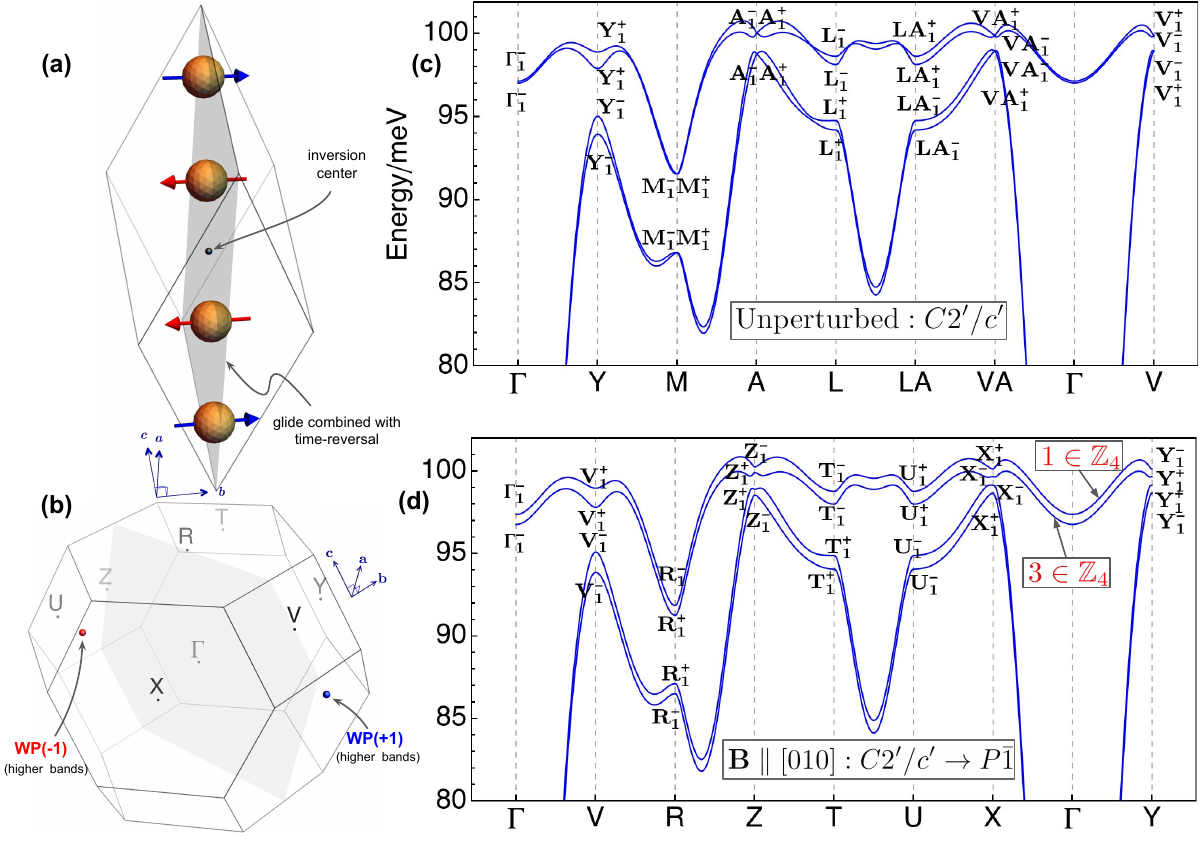}
  \caption{\textbf{Magnetic field-induced Weyl magnons in $\alpha$-Fe$_2$O$_3$.} \textbf{(a)}-\textbf{(b)}, Magnetic unit cell of Fe$_2$O$_3$ (only Fe atoms shown) and the Brillouin zone. \textbf{(c)}, Unperturbed magnon spectrum with two pairs of bands separated at all high-symmetry momentum. The spectrum is calculated using the first 10 nearest-neighbor Heisenberg interactions obtained from inelastic neutron scattering data, which we supplemented with small anisotropy terms (see the Supplementary Materials). \textbf{(d)}, Magnon spectrum upon the application of $\bm{B}$-field perturbation gapping out the degeneracies within each pair of bands. The inversion SI of the newly separated bands (illustrated in red) diagnose Weyl magnons thanks to their odd values. The WPs and their chirality are depicted in \textbf{(b)}.\label{fig:fe2o3}}
\end{figure*}

We now describe the workflow of our material search process, summarized in \figurename~1(a). Our starting point requires basic input about the structure and symmetry, which consists of the space group of the crystal structure, the positions of the magnetic atoms and their magnetic moments of the long range magnetic order. This information is sufficient to fix the MSG describing the magnetic structure, as well as the magnetic Wyckoff positions of the magnetic atoms. Using the theory of topological quantum chemistry, we are able to classify and determine all possible magnon band representations for a given MSG and the associated Wyckoff positions.

In the next step, following the strategy described previously, we need to identify (1) connected magnon bands with protected degeneracy in the magnetically ordered material; (2) separated magnon bands, when the degeneracy is split by external perturbations including electric field $\vec E$, magnetic (Zeeman) field $\vec B$, and mechanical strains $\sigma$; (3) nontrivial SIs of the separated bands, supported by the magnetic subgroup reduced from the original MSG by the symmetry-breaking perturbations. This step is implemented by two filters on the MSG and Wyckoff positions. First of all, we search for the MSGs and magnetic Wyckoff positions, at least one of whose magnetic subgroups do support a nontrivial SI group. In other words, we screen out those MSGs whose subgroups all have a trivial ($\mathbb{Z}_1$) symmetry indicator group. Among the 1651 magnetic
space groups, only 345 (including 50 gray groups) violate the first constraint (see the supplemental materials). Secondly, we require the MSG and Wyckoff positions to host symmetry-protected magnon band degeneracy, which can be split by perturbations that break the original MSG to a subgroup with nontrivial SIs. More precisely, we require the magnon bands constrained by the MSG and Wyckoff positions to have at least one multi-dimensional irreducible representations, which will split when the MSG is broken down to a subgroup. Out of all 9182
Wyckoff positions compatible with a commensurate magnetic order in all MSGs, only 1094 violate the second constraint.

We have applied these theoretical constraints on MSGs and Wyckoff positions to all magnetic materials in the BCS database. Among 1649 commensurate structures, we find that 1263 entries pass the first filter, and 1171 entries pass both filters, as shown in \figurename~1(b). These structures are summarized in the supplemental materials. In this paper, out of the 1171 compounds passing the group-theoretical filters, we focus on the 23 room-temperature magnetic insulators with $T_\textrm{c}>\SI{300}{\kelvin}$ and a charge gap.

The last step of the search process is to determine the SIs of magnon bands, after external perturbations are applied to the magnetic materials. We applied LSW theory to generic spin models, with dominant Heisenberg exchange interactions supplemented by other symmetry-allowed terms, to determine the SIs and topology of the resulting magnon bands. The search outcomes are 12 magnetic insulators that host either Weyl magnons~\cite{Li2016} or the magnonic analog of axion insulators~\cite{Wan2011,Hughes2011} (labeled as ``magnon axion insulators''),  listed in \tablename~\ref{tab:results}. We categorize the 12 candidate materials into three types, (I), (II) and (III) as labeled in Table \ref{tab:results}:

(I) Topological magnons can be obtained irrespective of the form of the perturbations, as long as the original MSG is broken down to a certain subgroup. The examples include the rare-earth perovskites RMO$_3$ in the top row of Table \ref{tab:results}.

(II) The type of topological magnons, or more precisely, the SIs of separated magnon bands depend on the type of external perturbations applied to the system. The examples include FeF$_3$ and MnTe in Table \ref{tab:results}, where different perturbations can give rise to either Weyl magnons or magnon axion insulators.

(III) The SIs of magnon bands depend not only on the form of external perturbations but also on the unperturbed magnon band structure. In this case, we make predictions based on microscopic spin models, with dominant Heisenberg exchange interactions, and other symmetry-allowed terms with small and randomly chosen coupling constants. For example, we use a $J_1$-$J_2$-$J_3$ Heisenberg model plus perturbations to predict the magnon band topology in the case of $\alpha$-Fe$_2$O$_3$ (\figurename~2).

{
Type (I) topological magnons are the most robust ones among our search results, for their presence depends neither on the form of the unperturbed Hamiltonian nor the perturbations, but only on the magnetic space groups before and after external perturbations are applied. Below we present two candidate materials as examples of our search results in Table \ref{tab:results}.}

\subsection*{Examples}

We highlight two examples of room temperature topological magnons in magnetic insulators. The first example is G-type antiferromagnetic perovskite TbFeO$_3$, as depicted in \figurename~2{(a)}, with four ordered Fe$^{3+}$ moments at the Wyckoff position $4b$ of its magnetic space group $Pn'ma'~(62.448)$. The Tb$^{3+}$ ions remain paramagnetic until cooled down to 8.5~K~\cite{Bertaut1967,Artyukhin2012}. These basic inputs dictate a symmetry-enforced 4-fold degeneracy at $R$ and $S$, and two 2-fold degeneracies at each of $T$, $U$, $X$, $Y$ and $Z$ in the magnon dispersion in \figurename~2(b).

An external magnetic field parallel to the moment direction reduces its symmetry group down to $P2_1'/c'~(14.79)$, as does any uni-axial strain in $\bm{a}$-$\bm{b}$ or $\bm{b}$-$\bm{c}$ plane. \figurename~2(c) illustrates the effect of a magnetic (Zeeman) field on the magnon spectrum and magnon band representations. Note that each 4-fold degeneracy in \figurename~2(b) is split into two two-dimensional (2D) irreps of $P2_1'/c'$. The magnetic subgroup $P2_1'/c'$ has a $\mathbb{Z}_4$ symmetry indicator group due to the inversion symmetry~\cite{Fu2007,Kruthoff2017,Po2017}, characterized by the number of negative inversion eigenvalues at the eight inversion-symmetric $k$ points modulo 4.

Importantly, the $\mathbb{Z}_4$ indicator of the higher (or lower) two bands in \figurename~2{(d)} must be an odd integer ($\nu=1,3\in\mathbb{Z}_4$), as long as the four connected bands of group $Pn'ma'$ are split into two 2D irreps of group $P2_1'/c'$. Therefore, the presence of Weyl magnons is robust, independent of either the microscopic spin model or the form of external perturbations. In particular, a magnetic field along $c$ axis creates a pair of Weyl magnons shown in \figurename~2(b). A similar analysis applies to all six rare earth perovskite in the top row of Table \ref{tab:results}, which share the same MSG and Wyckoff positions of magnetically ordered ions.

The second example is $\alpha$-Fe$_2$O$_3$, an antiferromagnetic insulator with a monoclinic magnetic space group $C2'/c'$~\cite{Hill2008}, shown in \figurename~3. The four sublattices of Fe$^{3+}$ ions give rise to four magnon bands, and a generic spin model with dominant $J_1$-$J_2$-$J_3$ Heisenberg terms leads to a gap at all inversion-symmetric $k$-points between the higher and lower pairs of bands, shown in \figurename~3(c). If an external magnetic field is additionally applied in the $[010]$ direction, the magnetic space group symmetry is lowered to $P\bar{1}$, lifting all the symmetry protected degeneracies within the higher two bands and creating a non-trivial gap. More precisely, each of the higher two bands acquires a non-trivial inversion symmetry indicator $\nu=1,3\in\mathbb{Z}_4$, corresponding to bulk Weyl magnons and surface magnon arcs~\cite{Li2016}. In contrast to the first example above, the magnetic interactions in $\alpha$-Fe$_2$O$_3$ have been estimated with inelastic neutron scattering where up to the 10\textsuperscript{th} NN Heisenberg parameters have been fitted to the magnon spectrum~\cite{fe2o3-spectrum}. The magnon spectrum in Figure 3(c) is reproduced using the reported Heisenberg terms in addition to other small symmetry-allowed terms to life accidental degeneracies in the spectrum (see the supplemental materials). The perturbed spectrum and its symmetry indicators are shown in Figure 3(d), and the Weyl points are illustrated in Figure 3(b).

{ While we focused on the bulk magnon spectra and their band topology in the main text, their associated topological magnon surface states are discussed and illustrated in the Supplemental Material S3. As a brief summary, while the bulk Weyl magnons always correspond to surface magnon arcs, magnon axion insulators in the bulk can exhibit two types of protected surface states. When the Chern number of the bulk magnon bands is nonzero in a 2D plane of the 3D Brillouin zone, the magnon axion insulator hosts chiral surface magnons\cite{Wan2011}. On the other hand, if the magnon Chern number vanishes in any 2D plane of the 3D Brillouin zone, the magnon axion insulator exhibits chiral hinge magnons\cite{Yue2019}.}

\subsection*{Synthetic accessibility of candidates}
Since the proposed realization of topological magnons requires the application of magnetic/electric fields and/or mechanical strains along specific crystallographic directions, the preparation of crystals or thin films with macroscopic dimensions is essential. The synthesis of pure, high-quality crystals, and thin films of the identified compounds have been well-established. For hematite, chemical vapor transport has been used to produce mm$\times$mm scale $\alpha$-Fe$_2$O$_3$~\cite{Rettie2016}.  In the case of LaCrO$_3$ thin films, the pseudocubic phase has been grown epitaxially on SrTiO$_3$ (001) to use strain to tune magnetic and optical properties~\cite{Sharma2021}. Furthermore, substitutions in $M_{x}$La$_{1-x}$CrO$_3$ (i.e. $M$ = Sr, $x < 0.25$) has been utilized to control $p$-type doping and induce in-plane $\approx 2\%$ compressive to $\approx 2\%$ tensile strain~\cite{Han2021}.  For $R$FeO$_3$ ($R$ = Nd, Ce, Tb, Sm), high quality crystals on the order of cm (length) and mm (diameter) have been grown with floating zone techniques~\cite{Cao2016,Yuan2011}.

\subsection*{Conclusion and outlook}

In summary, in a symmetry-based approach utilizing the theory of topological quantum chemistry and symmetry indicators, we carry out an efficient semi-automatic search for topological magnons in all high-transition-temperature ($T_c>300$~K) magnetic materials in BCS database. We identified 12 magnetic insulator candidate materials, which host room-temperature topological magnons induced by external perturbations, including electric/magnetic fields and mechanical strains. They feature Weyl magnons with surface magnon arcs, and magnonic axion insulators with either chiral surface or hinge magnon modes~\cite{Mook2021,Park2021}.

{ 
Now we briefly comment on the limitations of this work. While our symmetry-based approach can reliably predict the presence/absence of a magnon spectrum gap and the associated topological properties, it does not give a quantitative prediction on the size of the magnon spectrum gap that protects the topological magnons. In reality, both the finite temperature and magnon-magnon interactions give rise to a broadening of the magnon spectrum and a finite lifetime of magnons\cite{Zhitomirsky2013,Chernyshev2016,Hong2017,Mourigal2010}. We expect the topological magnons to survive when the magnon spectrum gap is much larger than the broadening. Another effect that exists in real materials but not captured in our approach is disorder. It has been shown that topology is well-defined even in the presence of disorders, if the disorders on average preserve the symmetry which protects the topology\cite{Fu2012,Ringel2012,Ma2022}. Most of the topological magnons predicted by our search algorithm are diagnosed by the inversion symmetry indicator, therefore we expect the topological magnons to be robust in the presence of weak disorders which obey an average inversion symmetry. 
}

The results of this work pave the road for future material synthesis and experimental discovery of room temperature topological magnons in magnetic insulators. A complete search through the whole BCS database, irrespective of the magnetic transition temperature, is a natural next step of the current work. Moreover, our search strategy can be naturally applied to other bosonic topological excitations, such as topological phonons~\cite{Mao2022,Li2021}. In particular, phonons in magnetic materials naturally fit into our search algorithm, where the long-range magnetic orders at low temperatures can be viewed as perturbations to the high temperature paramagnetic phase. We leave these interesting questions for future works.


\begin{thebibliography}{10}
\expandafter\ifx\csname url\endcsname\relax
  \def\url#1{\texttt{#1}}\fi
\expandafter\ifx\csname urlprefix\endcsname\relax\def\urlprefix{URL }\fi
\providecommand{\bibinfo}[2]{#2}
\providecommand{\eprint}[2][]{\url{#2}}

\bibitem{Kane2005}
\bibinfo{author}{Kane, C.~L.} \& \bibinfo{author}{Mele, E.~J.}
\newblock \bibinfo{title}{${Z}_{2}$ topological order and the quantum spin hall
  effect}.
\newblock \emph{\bibinfo{journal}{Phys. Rev. Lett.}}
  \textbf{\bibinfo{volume}{95}}, \bibinfo{pages}{146802}
  (\bibinfo{year}{2005}).
\newblock
  \urlprefix\url{https://link.aps.org/doi/10.1103/PhysRevLett.95.146802}.

\bibitem{Bernevig2006}
\bibinfo{author}{Bernevig, B.~A.} \& \bibinfo{author}{Zhang, S.-C.}
\newblock \bibinfo{title}{Quantum spin hall effect}.
\newblock \emph{\bibinfo{journal}{Phys. Rev. Lett.}}
  \textbf{\bibinfo{volume}{96}}, \bibinfo{pages}{106802}
  (\bibinfo{year}{2006}).
\newblock
  \urlprefix\url{https://link.aps.org/doi/10.1103/PhysRevLett.96.106802}.

\bibitem{Markus2007}
\bibinfo{author}{Markus, K.} \emph{et~al.}
\newblock \bibinfo{title}{Quantum spin hall insulator state in hgte quantum
  wells}.
\newblock \emph{\bibinfo{journal}{Science}} \textbf{\bibinfo{volume}{318}},
  \bibinfo{pages}{766--770} (\bibinfo{year}{2007}).
\newblock \urlprefix\url{https://doi.org/10.1126/science.1148047}.

\bibitem{Hasan2010}
\bibinfo{author}{Hasan, M.~Z.} \& \bibinfo{author}{Kane, C.~L.}
\newblock \bibinfo{title}{Colloquium: Topological insulators}.
\newblock \emph{\bibinfo{journal}{RMP}} \textbf{\bibinfo{volume}{82}},
  \bibinfo{pages}{3045--3067} (\bibinfo{year}{2010}).
\newblock \urlprefix\url{https://link.aps.org/doi/10.1103/RevModPhys.82.3045}.

\bibitem{Qi2011}
\bibinfo{author}{Qi, X.-L.} \& \bibinfo{author}{Zhang, S.-C.}
\newblock \bibinfo{title}{Topological insulators and superconductors}.
\newblock \emph{\bibinfo{journal}{RMP}} \textbf{\bibinfo{volume}{83}},
  \bibinfo{pages}{1057--1110} (\bibinfo{year}{2011}).
\newblock \urlprefix\url{https://link.aps.org/doi/10.1103/RevModPhys.83.1057}.

\bibitem{Armitage2018}
\bibinfo{author}{Armitage, N.~P.}, \bibinfo{author}{Mele, E.~J.} \&
  \bibinfo{author}{Vishwanath, A.}
\newblock \bibinfo{title}{Weyl and dirac semimetals in three-dimensional
  solids}.
\newblock \emph{\bibinfo{journal}{RMP}} \textbf{\bibinfo{volume}{90}},
  \bibinfo{pages}{015001} (\bibinfo{year}{2018}).
\newblock
  \urlprefix\url{https://link.aps.org/doi/10.1103/RevModPhys.90.015001}.

\bibitem{Chiu2016}
\bibinfo{author}{Chiu, C.-K.}, \bibinfo{author}{Teo, J. C.~Y.},
  \bibinfo{author}{Schnyder, A.~P.} \& \bibinfo{author}{Ryu, S.}
\newblock \bibinfo{title}{Classification of topological quantum matter with
  symmetries}.
\newblock \emph{\bibinfo{journal}{RMP}} \textbf{\bibinfo{volume}{88}},
  \bibinfo{pages}{035005} (\bibinfo{year}{2016}).
\newblock
  \urlprefix\url{https://link.aps.org/doi/10.1103/RevModPhys.88.035005}.

\bibitem{Bradlyn2017}
\bibinfo{author}{Bradlyn, B.} \emph{et~al.}
\newblock \bibinfo{title}{Topological quantum chemistry}.
\newblock \emph{\bibinfo{journal}{Nature}} \textbf{\bibinfo{volume}{547}},
  \bibinfo{pages}{298--305} (\bibinfo{year}{2017}).
\newblock \urlprefix\url{https://doi.org/10.1038/nature23268}.

\bibitem{Tang2019}
\bibinfo{author}{Tang, F.}, \bibinfo{author}{Po, H.~C.},
  \bibinfo{author}{Vishwanath, A.} \& \bibinfo{author}{Wan, X.}
\newblock \bibinfo{title}{Comprehensive search for topological materials using
  symmetry indicators}.
\newblock \emph{\bibinfo{journal}{Nature}} \textbf{\bibinfo{volume}{566}},
  \bibinfo{pages}{486--489} (\bibinfo{year}{2019}).
\newblock \urlprefix\url{https://doi.org/10.1038/s41586-019-0937-5}.

\bibitem{Vergniory2019}
\bibinfo{author}{Vergniory, M.~G.} \emph{et~al.}
\newblock \bibinfo{title}{A complete catalogue of high-quality topological
  materials}.
\newblock \emph{\bibinfo{journal}{Nature}} \textbf{\bibinfo{volume}{566}},
  \bibinfo{pages}{480--485} (\bibinfo{year}{2019}).
\newblock \urlprefix\url{https://doi.org/10.1038/s41586-019-0954-4}.

\bibitem{Zhang2019}
\bibinfo{author}{Zhang, T.} \emph{et~al.}
\newblock \bibinfo{title}{Catalogue of topological electronic materials}.
\newblock \emph{\bibinfo{journal}{Nature}} \textbf{\bibinfo{volume}{566}},
  \bibinfo{pages}{475--479} (\bibinfo{year}{2019}).
\newblock \urlprefix\url{https://doi.org/10.1038/s41586-019-0944-6}.

\bibitem{Xu2020b}
\bibinfo{author}{Xu, Y.} \emph{et~al.}
\newblock \bibinfo{title}{High-throughput calculations of magnetic topological
  materials}.
\newblock \emph{\bibinfo{journal}{Nature}} \textbf{\bibinfo{volume}{586}},
  \bibinfo{pages}{702--707} (\bibinfo{year}{2020}).
\newblock \urlprefix\url{https://doi.org/10.1038/s41586-020-2837-0}.

\bibitem{McClarty2022}
\bibinfo{author}{McClarty, P.~A.}
\newblock \bibinfo{title}{Topological magnons: A review}.
\newblock \emph{\bibinfo{journal}{Annu. Rev. Condens. Matter Phys.}}
  \textbf{\bibinfo{volume}{13}}, \bibinfo{pages}{171--190}
  (\bibinfo{year}{2022}).
\newblock
  \urlprefix\url{https://doi.org/10.1146/annurev-conmatphys-031620-104715}.

\bibitem{Chumak2015}
\bibinfo{author}{Chumak, A.~V.}, \bibinfo{author}{Vasyuchka, V.},
  \bibinfo{author}{Serga, A.} \& \bibinfo{author}{Hillebrands, B.}
\newblock \bibinfo{title}{Magnon spintronics}.
\newblock \emph{\bibinfo{journal}{Nature Physics}}
  \textbf{\bibinfo{volume}{11}}, \bibinfo{pages}{453--461}
  (\bibinfo{year}{2015}).
\newblock \urlprefix\url{https://doi.org/10.1038/nphys3347}.

\bibitem{Wang2018}
\bibinfo{author}{Wang, X.~S.}, \bibinfo{author}{Zhang, H.~W.} \&
  \bibinfo{author}{Wang, X.~R.}
\newblock \bibinfo{title}{Topological magnonics: A paradigm for spin-wave
  manipulation and device design}.
\newblock \emph{\bibinfo{journal}{Phys. Rev. Applied}}
  \textbf{\bibinfo{volume}{9}}, \bibinfo{pages}{024029} (\bibinfo{year}{2018}).
\newblock
  \urlprefix\url{https://link.aps.org/doi/10.1103/PhysRevApplied.9.024029}.

\bibitem{Corticelli2022}
\bibinfo{author}{Corticelli, A.}, \bibinfo{author}{Moessner, R.} \&
  \bibinfo{author}{McClarty, P.~A.}
\newblock \bibinfo{title}{Spin-space groups and magnon band topology}.
\newblock \emph{\bibinfo{journal}{Phys. Rev. B}}
  \textbf{\bibinfo{volume}{105}}, \bibinfo{pages}{064430}
  (\bibinfo{year}{2022}).
\newblock \urlprefix\url{https://link.aps.org/doi/10.1103/PhysRevB.105.064430}.

\bibitem{Corticelli2022a}
\bibinfo{author}{{Corticelli}, A.}, \bibinfo{author}{{Moessner}, R.} \&
  \bibinfo{author}{{McClarty}, P.~A.}
\newblock \bibinfo{title}{{Identifying, and constructing, complex magnon band
  topology}}.
\newblock \emph{\bibinfo{journal}{arXiv e-prints}}
  \bibinfo{pages}{arXiv:2203.06678} (\bibinfo{year}{2022}).
\newblock \eprint{2203.06678}.

\bibitem{Po2017}
\bibinfo{author}{Po, H.~C.}, \bibinfo{author}{Vishwanath, A.} \&
  \bibinfo{author}{Watanabe, H.}
\newblock \bibinfo{title}{Symmetry-based indicators of band topology in the 230
  space groups}.
\newblock \emph{\bibinfo{journal}{Nature Communications}}
  \textbf{\bibinfo{volume}{8}}, \bibinfo{pages}{50} (\bibinfo{year}{2017}).
\newblock \urlprefix\url{https://doi.org/10.1038/s41467-017-00133-2}.

\bibitem{Haruki2022}
\bibinfo{author}{Haruki, W.}, \bibinfo{author}{Chun, P.~H.} \&
  \bibinfo{author}{Ashvin, V.}
\newblock \bibinfo{title}{Structure and topology of band structures in the 1651
  magnetic space groups}.
\newblock \emph{\bibinfo{journal}{Science Advances}}
  \textbf{\bibinfo{volume}{4}}, \bibinfo{pages}{eaat8685}
  (\bibinfo{year}{2022}).
\newblock \urlprefix\url{https://doi.org/10.1126/sciadv.aat8685}.

\bibitem{Elcoro2021}
\bibinfo{author}{Elcoro, L.} \emph{et~al.}
\newblock \bibinfo{title}{Magnetic topological quantum chemistry}.
\newblock \emph{\bibinfo{journal}{Nature Communications}}
  \textbf{\bibinfo{volume}{12}}, \bibinfo{pages}{5965} (\bibinfo{year}{2021}).
\newblock \urlprefix\url{https://doi.org/10.1038/s41467-021-26241-8}.

\bibitem{Colpa1978}
\bibinfo{author}{Colpa, J. H.~P.}
\newblock \bibinfo{title}{Diagonalization of the quadratic boson hamiltonian}.
\newblock \emph{\bibinfo{journal}{Physica A: Statistical Mechanics and its
  Applications}} \textbf{\bibinfo{volume}{93}}, \bibinfo{pages}{327--353}
  (\bibinfo{year}{1978}).
\newblock
  \urlprefix\url{https://www.sciencedirect.com/science/article/pii/0378437178901607}.

\bibitem{Lu2018}
\bibinfo{author}{Lu, F.} \& \bibinfo{author}{Lu, Y.-M.}
\newblock \bibinfo{title}{Magnon band topology in spin-orbital coupled magnets:
  classification and application to {$\alpha$-RuCl$_3$}}.
\newblock \emph{\bibinfo{journal}{arXiv e-prints}}
  \bibinfo{pages}{arXiv:1807.05232} (\bibinfo{year}{2018}).
\newblock
  \urlprefix\url{https://ui.adsabs.harvard.edu/abs/2018arXiv180705232L}.
\newblock \eprint{1807.05232}.

\bibitem{Xu2020a}
\bibinfo{author}{Xu, Q.-R.} \emph{et~al.}
\newblock \bibinfo{title}{Squaring the fermion: The threefold way and the fate
  of zero modes}.
\newblock \emph{\bibinfo{journal}{PRB}} \textbf{\bibinfo{volume}{102}},
  \bibinfo{pages}{125127} (\bibinfo{year}{2020}).
\newblock \urlprefix\url{https://link.aps.org/doi/10.1103/PhysRevB.102.125127}.

\bibitem{Fu2007}
\bibinfo{author}{Fu, L.} \& \bibinfo{author}{Kane, C.~L.}
\newblock \bibinfo{title}{Topological insulators with inversion symmetry}.
\newblock \emph{\bibinfo{journal}{Phys. Rev. B}} \textbf{\bibinfo{volume}{76}},
  \bibinfo{pages}{045302} (\bibinfo{year}{2007}).
\newblock \urlprefix\url{https://link.aps.org/doi/10.1103/PhysRevB.76.045302}.

\bibitem{Kruthoff2017}
\bibinfo{author}{Kruthoff, J.}, \bibinfo{author}{de~Boer, J.},
  \bibinfo{author}{van Wezel, J.}, \bibinfo{author}{Kane, C.~L.} \&
  \bibinfo{author}{Slager, R.-J.}
\newblock \bibinfo{title}{Topological classification of crystalline insulators
  through band structure combinatorics}.
\newblock \emph{\bibinfo{journal}{Phys. Rev. X}} \textbf{\bibinfo{volume}{7}},
  \bibinfo{pages}{041069} (\bibinfo{year}{2017}).
\newblock \urlprefix\url{https://link.aps.org/doi/10.1103/PhysRevX.7.041069}.

\bibitem{Peng2021}
\bibinfo{author}{Peng, B.}, \bibinfo{author}{Jiang, Y.}, \bibinfo{author}{Fang,
  Z.}, \bibinfo{author}{Weng, H.} \& \bibinfo{author}{Fang, C.}
\newblock \bibinfo{title}{Topological classification and diagnosis in
  magnetically ordered electronic materials}.
\newblock \emph{\bibinfo{journal}{arXiv e-prints}}
  \bibinfo{pages}{arXiv:2102.12645} (\bibinfo{year}{2021}).
\newblock
  \urlprefix\url{https://ui.adsabs.harvard.edu/abs/2021arXiv210212645P}.

\bibitem{Li2016}
\bibinfo{author}{Li, F.-Y.} \emph{et~al.}
\newblock \bibinfo{title}{Weyl magnons in breathing pyrochlore
  antiferromagnets}.
\newblock \emph{\bibinfo{journal}{Nature Communications}}
  \textbf{\bibinfo{volume}{7}}, \bibinfo{pages}{12691} (\bibinfo{year}{2016}).
\newblock \urlprefix\url{https://doi.org/10.1038/ncomms12691}.

\bibitem{Wan2011}
\bibinfo{author}{Wan, X.}, \bibinfo{author}{Turner, A.~M.},
  \bibinfo{author}{Vishwanath, A.} \& \bibinfo{author}{Savrasov, S.~Y.}
\newblock \bibinfo{title}{Topological semimetal and fermi-arc surface states in
  the electronic structure of pyrochlore iridates}.
\newblock \emph{\bibinfo{journal}{Phys. Rev. B}} \textbf{\bibinfo{volume}{83}},
  \bibinfo{pages}{205101} (\bibinfo{year}{2011}).
\newblock \urlprefix\url{https://link.aps.org/doi/10.1103/PhysRevB.83.205101}.

\bibitem{Hughes2011}
\bibinfo{author}{Hughes, T.~L.}, \bibinfo{author}{Prodan, E.} \&
  \bibinfo{author}{Bernevig, B.~A.}
\newblock \bibinfo{title}{Inversion-symmetric topological insulators}.
\newblock \emph{\bibinfo{journal}{Phys. Rev. B}} \textbf{\bibinfo{volume}{83}},
  \bibinfo{pages}{245132} (\bibinfo{year}{2011}).
\newblock \urlprefix\url{https://link.aps.org/doi/10.1103/PhysRevB.83.245132}.

\bibitem{Bertaut1967}
\bibinfo{author}{Bertaut, E.~F.}, \bibinfo{author}{Chappert, J.},
  \bibinfo{author}{Mareschal, J.}, \bibinfo{author}{Rebouillat, J.~P.} \&
  \bibinfo{author}{Sivardière, J.}
\newblock \bibinfo{title}{Structures magnetiques de tbfeo3}.
\newblock \emph{\bibinfo{journal}{Solid State Communications}}
  \textbf{\bibinfo{volume}{5}}, \bibinfo{pages}{293--298}
  (\bibinfo{year}{1967}).
\newblock
  \urlprefix\url{https://www.sciencedirect.com/science/article/pii/0038109867902761}.

\bibitem{Artyukhin2012}
\bibinfo{author}{Artyukhin, S.} \emph{et~al.}
\newblock \bibinfo{title}{Solitonic lattice and yukawa forces in the rare-earth
  orthoferrite tbfeo3}.
\newblock \emph{\bibinfo{journal}{Nature Materials}}
  \textbf{\bibinfo{volume}{11}}, \bibinfo{pages}{694--699}
  (\bibinfo{year}{2012}).
\newblock \urlprefix\url{https://doi.org/10.1038/nmat3358}.

\bibitem{Hill2008}
\bibinfo{author}{Hill, A.~H.} \emph{et~al.}
\newblock \bibinfo{title}{Neutron diffraction study of mesoporous and bulk
  hematite, $\alpha$-{Fe2O3}}.
\newblock \emph{\bibinfo{journal}{Chem. Mater.}} \textbf{\bibinfo{volume}{20}},
  \bibinfo{pages}{4891--4899} (\bibinfo{year}{2008}).
\newblock \urlprefix\url{https://doi.org/10.1021/cm800009s}.

\bibitem{fe2o3-spectrum}
\bibinfo{author}{Samuelsen, E.~J.} \& \bibinfo{author}{Shirane, G.}
\newblock \bibinfo{title}{Inelastic neutron scattering investigation of spin
  waves and magnetic interactions in $\alpha$-{Fe2O3}}.
\newblock \emph{\bibinfo{journal}{physica status solidi (b)}}
  \textbf{\bibinfo{volume}{42}}, \bibinfo{pages}{241--256}
  (\bibinfo{year}{1970}).
\newblock
  \urlprefix\url{https://onlinelibrary.wiley.com/doi/abs/10.1002/pssb.19700420125}.
\newblock
  \eprint{https://onlinelibrary.wiley.com/doi/pdf/10.1002/pssb.19700420125}.

\bibitem{Yue2019}
\bibinfo{author}{Yue, C.} \emph{et~al.}
\newblock \bibinfo{title}{Symmetry-enforced chiral hinge states and surface
  quantum anomalous hall effect in the magnetic axion insulator bi2-xsmxse3}.
\newblock \emph{\bibinfo{journal}{Nature Physics}}
  \textbf{\bibinfo{volume}{15}}, \bibinfo{pages}{577--581}
  (\bibinfo{year}{2019}).
\newblock \urlprefix\url{https://doi.org/10.1038/s41567-019-0457-0}.

\bibitem{Rettie2016}
\bibinfo{author}{Rettie, A. J.~E.} \emph{et~al.}
\newblock \bibinfo{title}{Synthesis, electronic transport and optical
  properties of si:$\alpha$-fe$_2$o$_3$ single crystals}.
\newblock \emph{\bibinfo{journal}{Journal of Materials Chemistry C}}
  \textbf{\bibinfo{volume}{4}}, \bibinfo{pages}{559--567}
  (\bibinfo{year}{2016}).
\newblock \urlprefix\url{https://doi.org/10.1039/c5tc03368c}.

\bibitem{Sharma2021}
\bibinfo{author}{Sharma, Y.} \emph{et~al.}
\newblock \bibinfo{title}{Tuning magnetic and optical properties through strain
  in epitaxial {LaCrO}$_3$ thin films}.
\newblock \emph{\bibinfo{journal}{Applied Physics Letters}}
  \textbf{\bibinfo{volume}{119}}, \bibinfo{pages}{071902}
  (\bibinfo{year}{2021}).
\newblock \urlprefix\url{https://doi.org/10.1063/5.0058678}.

\bibitem{Han2021}
\bibinfo{author}{Han, D.} \emph{et~al.}
\newblock \bibinfo{title}{{Giant Tuning of Electronic and Thermoelectric
  Properties by Epitaxial Strain in p-Type Sr-Doped LaCrO3 Transparent Thin
  Films}}.
\newblock \emph{\bibinfo{journal}{ACS Applied Electronic Materials}}
  \textbf{\bibinfo{volume}{3}}, \bibinfo{pages}{3461--3471}
  (\bibinfo{year}{2021}).
\newblock \urlprefix\url{https://doi.org/10.1021/acsaelm.1c00425}.

\bibitem{Cao2016}
\bibinfo{author}{Cao, Y.} \emph{et~al.}
\newblock \bibinfo{title}{Magnetic phase transition and giant anisotropic
  magnetic entropy change in {TbFeO}$_3$ single crystal}.
\newblock \emph{\bibinfo{journal}{Journal of Applied Physics}}
  \textbf{\bibinfo{volume}{119}}, \bibinfo{pages}{063904}
  (\bibinfo{year}{2016}).
\newblock \urlprefix\url{https://doi.org/10.1063/1.4941105}.

\bibitem{Yuan2011}
\bibinfo{author}{Yuan, S.} \emph{et~al.}
\newblock \bibinfo{title}{Magnetic properties of {NdFeO}$_3$ single crystal in
  the spin reorientation region}.
\newblock \emph{\bibinfo{journal}{Journal of Applied Physics}}
  \textbf{\bibinfo{volume}{109}}, \bibinfo{pages}{07E141}
  (\bibinfo{year}{2011}).
\newblock \urlprefix\url{https://doi.org/10.1063/1.3562259}.

\bibitem{Mook2021}
\bibinfo{author}{Mook, A.}, \bibinfo{author}{D\'{\i}az, S.~A.},
  \bibinfo{author}{Klinovaja, J.} \& \bibinfo{author}{Loss, D.}
\newblock \bibinfo{title}{Chiral hinge magnons in second-order topological
  magnon insulators}.
\newblock \emph{\bibinfo{journal}{Phys. Rev. B}}
  \textbf{\bibinfo{volume}{104}}, \bibinfo{pages}{024406}
  (\bibinfo{year}{2021}).
\newblock \urlprefix\url{https://link.aps.org/doi/10.1103/PhysRevB.104.024406}.

\bibitem{Park2021}
\bibinfo{author}{Park, M.~J.}, \bibinfo{author}{Lee, S.} \&
  \bibinfo{author}{Kim, Y.~B.}
\newblock \bibinfo{title}{Hinge magnons from noncollinear magnetic order in a
  honeycomb antiferromagnet}.
\newblock \emph{\bibinfo{journal}{Phys. Rev. B}}
  \textbf{\bibinfo{volume}{104}}, \bibinfo{pages}{L060401}
  (\bibinfo{year}{2021}).
\newblock
  \urlprefix\url{https://link.aps.org/doi/10.1103/PhysRevB.104.L060401}.

\bibitem{Zhitomirsky2013}
\bibinfo{author}{Zhitomirsky, M.~E.} \& \bibinfo{author}{Chernyshev, A.~L.}
\newblock \bibinfo{title}{Colloquium: Spontaneous magnon decays}.
\newblock \emph{\bibinfo{journal}{Rev. Mod. Phys.}}
  \textbf{\bibinfo{volume}{85}}, \bibinfo{pages}{219--242}
  (\bibinfo{year}{2013}).
\newblock \urlprefix\url{https://link.aps.org/doi/10.1103/RevModPhys.85.219}.

\bibitem{Chernyshev2016}
\bibinfo{author}{Chernyshev, A.~L.} \& \bibinfo{author}{Maksimov, P.~A.}
\newblock \bibinfo{title}{Damped topological magnons in the kagome-lattice
  ferromagnets}.
\newblock \emph{\bibinfo{journal}{Phys. Rev. Lett.}}
  \textbf{\bibinfo{volume}{117}}, \bibinfo{pages}{187203}
  (\bibinfo{year}{2016}).
\newblock
  \urlprefix\url{https://link.aps.org/doi/10.1103/PhysRevLett.117.187203}.

\bibitem{Hong2017}
\bibinfo{author}{Hong, T.} \emph{et~al.}
\newblock \bibinfo{title}{Field induced spontaneous quasiparticle decay and
  renormalization of quasiparticle dispersion in a quantum antiferromagnet}.
\newblock \emph{\bibinfo{journal}{Nature Communications}}
  \textbf{\bibinfo{volume}{8}}, \bibinfo{pages}{15148} (\bibinfo{year}{2017}).
\newblock \urlprefix\url{https://doi.org/10.1038/ncomms15148}.

\bibitem{Mourigal2010}
\bibinfo{author}{Mourigal, M.}, \bibinfo{author}{Zhitomirsky, M.~E.} \&
  \bibinfo{author}{Chernyshev, A.~L.}
\newblock \bibinfo{title}{Field-induced decay dynamics in square-lattice
  antiferromagnets}.
\newblock \emph{\bibinfo{journal}{Phys. Rev. B}} \textbf{\bibinfo{volume}{82}},
  \bibinfo{pages}{144402} (\bibinfo{year}{2010}).
\newblock \urlprefix\url{https://link.aps.org/doi/10.1103/PhysRevB.82.144402}.

\bibitem{Fu2012}
\bibinfo{author}{Fu, L.} \& \bibinfo{author}{Kane, C.~L.}
\newblock \bibinfo{title}{Topology, delocalization via average symmetry and the
  symplectic anderson transition}.
\newblock \emph{\bibinfo{journal}{Phys. Rev. Lett.}}
  \textbf{\bibinfo{volume}{109}}, \bibinfo{pages}{246605}
  (\bibinfo{year}{2012}).
\newblock
  \urlprefix\url{https://link.aps.org/doi/10.1103/PhysRevLett.109.246605}.

\bibitem{Ringel2012}
\bibinfo{author}{Ringel, Z.}, \bibinfo{author}{Kraus, Y.~E.} \&
  \bibinfo{author}{Stern, A.}
\newblock \bibinfo{title}{Strong side of weak topological insulators}.
\newblock \emph{\bibinfo{journal}{Phys. Rev. B}} \textbf{\bibinfo{volume}{86}},
  \bibinfo{pages}{045102} (\bibinfo{year}{2012}).
\newblock \urlprefix\url{https://link.aps.org/doi/10.1103/PhysRevB.86.045102}.

\bibitem{Ma2022}
\bibinfo{author}{{Ma}, R.} \& \bibinfo{author}{{Wang}, C.}
\newblock \bibinfo{title}{{Average Symmetry-Protected Topological Phases}}.
\newblock \emph{\bibinfo{journal}{arXiv e-prints}}
  \bibinfo{pages}{arXiv:2209.02723} (\bibinfo{year}{2022}).
\newblock \eprint{2209.02723}.

\bibitem{Mao2022}
\bibinfo{author}{Mao, X.} \& \bibinfo{author}{Lubensky, T.~C.}
\newblock \bibinfo{title}{Maxwell lattices and topological mechanics}.
\newblock \emph{\bibinfo{journal}{Annu. Rev. Condens. Matter Phys.}}
  \textbf{\bibinfo{volume}{9}}, \bibinfo{pages}{413--433}
  (\bibinfo{year}{2022}).
\newblock
  \urlprefix\url{https://doi.org/10.1146/annurev-conmatphys-033117-054235}.

\bibitem{Li2021}
\bibinfo{author}{Li, J.} \emph{et~al.}
\newblock \bibinfo{title}{Computation and data driven discovery of topological
  phononic materials}.
\newblock \emph{\bibinfo{journal}{Nature Communications}}
  \textbf{\bibinfo{volume}{12}}, \bibinfo{pages}{1204} (\bibinfo{year}{2021}).
\newblock \urlprefix\url{https://doi.org/10.1038/s41467-021-21293-2}.

\bibitem{supp}
\bibinfo{title}{The details can be found in the supplemental materials} .

\end{thebibliography}

\clearpage
\twocolumngrid
\section*{Methods}
\subsection*{Review of symmetry indicators}
Consider a set of bands that is energetically separated from other bands at all high-symmetry momenta. For each high-symmetry momentum $\bm{k}$, define the integer $n_{\bm{k}}^{\alpha}$ as the number of times the little-group irreducible representation $\rho_{\bm{k}}^{\alpha}$ appears in the bands. These integers are well-defined thanks to the requirement that the set of bands are isolated from other bands. The collection of these integers for all pairs of distinct high-symmetry momenta and irreducible representations forms an integer-valued ``vector'' $\bm{b}=\{n_{\bm{k}}^{\alpha}\}$, and is dubbed a band structure~\cite{Po2017}.

Any band structure $\bm{b}$ must satisfy the so-called compatibility relations, which are a set of constraints needed to consistently patch together the representations while maintaining the gap at high-symmetry momenta. The set of all valid band structures is denoted by $\{\mathrm{BS}\}$.

A special kind of band structures arises from considering localized atomic orbitals with vanishing hopping. In this case, the irrep multiplicities $\{n_{\bm{k}}^{\alpha}\}$ trivially satisfy the compatibility relations and thus constitute a band structure. The set $\{\mathrm{AI}\}$ denotes the collection of all such atomic insulators. This set can be obtained by listing all the band structures arising from all pairs of sites and site-symmetry group irreducible representations, and by considering all combination thereof.

The topology of a given set of bands can be diagnosed based on symmetry information by considering its band structure $\bm{b}$. If $\bm{b}\in\{\mathrm{BS}\}$ but $\bm{b}\notin\{\mathrm{AI}\}$, it is not possible to adiabatically deform the bands to any atomic insulator; they must be topologically non-trivial. If $\bm{b}\notin\{\mathrm{BS}\}$ (and thus necessarily $\bm{b}\notin\{\mathrm{AI}\}$), the bands violate the compatibility relations, signalling a violation of the gap condition at high-symmetry momenta. Finally, non-trivial topology is not ruled out by $\bm{b}\in\{\mathrm{AI}\}$; it is just not detectable by symmetry indicators.

While the component-wise addition of two band structures corresponds to the direct sum of the representations of two bands, we can formally define the component-wise subtraction as the inverse operation. This allows the components of $\bm{b}$ to run negative, and the collections $\{\mathrm{AI}\}$ and $\{\mathrm{BS}\}$ become abelian groups of the same rank in all (magnetic) space groups~\cite{Po2017,Haruki2022}. Here we do not consider the case of fragile topology~\cite{Po2017,Bradlyn2017}.

Finally, the symmetry indicator group is defined as the quotient
\begin{equation}
  X_{\textrm{BS}}=\frac{\{\textrm{BS}\}}{\{\textrm{AI}\}}=\mathbb{Z}_{n_1}\times\mathbb{Z}_{n_2}\times\cdots.
\end{equation}

\subsection*{Spin-wave Hamiltonian, symmetry, and band representation}
For a bilinear exchange Hamiltonian
\begin{equation}
  H=\frac{1}{2}\sum_{i,j}\sum_{\alpha,\beta\in\{x,y,z\}}\mathsf{J}_{i,j}^{\alpha,\beta}S_{i}^{\alpha}S_{j}^{\beta},
\end{equation}
and a magnetic order $\langle S_{i}^{x}\rangle=\langle S_{i}^{y}\rangle=0$, $\langle S_{i}^{z}\rangle=S_{i}$ (written in a local frame by locally rotating the ordered moments to $z$-axis), one obtains the spin-wave dynamics by expanding $H$ around the order for small spin fluctuations ($S_{i}^{x}$ and $S_{i}^{y}$). Omitting an irrelevant constant term, and up to quadratic order in $S_{i}^{x}$ and $S_{i}^{y}$, the expansion yields
\begin{equation}
  H = \frac{1}{2}\sum_{i,j}\sum_{\alpha,\beta\in\{x,y\}}S_{i}^{\alpha}R_{i\alpha,j\beta}S_{j}^{\beta},
\end{equation}
where the real, symmetric matrix $R$ is required to be non-negative definite for the stability of the magnetic order.
Rescaling the transverse variables $S_{i}^{\alpha=x,y}$ by $S_{i}^{\alpha}\rightarrow S_{i}^{\alpha}/\sqrt{S_i}$, one obtains the commutation relations
\begin{equation}\label{eq:commrel}
  [S_{i}^{\alpha},S_{j}^{\beta}]=-\delta_{i,j}(\sigma_{y})^{\alpha,\beta}
\end{equation}
where $\sigma_{y}$ is the second Pauli matrix.

The Heisenberg equation of motion yields
\begin{equation}
  -i\frac{\mathrm{d}S_{i}^{\alpha}}{\mathrm{d}t}=\sum_{j,\beta}{(Y\cdot R)}_{i\alpha,j\beta}S_{j}^{\beta},
\end{equation}
where we define matrix $Y_{i\alpha,j\beta}=\delta_{i,j}(\sigma_y)^{\alpha,\beta}$, and the magnon spectrum is given by the eigenvalues of the non-Hermitian ``dynamical matrix'' $Y\cdot R$.

The spin-wave system inherits all the symmetries of the exchange Hamiltonian as long as they preserve the magnetic order. Under any symmetry $g$, $[O_g,R]=0$, where $O_g\in SO(2N)$. Being an imaginary matrix, note that the right hand side of Eq.~\ref{eq:commrel} acquires a minus sign under the action of an anti-unitary symmetry $g$. This corresponds to flipping the handedness of the local spin frame by time-reversal. Thus, $O_g$ anti-commutes (commutes) with $Y$ for an anti-unitary (unitary) symmetry $g$.

A similarity transformation of $YR$, defined as $H_{\mathrm{f}}\equiv R^{1/2}(Y R)R^{-1/2}$, maps the dynamical matrix to a Hermitian problem~\cite{,Colpa1978,Lu2018}. Importantly, note that $H_{\mathrm{f}}$ inherits the symmetry of the spin-wave system and its implementation, with $H_{\mathrm{f}}\rightarrow O_g H_{\mathrm{f}} O_{g}^{\dagger}$ for unitary $g$, and $H_{\mathrm{f}}\rightarrow -O_g H_{\mathrm{f}} O_{g}^{\dagger}=O_g H_{\mathrm{f}}^{\star}O_{g}^{\dagger}$ for anti-unitary $g$. Thus, the spin-wave problem and the Hermitian counterpart share the same band topology, and the implications of topological quantum chemistry and symmetry indicators carry over to the linear spin-wave problem.

In particular, the magnon system is mapped to a spinless (i.e., no spin-orbit coupling) electronic counterpart, and the spin-wave variables $S_{i}^{\alpha=x,y}$ pick up a $+1$ sign upon a $2\pi$-rotation or $\mathcal{T}^{2}$ (time-reversal squared.) Additionally, the ``atomic orbitals'' (at a given site $i$) from which the band representation can be induced are the site-symmetry group representation of the spin-wave variables $S_i^{\alpha=x,y}$. For all site-symmetry groups compatible with a magnetic order, this is always a direct-sum of two 1-dimensional representations~\cite{supp}. The Bilbao Crystallographic Server MBANDREP tool~\cite{Elcoro2021,Xu2020b} provides a complete tabulation of induced band representations.

\subsection*{Symmetry constraints on exchange interactions induced by an electric field or strain}

Now that an external magnetic field couples to the spin moment by a Zeeman term in the lowest order, here we consider perturbations to the spin Hamiltonian induced by an external electric field, and mechanical strains.

The electric field couples to electric polarization operator in a material. Odd powers of spin operators violate time-reversal symmetry, and thus the polarization operators $\hat{P}^{\alpha=x,y,z}$, which are time-reversal invariant, are bilinear in spins at the lowest order, and read
\begin{equation}\label{polarization}
  \hat{P}^\alpha=\frac{1}{2}\sum_{i,j}S_i^T P_{i,j}^\alpha S_j,
\end{equation}
where $S_i^T=(S_i^x,S_i^y,S_i^z)$ is composed of the spin operators at site $i$, and $P_{i,j}^\alpha$ are 3$\times$3-matrices of exchange coefficients.

$P_{i,j}^\alpha$ are further constrained by the space group $G$. This is because the polarization is a vector operator, and under any symmetry $g=\{O_g|\bm{t}_g\}\in G$ composed of a proper/improper rotational part $O_g\in O(3)$ and a translation $\bm{t}$, $\hat{P}^\alpha$ must transform like the components of a vector:
\begin{equation*}
  \hat{P}^\alpha\rightarrow O_g^{\alpha\alpha'}\hat{P}^{\alpha'}.
\end{equation*}

On the other hand, the transformation of the right hand side of Eq.~\ref{polarization} is
\begin{equation*}
  \frac{1}{2}\sum_{i,j}S_i^T P_{i,j}^\alpha S_j
  \rightarrow
  \frac{1}{2}\sum_{i,j}S_{g^{-1}i}^T O_g^T P_{i,j}^\alpha O_g S_{g^{-1}j}.
\end{equation*}
(Note that in the last equation the spin operators should transform like a pseudo-vector rather than a vector. However, this distinction should not matter for the bilinear terms.)

Thus $P_{i,j}^\alpha$ must satisfy the constraints
\begin{equation*}
  O_g^{\alpha\alpha'}\sum_{i,j}S_i^T P_{i,j}^{\alpha'} S_j
  =
  \sum_{i,j}S_{g^{-1}i}^T O_g^T P_{i,j}^\alpha O_g S_{g^{-1}j}.
\end{equation*}

In practice, one can impose these constraints for a given exchange path $i\leftrightarrow j$ by considering all the elements $g\in G$ that leave sites $i$ and $j$ invariant (up to swapping the two sites). These elements impose constraints only on three matrices $P_{i,j}^{\alpha=x,y,z}$. Subsequently, for all other symmetry-related bonds $i'\leftrightarrow j'$ (with $i'=h^{-1}i$, $j'=h^{-1}j$, $h\in G$), the corresponding matrices are obtained with
\begin{equation*}
  P_{i',j'}^\alpha=\sum_{\alpha'}O_h^{\alpha'\alpha} O_h^T P_{i,j}^{\alpha'} O_h.
\end{equation*}
Therefore we have $H(\bm{E})=-\bm{E}\cdot\hat{\bm{P}}$ as the most general bilinear perturbation induced by an electric field $\bm{E}$.

A strain-induced perturbation for a given strain tensor $\sigma^{\alpha,\beta=x,y,z}$ reads
\begin{equation*}
  H(\{\sigma^{\alpha,\beta}\})=\sum_{\alpha,\beta}\sigma^{\alpha,\beta}\hat{\Sigma}^{\alpha,\beta},
\end{equation*}
where the operators $\hat{\Sigma}^{\alpha,\beta}$ are time-reversal- as well as inversion-invariant, and transform like a rank-two tensor under a proper rotation $O_g\in SO(3)$,
\begin{equation*}
  \hat{\Sigma}^{\alpha,\beta}\rightarrow O_g^{\alpha\alpha'}O_g^{\beta\beta'}\hat{\Sigma}^{\alpha',\beta'},
\end{equation*}
and at the bilinear level have the form
\begin{equation*}
  \hat{\Sigma}^{\alpha,\beta}=\frac{1}{2}\sum_{i,j}S_i^T \Sigma_{i,j}^{\alpha,\beta} S_j,
\end{equation*}
where the 3$\times$3-matrices $\Sigma_{i,j}^{\alpha,\beta}$ determine the exchange coefficients. Analogous to the electric field perturbation above, these matrices must satisfy
\begin{equation*}
  O_g^{\alpha\alpha'}O_g^{\beta\beta'}\sum_{i,j}S_i^T \Sigma_{i,j}^{\alpha',\beta'} S_j
  =
  \sum_{i,j}S_{g^{-1}i}^T O_g^T \Sigma_{i,j}^{\alpha,\beta} O_g S_{g^{-1}j}
\end{equation*}
for any $g\in G$.

\section{Author contribution}
MK designed and carried out the search algorithm. MK and XY carried out calculations for surface and bulk magnon spectra. MK, XY and YML analyzed the search results and magnon spectra data. AW, MN, VDN and JG carried out the literature search and the screening of magnetic materials. YML designed and supervised the research. MK and YML wrote the manuscript. 

\section*{Acknowledgements}
This work is supported by Center for Emergent Materials at The Ohio State University, a National Science Foundation (NSF) MRSEC through NSF Award No. DMR-2011876.



%

\end{document}


\title{Supplementary Materials}
\author{Mohammed J. Karaki}
\email{karaki.4@osu.edu}
\affiliation{Department of Physics, The Ohio State University, Columbus, OH 43210, USA}
\author{Xu Yang}
\affiliation{Department of Physics, The Ohio State University, Columbus, OH 43210, USA}
\author{Archibald J. Williams}
\affiliation{Department of Chemistry and Biochemistry, The Ohio State University, Columbus, OH 43210, USA }
\author{Mohamed Nawwar}
\affiliation{Department of Materials Science and Engineering, The Ohio State University, Columbus, OH}
\author{Vicky Doan-Nguyen}
\affiliation{Department of Materials Science and Engineering, The Ohio State University, Columbus, OH}
\author{Joshua E. Goldberger}
\affiliation{Department of Chemistry and Biochemistry, The Ohio State University, Columbus, OH 43210, USA }
\author{Yuan-Ming Lu}
\email{lu.1435@osu.edu}
\affiliation{Department of Physics, The Ohio State University, Columbus, OH 43210, USA}
\maketitle

\onecolumngrid
\tableofcontents

\renewcommand{\theequation}{S\arabic{equation}}
\renewcommand{\figurename}{Supplementary Fig.}
\renewcommand{\tablename}{Supplementary Table}
\setcounter{equation}{0}
\setcounter{figure}{0}
\setcounter{table}{0}

\section{Spin-wave details for topological magnon candidates}
In this section we provide details about the spin-wave calculations for the systems mentioned in the main text. For each material, we describe the crystalline/magnetic structure and symmetries, the most generic symmetry-constrained exchange interactions for the first few nearest-neighbor paths, the little group irreducible representation (irrep) content of the induced magnon bands (and thus the symmetry-enforced degeneracies) at the high-symmetry $k$-points, and the effects of external perturbations on the magnetic space group symmetry as well as on the magnon spectrum, including the SI-based prediction of topological bands.

For each system, we use $\bm{r}_s$ to denote the position of magnetic sublattice $s=\mathrm{A},\mathrm{B},\cdots$, within a structural unit cell. Using 3 integers $\{x_i\}$, we label each lattice site with the coordinates $(x_1,x_2,x_3,s)$, and its position is given by
\begin{equation}
  \bm{R}(x_1,x_2,x_3,s)=\sum_{i=1}^{3}x_{i}\bm{a}_{i}+\bm{r}_{s},
\end{equation}
where $\bm{a}_1$, $\bm{a}_2$ and $\bm{a}_3$ are the primitive Bravais translations of the crystal.

A general interaction Hamiltonian bilinear in spins reads
\begin{equation}
  H=\frac{1}{2}\sum_{\bm{R}_1,\bm{R}_2}\sum_{\alpha,\beta}\mathsf{J}_{\bm{R}_1,\bm{R}_2}^{\alpha,\beta}S_{\bm{R}_1}^{\alpha}S_{\bm{R}_2}^{\beta},
\end{equation}
where $\mathsf{J}_{\bm{R}_1,\bm{R}_2}={[\mathsf{J}_{\bm{R}_2,\bm{R}_1}]}^T$ is a 3\texttimes3 real matrix of exchange couplings. For example, a Heisenberg interaction corresponds to an exchange matrix proportional to the identity. A general exchange matrix $\mathsf{J}$ can be decomposed as
\begin{equation}
  \mathsf{J}=\begin{pmatrix}
    J & \Gamma^{xy}+D^{z} & \Gamma^{xz}-D^{y}\\
    \Gamma^{xy}-D^{z} & J+J^{y} & \Gamma^{yz} + D^{x}\\
    \Gamma^{xz}+D^{y} & \Gamma^{yz}-D^{x} & J + J^{z}\\
  \end{pmatrix},
\end{equation}
where the parameters $\Gamma^{\alpha,\beta}$ ($D^{\alpha}$) describe the symmetric (anti-symmetric) exchange. In particular, $D^{\alpha}$ correspond to the Dzyaloshinskii–Moriya interaction, of the form $\bm{D}\cdot\bm{S}_{1}\times\bm{S}_{2}$.

Our conventions for labelling the Brillouin zone $k$-points as well as  the band irreps follow the notation used in the Bilbao Crystallographic Server (BCS) tables~\cite{Elcoro2021,Xu2020b}. In this notation, a band irrep label at a given $k$-point always contains the $k$-point label. For example, in $P4/mmm$ the irreps at $k=X$ are labelled $X_1^+,X_1^-,X_2^+,\cdots$.

Note that in two magnetic space groups $G_2\subset G_1$, the label of the same $k$-point is in general different. To maintain consistency with the BCS tables of little group irreps, we choose to switch our labelling accordingly when we consider symmetry-lowering perturbations. In particular, plots illustrating the effects of external fields on the magnon spectrum are affected by this relabelling.

\subsection{\texorpdfstring{TbFeO$_3$}{TbFeO3}}
\subsubsection{The spin model}

The crystal structure of TbFeO$_3$ is described by the orthorhombic space group $Pnma~(62)$, and the primitive unit cell hosts 4 Fe atoms. In Cartesian coordinates, the primitive lattice vectors are
\begin{equation}
  \bm{a}_1=(a,0,0),\qquad
  \bm{a}_2=(0,b,0),\qquad
  \bm{a}_3=(0,0,c).
\end{equation}

The four Fe sublattices are coordinated at
\begin{align*}
  \bm{r}_{\mathrm{A}} &=(0,0,c/2),\\
  \bm{r}_{\mathrm{B}} &=(a/2, b/2, 0),\\
  \bm{r}_{\mathrm{C}} &=(0, b/2, c/2),\\
  \bm{r}_{\mathrm{D}} &=(a/2, 0, 0).
\end{align*}

The space group $Pnma$ is generated by
\begin{itemize}
  \item $G_{100}^{\parallel\bm{a}_2+\bm{a}_3}$, glide plane $\bm{a_1}/4+y\bm{a}_2+z\bm{a}_3$ with a fractional translation $\bm{a}_2/2+\bm{a}_3/2$. The Fe atoms transform as
    \begin{align*}
      (x_1,x_2,x_3,s)&\longrightarrow
      \begin{cases}
        (-x_1,x_2,1+x_3,\mathrm{B}),&s=\mathrm{A}\\
        (-x_1,1+x_2,x_3,\mathrm{A}),&s=\mathrm{B}\\
        (-x_1,1+x_2,1+x_3,\mathrm{D}),&s=\mathrm{C}\\
        (-x_1,x_2,x_3,\mathrm{C}),&s=\mathrm{D}
      \end{cases}
    \end{align*}
  \item $M_{010}$, mirror plane $x\bm{a_1}+\bm{a}_2/4+z\bm{a}_3$:
    \begin{align*}
      (x_1,x_2,x_3,s)&\longrightarrow
      \begin{cases}
        (x_1,-x_2,x_3,\mathrm{C}),&s=\mathrm{A}\\
        (x_1,-x_2,x_3,\mathrm{D}),&s=\mathrm{B}\\
        (x_1,-x_2,x_3,\mathrm{A}),&s=\mathrm{C}\\
        (x_1,-x_2,x_3,\mathrm{B}),&s=\mathrm{D}
      \end{cases}
    \end{align*}
  \item $G_{001}^{\parallel\bm{a}_1}$, glide plane $x\bm{a}_1+y\bm{a}_2+\bm{a}_3/4$ with a fractional translation $\bm{a}_{1}/2$:
    \begin{align*}
      (x_1,x_2,x_3,s)&\longrightarrow
      \begin{cases}
        (x_1,x_2,-x_3,\mathrm{D}),&s=\mathrm{A}\\
        (1+x_1,x_2,-x_3,\mathrm{C}),&s=\mathrm{B}\\
        (x_1,x_2,-x_3,\mathrm{B}),&s=\mathrm{C}\\
        (1+x_1,x_2,-x_3,\mathrm{A}),&s=\mathrm{D}
      \end{cases}
    \end{align*}
\end{itemize}

The 1\textsuperscript{st} nearest-neighbor (NN) exchange coupling 
\begin{equation}
  \mathsf{J}_1:(0,0,0,\mathrm{A})\leftrightarrow(0,0,0,\mathrm{C})
\end{equation}
is constrained only by the mirror symmetry $M_{010}$, and it has the following generic form
\begin{equation}
  \mathsf{J}_1=\begin{pmatrix}
    J_{1} & D_{1}^{z} & \Gamma_{1}^{xz}\\
    -D_{1}^{z} & J_{1}+J_{1}^{y} & D_{1}^{x}\\
    \Gamma_{1}^{xz} & -D_{1}^{x} & J_{1} + J_{1}^{z}\\
  \end{pmatrix}
\end{equation}

The 2\textsuperscript{nd} NN exchange coupling is
\begin{equation}
  \mathsf{J}_2:(0,0,0,\mathrm{A})\leftrightarrow(0,0,0,\mathrm{D})
\end{equation}
Having no symmetry constraints, it is an arbitrary 3\texttimes3 matrix,
\begin{equation}
  \mathsf{J}_2=\begin{pmatrix}
    J_{2} & \Gamma_{2}^{xy}+D_{2}^{z} & \Gamma_{2}^{xz}-D_{2}^{y}\\
    \Gamma_{2}^{xy}-D_{2}^{z} & J_{2}+J_{2}^{y} & \Gamma_{2}^{yz} + D_{2}^{x}\\
    \Gamma_{2}^{xz}+D_{2}^{y} & \Gamma_{2}^{yz}-D_{2}^{x} & J_{2} + J_{2}^{z}\\
  \end{pmatrix}
\end{equation}

The 3\textsuperscript{rd} NN exchange coupling is
\begin{equation}
  \mathsf{J}_3:(0,0,0,\mathrm{D})\leftrightarrow(0,0,1,\mathrm{D})
\end{equation}
which is constrained only by inversion symmetry, and thus it is a symmetric matrix,
\begin{equation}
  \mathsf{J}_3=\begin{pmatrix}
    J_{3} & \Gamma_{3}^{xy} & \Gamma_{3}^{xz}\\
    \Gamma_{3}^{xy} & J_{3}+J_{3}^{y} & \Gamma_{3}^{yz} \\
    \Gamma_{3}^{xz} & \Gamma_{3}^{yz}& J_{3} + J_{3}^{z}\\
  \end{pmatrix}
\end{equation}

The 4\textsuperscript{th} NN exchange coupling is
\begin{equation}
  \mathsf{J}_4:(0,0,0,\mathrm{A})\leftrightarrow(0,0,0,\mathrm{B})
\end{equation}
and is subject to no constraints,
\begin{equation}
  \mathsf{J}_4=\begin{pmatrix}
    J_{4} & \Gamma_{4}^{xy}+D_{4}^{z} & \Gamma_{4}^{xz}-D_{4}^{y}\\
    \Gamma_{4}^{xy}-D_{4}^{z} & J_{4}+J_{4}^{y} & \Gamma_{4}^{yz} + D_{4}^{x}\\
    \Gamma_{4}^{xz}+D_{4}^{y} & \Gamma_{4}^{yz}-D_{4}^{x} & J_{4} + J_{4}^{z}\\
  \end{pmatrix}
\end{equation}

\subsubsection{Linear spin-wave theory and field-induced magnon topology}
\begin{figure*}[htb!]
  \centering
  \includegraphics[width=0.9\linewidth]{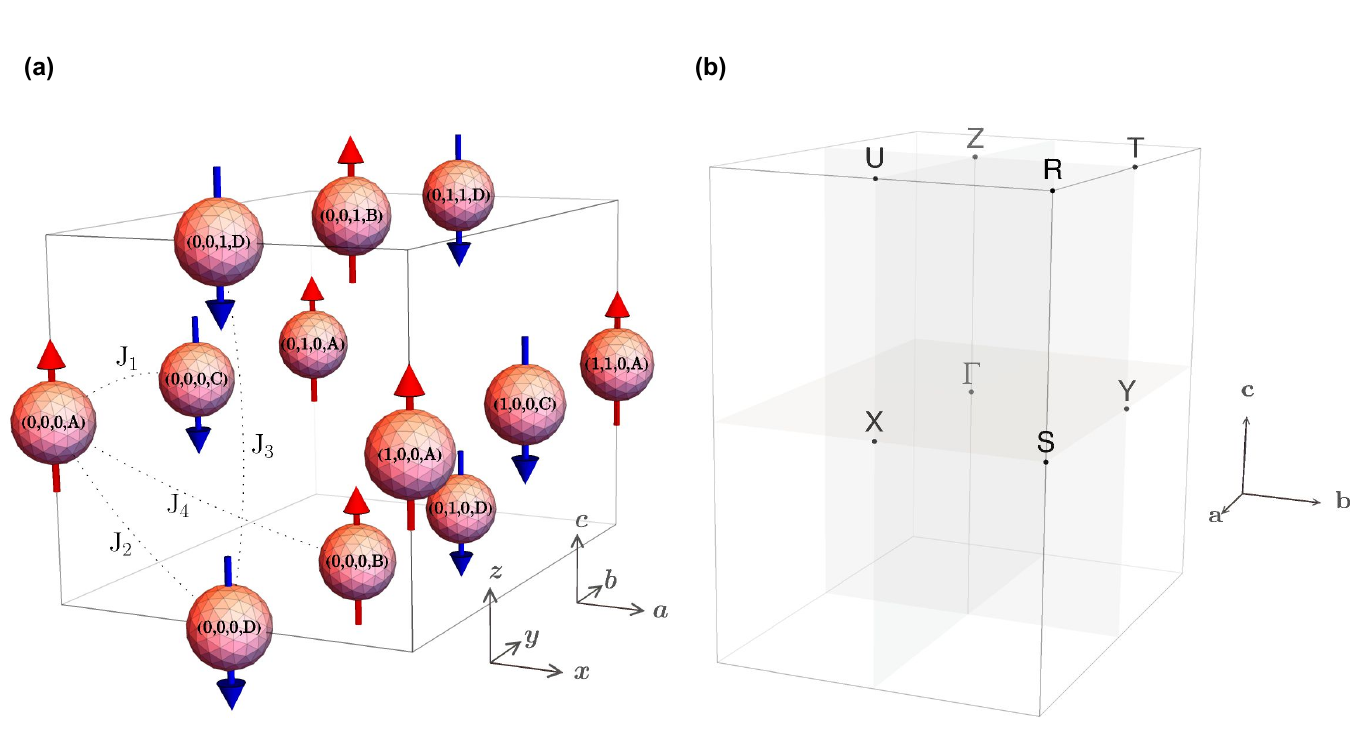}
  \caption{\textbf{The magnetic structure of TbFeO$_3$.} \textbf{a}, The magnetic unit cell with with four nearest-neighbor exchange paths illustrated. \textbf{b}, The magnetic Brillouin zone.\label{sup-fig:tbfeo3}}
\end{figure*}

At a transition temperature $T_{\mathrm{N}}=\SI{681}{\kelvin}$, the Fe atoms order antiferromagnetically as shown in \figurename~\ref{sup-fig:tbfeo3}. The moments of the A- and B-sublattices are parallel to $z$-axis, whereas the C- and D-sublattices point in the opposite direction:

\begin{equation}
  \langle\bm{S}(x_i,s)\rangle\propto \begin{cases}
    (0,0,+1)&s=\textrm{A or B}\\
    (0,0,-1)&s=\textrm{C or D}
  \end{cases}
\end{equation}

The magnetic order is invariant under the mirror symmetry $M_{010}$ but it breaks both glide symmetries, $G_{100}^{\parallel\bm{a}_2+\bm{a}_3}$ and $G_{001}^{\parallel\bm{a}_1}$. However, the combination of time-reversal with $G_{100}^{\parallel\bm{a}_2+\bm{a}_3}$ or $G_{001}^{\parallel\bm{a}_1}$ still preserves the magnetic order. Thus, the magnetic space group is $Pn'ma'~(62.448)$, with magnetic Bravais translations
\begin{equation}
  \bm{a}=\bm{a}_1,\qquad \bm{b}=\bm{a}_2,\qquad \bm{c}=\bm{a}_3,
\end{equation}
and the Brillouin zone is shown in \figurename~\ref{sup-fig:tbfeo3}.

As mentioned in the main text, to study the field-induced topology of magnon bands, our starting point is identify symmetry-protected degeneracies.

Fe$^{3+}$ ions occupy the $4b$ Wyckoff position of $Pn'ma'$, and the site-symmetry group is $\bar{1}$. The spin-wave variables transform trivially under this group, and therefore they decompose into two copies of the trivial 1D irrep $A_g$ (see \tablename~\ref{sup-tab:irreps} for the spin-wave representation in all magnetic point groups.) Using the MBANDREP tool on the Bilbao Crystallographic Server, one obtains the band representation (\tablename~\ref{sup-tab:tbfeo3-bandrep}).

\begin{table}[htb!]
  \centering
  \begin{tabular}{cc}
    \toprule
    $k$ point & $A_g\uparrow$ MSG $\downarrow$ Little Group\\
    \midrule
    $\Gamma(0,0,0)$ & $2\Gamma_1^+(1)\oplus 2\Gamma_2^+(1)$ \\[2pt]
    $R(\frac{1}{2},\frac{1}{2},\frac{1}{2})$ & $R_1R_1(4)$ \\[2pt]
    $S(\frac{1}{2},\frac{1}{2},0)$ & $S_1S_1(4)$ \\[2pt]
    $T(0,\frac{1}{2},\frac{1}{2})$ & $2T_1(2)$ \\[2pt]
    $U(\frac{1}{2},0,\frac{1}{2})$ & $U_1^-U_1^-(2)\oplus U_2^-U_2^-(2)$ \\[2pt]
    $X(\frac{1}{2},0,0)$ & $X_1^+X_2^-(2)\oplus X_1^-X_2^+(2)$ \\[2pt]
    $Y(0,\frac{1}{2},0)$ & $2Y_1(2)$ \\[2pt]
    $Z(0,0,\frac{1}{2})$ & $Z_1^+Z_2^-(2)\oplus Z_1^-Z_2^+(2)$ \\[2pt]
    \bottomrule
  \end{tabular}
  \caption{Band representation induced from the $A_g$ irrep of $\bar{1}$, the site-symmetry group of the $4b$ Wyckoff position in $Pn'ma'$. The magnon band representation in TbFeO$_3$ is induced from two copies of $A_g$ (see \tablename~\ref{sup-tab:irreps}.) The integers in parentheses indicate the dimensionality of an irrep. The $\pm$ superscripts indicate the inversion eigenvalue.\label{sup-tab:tbfeo3-bandrep}}
\end{table}

To study the field-induced topology of magnon bands, we start from protected degeneracies. From \tablename~\ref{sup-tab:tbfeo3-bandrep}, we find that the bands are four-fold degenerate at both $R$ and $S$. This degeneracy is partially protected by $\tilde{T}\cdot G_{001}^{\parallel\bm{a}_1}$, a combination of time-reversal $\tilde{T}$ and the glide symmetry $G_{001}^{\parallel\bm{a}_1}$. This anti-unitary symmetry squares to $-1$ at $R$ and $S$, resulting in effective Kramers pairs.

Next, we seek to isolate the bands by reducing the magnetic space group symmetry and evaluate the SIs of the separated bands. Using the subgroup tables in the auxiliary supplementary materials, we find that an external magnetic field along the $\bm{c}$ axis reduces the symmetry
\begin{equation}
  Pn'ma'~(62.448) \xlongrightarrow{\bm{B}\parallel [001]} P2_1'/c'~(14.79)
\end{equation}
into a subgroup with a non-trivial classification $\mathbb{Z}_2\times\mathbb{Z}_4$, in which $\mathbb{Z}_4$ is the inversion indicator.
The monoclinic axes of the subgroup $P2_1'/c'$ are obtained from the orthorhombic axes of $Pn'ma'$ by
\begin{equation}
  \begin{pmatrix}
    \bm{a}\\
    \bm{b}\\
    \bm{c}
  \end{pmatrix}
  \rightarrow
  \begin{pmatrix}
    &&1\\
    1&&\\
    &1&-1
  \end{pmatrix}
  \begin{pmatrix}
    \bm{a}\\
    \bm{b}\\
    \bm{c}
  \end{pmatrix}
\end{equation}

Fig.~\ref{sup-fig:tbfeo3}c-d show the magnon spectrum before and after an applied field, using parameters $J_1=1.00$, $J_2=0.44$, $J_4=-0.30$, $J_1^y=-0.19$, $J_1^z=0.32$, $J_2^y=-0.21$, $\Gamma_2^{xy}=0.10$, $\Gamma_4^{xy}=0.05$, $\Gamma_6^{xy}=0.04$ and magnetic field $h=0.5$. The separated bands acquire an odd value of $\mathbb{Z}_4$, signaling Weyl magnons in the bulk. This is because the upper two bands started with an even number of inversion-odd eigenvalues at the gapped TRIM points without an external field, and the field-induced degeneracy splitting (at $R$ and $S$) can only change this count by an odd number, see \tablename~\ref{sup-tab:tbfeo3-splitting}.

\begin{table}[t!]
  \centering
  \begin{tabular}{ccc}
    \toprule
    $k$-label &  Effect of $\bm{B}\parallel[001]$ & Contribution to $\mathbb{Z}_{4}$ \\
    ($Pn'ma'$) & on irreps & of higher two bands\\
    \midrule
    $R$ & $R_1R_1\longrightarrow C_1^-C_1^+\oplus C_1^-C_1^+$ & 1 (always)\\
    $S$ & $S_1S_1\longrightarrow D_1^+D_1^+\oplus D_1^-D_1^-$ & 0 or 2\\
    \bottomrule
  \end{tabular}
  \caption{Effect of a magnetic field perturbation along the $\bm{c}$ axis of $Pn'ma'$ on the four-fold magnon band degeneracies at $k=R$ and $S$, as well as on the $\mathbb{Z}_4$ inversion SI.\label{sup-tab:tbfeo3-splitting}}
\end{table}

This conclusion resulted purely from symmetry, and it is valid for generic spin models and any perturbation leading to the same subgroup $P2_1'/c'$, such as a uniaxial strain in the $\bm{b}-\bm{c}$ plane.

\subsection{\texorpdfstring{$\alpha$-Fe$_2$O$_3$\label{sec:fe2o3}}{alpha-Fe2O3}}
\subsubsection{The spin model}
The crystal structure of $\alpha$-Fe$_2$O$_3$ is described by the trigonal space group $R\bar{3}c~(167)$, and its primitive rhombohedral lattice vectors can be chosen as
\begin{equation}
  \begin{pmatrix}
    \bm{a}_1\\
    \bm{a}_2\\
    \bm{a}_3
  \end{pmatrix}
  =
  \frac{1}{3}\begin{pmatrix}
    2&1&1\\
    -1&1&1\\
    -1&-2&1\\
  \end{pmatrix}
  \begin{pmatrix}
    \bm{a}\\
    \bm{b}\\
    \bm{c}
  \end{pmatrix},
\end{equation}
where
\begin{equation}
  \bm{a}=a(1,0,0),\qquad
  \bm{b}=\frac{a}{2}(-1,\sqrt{3},0),\qquad
  \bm{c}=c(0,0,1).
\end{equation}
are the conventional hexagonal axes. This choice leads to a simple transformation under a $3$-fold rotation:
\begin{equation}
  \bm{a}_1\rightarrow\bm{a}_2\rightarrow\bm{a}_3\rightarrow\bm{a}_1.
\end{equation}

With $\alpha\approx 0.355322$, the four Fe$^{3+}$ ions are coordinated at
\begin{align*}
  \bm{r}_{\mathrm{A}} &=\alpha(\bm{a}_1+\bm{a}_2+\bm{a}_3),\\
  \bm{r}_{\mathrm{B}} &=\Bigl(\frac{1}{2}-\alpha\Bigr)(\bm{a}_1+\bm{a}_2+\bm{a}_3),\\
  \bm{r}_{\mathrm{C}} &=(1-\alpha)(\bm{a}_1+\bm{a}_2+\bm{a}_3),\\
  \bm{r}_{\mathrm{D}} &=\Bigl(\frac{1}{2}+\alpha\Bigr)(\bm{a}_1+\bm{a}_2+\bm{a}_3).
\end{align*}

The space group $R\bar{3}c$ is generated by
\begin{itemize}
  \item $i$: inversion about the cell center $(\bm{a}_1+\bm{a}_2+\bm{a}_3)/2$. This swaps $\mathrm{A}\leftrightarrow\mathrm{C}$ and $\mathrm{B}\leftrightarrow\mathrm{D}$ within the unit cell.
  \item $3_{001}$: a three-fold axis along $\bm{c}=(\bm{a}_1+\bm{a}_2+\bm{a}_3)$:
    \begin{align*}
      (x_1,x_2,x_3,s)&\longrightarrow
      \begin{cases}
        (x_3,x_1,x_2,\mathrm{A}),&s=\mathrm{A}\\
        (x_3,x_1,x_2,\mathrm{B}),&s=\mathrm{B}\\
        (x_3,x_1,x_2,\mathrm{C}),&s=\mathrm{C}\\
        (x_3,x_1,x_2,\mathrm{D}),&s=\mathrm{D}
      \end{cases}
    \end{align*}
  \item $G_{100}^{\parallel\bm{c}}$: a glide plane $x(\bm{a_1}+\bm{a}_2)+z\bm{a}_3$, with a fractional translation $(\bm{a}_1+\bm{a}_2+\bm{a}_3)/2$:
    \begin{align*}
      (x_1,x_2,x_3,s)&\longrightarrow
      \begin{cases}
        (x_2,x_1,x_3,\mathrm{D}),&s=\mathrm{A}\\
        (x_2,x_1,x_3,\mathrm{C}),&s=\mathrm{B}\\
        (1+x_2,1+x_1,1+x_3,\mathrm{B}),&s=\mathrm{C}\\
        (1+x_2,1+x_1,1+x_3,\mathrm{A}),&s=\mathrm{D}
      \end{cases}
    \end{align*}
\end{itemize}

We consider the exchange paths for the first 10 nearest neighbor bonds:
\begin{align}
  \mathsf{J}_1&:(0,0,0,\mathrm{A})\leftrightarrow(0,0,0,\mathrm{B})\\
  \mathsf{J}_2&:(0,0,0,\mathrm{A})\leftrightarrow(0,0,-1,\mathrm{C})\\
  \mathsf{J}_3&:(0,0,0,\mathrm{C})\leftrightarrow(0,0,-1,\mathrm{D})\\
  \mathsf{J}_4&:(0,0,0,\mathrm{A})\leftrightarrow(0,0,-1,\mathrm{D})\\
  \mathsf{J}_5&:(0,0,0,\mathrm{A})\leftrightarrow(0,0,0,\mathrm{C})\\
  \mathsf{J}_6&:(0,0,0,\mathrm{A})\leftrightarrow(0,-1,1,\mathrm{A})\\
  \mathsf{J}_7&:(0,0,0,\mathrm{A})\leftrightarrow(0,0,-1,\mathrm{A})\\
  \mathsf{J}_8&:(0,0,0,\mathrm{A})\leftrightarrow(0,-1,1,\mathrm{B})\\
  \mathsf{J}_9&:(0,0,0,\mathrm{A})\leftrightarrow(-1,-1,1,\mathrm{C})\\
  \mathsf{J}_{10}&:(0,0,0,\mathrm{A})\leftrightarrow(0,-1,-1,\mathrm{C})
\end{align}

The 1\textsuperscript{st} NN exchange is constrained by 3-fold and 2-fold axes, parallel and perpendicular to the bond axis, respectively. The 2\textsuperscript{nd} NN bond is invariant only under inversion, and the exchange matrix is symmetric. The 3\textsuperscript{rd} NN exchange is constrianed by a 2-fold axis perpendicular to the bond. The 4\textsuperscript{th} NN exchange is subject to no constraints. Finally, the 5\textsuperscript{th} NN bond is constrained by inversion and a 3-fold axis.

The following is the generic form of the first five exchange matrices.
\begin{equation}
  \mathsf{J}_1=\begin{pmatrix}
    J_{1} & +D_{1}^{z} & \\
    -D_{1}^{z} & J_{1} & \\
    &  & J_{1} + J_{1}^{z}\\
  \end{pmatrix}\qquad
  \mathsf{J}_2=\begin{pmatrix}
    J_{2} & \Gamma_{2}^{xy} & \Gamma_{2}^{xz}\\
    \Gamma_{2}^{xy} & J_{2}+J_{2}^{y} & \Gamma_{2}^{yz}\\
    \Gamma_{2}^{xz} & \Gamma_{2}^{yz} & J_{2} + J_{2}^{z}\\
  \end{pmatrix}
\end{equation}

\begin{equation}
  \mathsf{J}_3=\begin{pmatrix}
    J_{3} & +D_{3}^{z} & -D_{3}^{y}\\
    -D_{3}^{z} & J_{3}+J_{3}^{y} & \Gamma_{3}^{yz}\\
    +D_{3}^{y} & \Gamma_{3}^{yz} & J_{3} + J_{3}^{z}\\
  \end{pmatrix}\qquad
  \mathsf{J}_4=\begin{pmatrix}
    J_{4} & \Gamma_{4}^{xy}+D_{4}^{z} & \Gamma_{4}^{xz}-D_{4}^{y}\\
    \Gamma_{4}^{xy}-D_{4}^{z} & J_{4}+J_{4}^{y} & \Gamma_{4}^{yz} + D_{4}^{x}\\
    \Gamma_{4}^{xz}+D_{4}^{y} & \Gamma_{4}^{yz}-D_{4}^{x} & J_{4} + J_{4}^{z}\\
  \end{pmatrix}
\end{equation}

\begin{equation}
  \mathsf{J}_5=\begin{pmatrix}
    J_{5} & &\\
    & J_{5} &\\
    & & J_{5} + J_{5}^{z}\\
  \end{pmatrix}
\end{equation}

\subsubsection{Linear spin-wave spectrum and topology}
\begin{figure*}[htb!]
  \centering
  \includegraphics[width=0.8\linewidth]{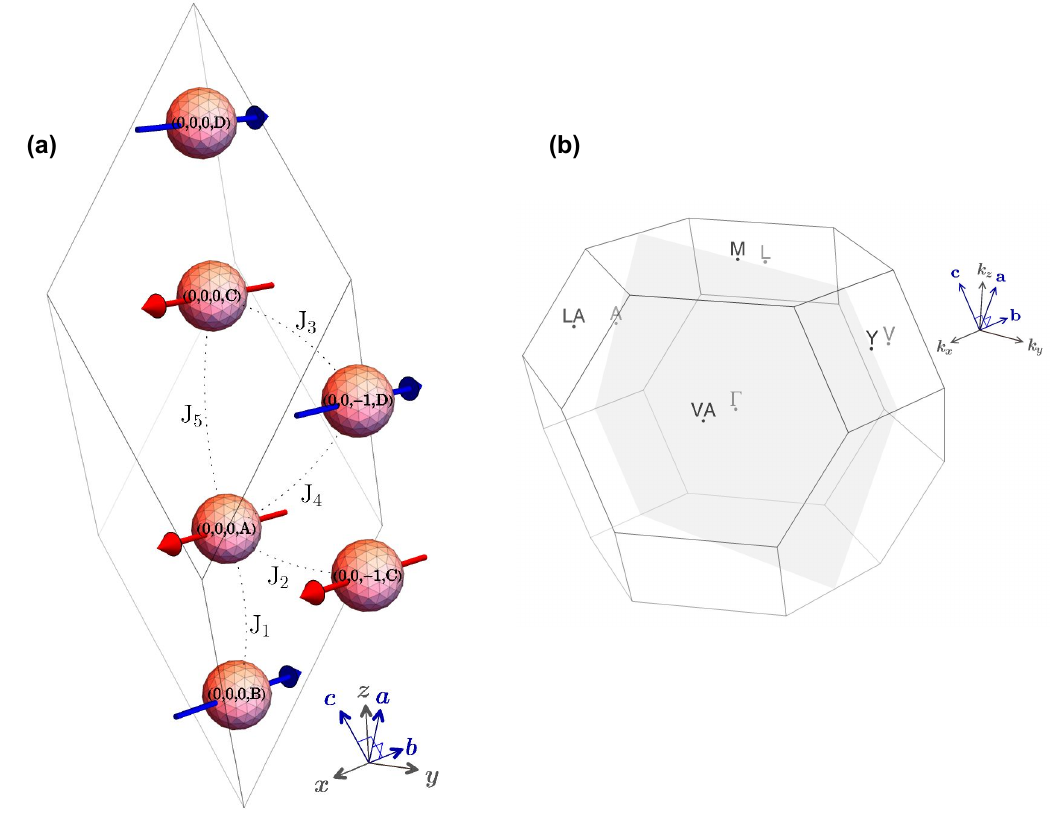}
  \caption{\textbf{The magnetic structure of Fe$_2$O$_3$.} \textbf{a}, The magnetic unit cell with with five nearest-neighbor exchange paths illustrated. \textbf{b}, The magnetic Brillouin zone.\label{sup-fig:fe2o3}}
\end{figure*}

The magnetic order is collinear AFM with magnetic moments along $\bm{a}=\bm{a}_1-\bm{a}_2\propto(1,0,0)$, with $T_{\mathrm{N}}=\SI{955}{\kelvin}$. More precisely,
\begin{equation}
  \langle\bm{S}(x_i,s)\rangle\propto \begin{cases}
    (+1,0,0)&s=\textrm{A or C}\\
    (-1,0,0)&s=\textrm{B or D}
  \end{cases}
\end{equation}
This is illustrated in \figurename~\ref{sup-fig:fe2o3}.

The magnetic order preserves inversion, an anti-unitary 2-fold axis and an anti-unitary glide. The magnetic space group is $C2'/c'~(15.89)$, whose conventional monoclinic axes are
\begin{equation}
  \bm{a}_m=\bm{a}_1+\bm{a}_2,\quad
  \bm{b}_m=-\bm{a}_1+\bm{a}_2,\quad
  \bm{c}_m=\bm{a}_3,
\end{equation}
with $\bm{b}_m$ along the 2-fold axis of $C2'/m'$. Here, we use the subscript $m$ to distinguish these magnetic lattice axes from those of the parent hexagonal lattice.

The four magnetic Fe sublattices occupy the $8f$ Wyckoff position of $C2'/c'$, which has a trivial site-symmetry group, and thus the spin-wave variables transform trivially on-site. 
\tablename~\ref{sup-tab:fe2o3-bandrep} summarizes the induced magnon band representation, restricted to the little groups of high-symmetry $k$-points.

\begin{table}[htb!]
  \centering
  \begin{tabular}{cc}
    \toprule
    $k$-point & $A\uparrow$ MSG $\downarrow$ Little Group\\
    \midrule
    $\Gamma(0,0,0)$ & $2\Gamma_1^+(1)\oplus 2\Gamma_1^-(1)$\\[2pt]
    $A(0,0,\frac{1}{2})$ & $2A_1^-A_1^+(2)$\\[2pt]
    $L(\frac{1}{2},\frac{1}{2},\frac{1}{2})$ & $2L_1^+(1)\oplus 2L_1^-(1)$\\[2pt]
    $LA(\frac{1}{2},-\frac{1}{2},\frac{1}{2})$ & $2LA_1^-(1)\oplus 2LA_1^+(1)$\\[2pt]
    $M(0,1,\frac{1}{2})$ & $2M_1^-M_1^+(2)$\\[2pt]
    $V(\frac{1}{2},\frac{1}{2},0)$ & $2V_1^+(1)\oplus 2V_1^-(1)$\\[2pt]
    $VA(\frac{1}{2},-\frac{1}{2},0)$ & $2VA_1^+(1)\oplus 2VA_1^-(1)$\\[2pt]
    $Y(0,1,0)$ & $2Y_1^+(1)\oplus 2Y_1^-(1)$\\
    \bottomrule
  \end{tabular}
  \caption{Band representation induced from the $A$ irrep of $1$, the site-symmetry group of the $8f$ Wyckoff position in $C2'/c'$. The magnon band representation in $\alpha$-Fe$_2$O$_3$ is induced from two copies of $A$ (see \tablename~\ref{sup-tab:irreps}.)\label{sup-tab:fe2o3-bandrep}}
\end{table}
According to \tablename~\ref{sup-tab:fe2o3-bandrep}, the highest degeneracy at the inversion-symmetric points is two-fold, and therefore we expect two pairs of bands that are energetically isolated at least at the high-symmetry points, for a generic Hamiltonian. It turns out that 2\textsuperscript{nd} NN Heisenberg term $J_2$ alone is sufficient open up this gap.

Respectively, $J_1$ through $J_{10}$ have the following experimental estimates~\cite{fe2o3-spectrum} in meV units: $-1.03$, $-0.28$, $5.12$, $4.00$, $0.17$, $-0.09$, $0.02$, $0.05$, $0.02$ and $0.00$. We supplement this with an on-site anisotropy $J_0^{zz}=0.12$ meV, a symmetric interaction $\Gamma_{3}^{yz}=0.10$ meV and an antisymmetric term $D_{3}^{y}=0.05$ meV to produce the spectrum in Figure 3(c). Finally, the perturbed spectrum in Figure 3(d) corresponds to magnetic field $B=2.5$ T and $g=2$.

\subsection{\texorpdfstring{CoF$_3$}{CoF3}\label{sup-subsec:cof3}}
\subsubsection{The spin model}
The crystal symmetry is described by the trigonal space group $R\bar{3}c~(167)$, and below we work with a rhombohedral primitive cell, and Bravais translations
\begin{equation}
  \begin{pmatrix}
    \bm{a}_1\\
    \bm{a}_2\\
    \bm{a}_3
  \end{pmatrix}
  =
  \frac{1}{3}\begin{pmatrix}
    2&1&1\\
    -1&1&1\\
    -1&-2&1\\
  \end{pmatrix}
  \begin{pmatrix}
    \bm{a}\\
    \bm{b}\\
    \bm{c}
  \end{pmatrix},
\end{equation}
where
\begin{equation}
  \bm{a}=\frac{a}{\sqrt{2}}(0,1,-1),\qquad
  \bm{b}=\frac{a}{\sqrt{2}}(-1,0,1),\qquad
  \bm{c}=\frac{c}{\sqrt{3}}(1,1,1).
\end{equation}
are the conventional hexagonal axes. The two Co atoms are located on the 3-fold axis. Specifically
\begin{align*}
  \bm{r}_{\mathrm{A}} &=(0,0,0),\\
  \bm{r}_{\mathrm{B}} &=\frac{1}{2}(\bm{a}_1+\bm{a}_2+\bm{a}_3).
\end{align*}

\begin{figure*}[b!]
  \centering
  \hfill
  \includegraphics[width=0.95\linewidth]{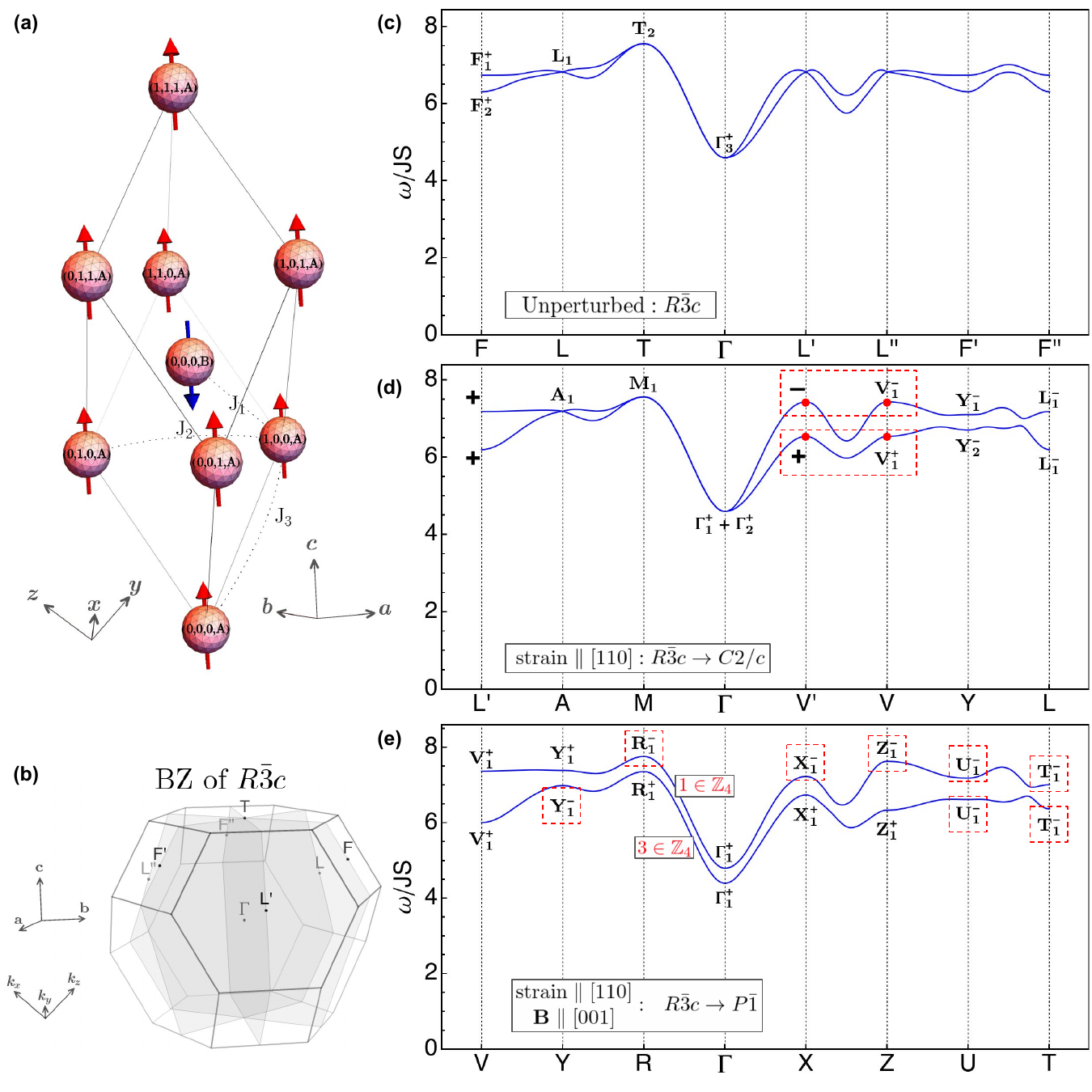}
  \hfill
  \caption{\textbf{Topological magnons in CoF$_3$ induced by a combination of strain and magnetic field.} \textbf{a}-\textbf{b}, The magnetic structure of CoF$_3$ and the Brillouin zone. \textbf{c}, Unperturbed magnon spectrum with $J_1=1.00$ and $\Gamma_2^{yz}=-0.26$. \textbf{d}, Magnon spectrum in the presence of Heisenberg interactions induced by uni-axial strain in the $[110]$ direction, reducing $R\bar{3}c$ into $C2/c$. At each of $V$ and $V'$ in \textbf{d}, there are two non-degenerate states with different contributions to $\mathbb{Z}_4$. However, thanks to the glide symmetry relating $k$-points $V$ and $V'$, their states show up in pairs of equal energy and identical inversion eigenvalues, rendering their contribution to $\mathbb{Z}_4$ independent of the spin model. \textbf{e}, Further applying a magnetic field in the $[001]$ direction gaps out the remaining degeneracies and results in two bands with $1,3\in\mathbb{Z}_4$ inversion SI, signaling Weyl magnons. (Perturbation parameters $\sigma=0.5$, $\beta_3=1.2$ and magnetic field $h=0.2$.)\label{sup-fig:cof3}}
\end{figure*}

We already discussed the generators of $R\bar{3}c$ in Sec.~\ref{sec:fe2o3}, and the following describes how they transform Co lattice sites.
\begin{itemize}
  \item Under cell-centered inversion $i$,
    \begin{align*}
      (x_1,x_2,x_3,s)&\longrightarrow
      \begin{cases}
        (1-x_1,1-x_2,1-x_3,\mathrm{A}),&s=\mathrm{A}\\
        (-x_1,-x_2,-x_3,\mathrm{B}),&s=\mathrm{B}
      \end{cases}
    \end{align*}
  \item Under three-fold axis $3_{\bm{c}}$,
    \begin{align*}
      (x_1,x_2,x_3,s)&\longrightarrow
      \begin{cases}
        (x_3,x_1,x_2,\mathrm{A}),&s=\mathrm{A}\\
        (x_3,x_1,x_2,\mathrm{B}),&s=\mathrm{B}
      \end{cases}
    \end{align*}
  \item Under glide $G_{\perp\bm{a}}^{\parallel\bm{c}}$,
    \begin{align*}
      (x_1,x_2,x_3,s)&\longrightarrow
      \begin{cases}
        (x_2,x_1,x_3,\mathrm{B}),&s=\mathrm{A}\\
        (1+x_2,1+x_1,1+x_3,\mathrm{A}),&s=\mathrm{B}
      \end{cases}
    \end{align*}
\end{itemize}

The three nearest-neighbor exchange paths
\begin{align}
  \mathsf{J}_1&:(1,0,0,\mathrm{A})\leftrightarrow(0,0,0,\mathrm{B})\\
  \mathsf{J}_2&:(1,0,0,\mathrm{A})\leftrightarrow(0,1,0,\mathrm{A})\\
  \mathsf{J}_3&:(0,0,0,\mathrm{B})\leftrightarrow(1,0,0,\mathrm{B})
\end{align}
have the generic form
\begin{equation}
  \mathsf{J}_1=\begin{pmatrix}
    J_{1} & \Gamma_{1}^{xy}+D_{1}^{z} & \Gamma_{1}^{xz}-D_{1}^{y}\\
    \Gamma_{1}^{xy}-D_{1}^{z} & J_{1}+J_{1}^{y} & \Gamma_{1}^{xy}+D_{1}^{z}\\
    \Gamma_{1}^{xz}+D_{1}^{y}& \Gamma_{1}^{xy}-D_{1}^{z}& J_{1}\\
  \end{pmatrix}
\end{equation}
\begin{equation}
  \mathsf{J}_2=\begin{pmatrix}
    J_{2} & \Gamma_{2}^{xy} & \Gamma_{2}^{xz}\\
    \Gamma_{2}^{xy} & J_{2}+J_{2}^{y} & \Gamma_{2}^{yz}\\
    \Gamma_{2}^{xz} & \Gamma_{2}^{yz} & J_{2} + J_{2}^{z}\\
  \end{pmatrix}
\end{equation}
\begin{equation}
  \mathsf{J}_3=\begin{pmatrix}
    J_{3} & \Gamma_{3}^{xy} & \Gamma_{3}^{xz}\\
    \Gamma_{3}^{xy} & J_{3}+J_{3}^{y} & \Gamma_{3}^{yz}\\
    \Gamma_{3}^{xz} & \Gamma_{3}^{yz} & J_{3} + J_{3}^{z}\\
  \end{pmatrix}
\end{equation}

\begin{table}[htb!]
  \centering
  \begin{tabular}{ccc}
    \toprule
    $k$ point & ${}^1E_g\uparrow$ MSG $\downarrow G_{\bm{k}}$ & ${}^1E_g\uparrow$ MSG $\downarrow G_{\bm{k}}$\\
    \midrule
    $\Gamma(0,0,0)$ & $\Gamma_3^+(2)$ & $\Gamma_3^+(2)$\\[2pt]
    $F(0,\frac{1}{2},1)$ & $F_1^+(1)\oplus F_2^+(1)$ & $F_1^+(1)\oplus F_2^+(1)$\\[2pt]
    $L(-\frac{1}{2},\frac{1}{2},\frac{1}{2})$ & $L_1(2)$ & $L_1(2)$\\[2pt]
    $T(0,0,\frac{3}{2})$ & $T_1(2)$ & $T_2(2)$\\
    \bottomrule
  \end{tabular}
  \caption{Band representation induced from the ${}^1E_g$ and ${}^2E_g$ irreps of $\bar{3}$, the site-symmetry group of the $6b$ Wyckoff position in $R\bar{3}c$. The magnon band representation in CoF$_3$ is induced from ${}^1E_g\oplus {}^2E_g$ (see \tablename~\ref{sup-tab:irreps}.)\label{sup-tab:cof3-bandrep}}
\end{table}

Additionally, we consider interactions induced by mechanical strain. With $\sigma_{\alpha,\beta}$ denoting the components of the strain matrix $\sigma=\sigma^T$, the generic form of the 2\textsuperscript{nd} NN Heisenberg interactions induced by strain reads
\begin{align}
  H'&=\sum_{\bm{x}}\biggl\{\\
  &\Bigl(\vec{S}_{\bm{x}+\bm{a}_1,\mathrm{A}}\cdot\vec{S}_{\bm{x}+\bm{a}_2,\mathrm{A}}+[\mathrm{A}\leftrightarrow\mathrm{B}]\Bigr)\Bigl(\alpha_1\sigma_{xx}+\alpha_2\sigma_{yy}+\alpha_3\sigma_{zz}+\beta_1\sigma_{xy}+\beta_2\sigma_{yz}+\beta_3\sigma_{zx}\Bigr)\\
  +&\Bigl(\vec{S}_{\bm{x}+\bm{a}_2,\mathrm{A}}\cdot\vec{S}_{\bm{x}+\bm{a}_3,\mathrm{A}}+[\mathrm{A}\leftrightarrow\mathrm{B}]\Bigr)\Bigl(\alpha_1\sigma_{yy}+\alpha_2\sigma_{zz}+\alpha_3\sigma_{xx}+\beta_1\sigma_{yz}+\beta_2\sigma_{zx}+\beta_3\sigma_{xy}\Bigr)\\
  +&\Bigl(\vec{S}_{\bm{x}+\bm{a}_3,\mathrm{A}}\cdot\vec{S}_{\bm{x}+\bm{a}_1,\mathrm{A}}+[\mathrm{A}\leftrightarrow\mathrm{B}]\Bigr)\Bigl(\alpha_1\sigma_{zz}+\alpha_2\sigma_{xx}+\alpha_3\sigma_{yy}+\beta_1\sigma_{zx}+\beta_2\sigma_{xy}+\beta_3\sigma_{yz}\Bigr)\biggr\}.
\end{align}

For the strain perturbation we consider below (uniaxial $\parallel[110]$), the 1NN Heisenberg terms do not lift any degeneracy at the TRIM points, and we do not include their explicit form here.

\subsubsection{Spin-wave spectrum and topology}
The AFM order of CoF$_3$ is illustrated in \figurename~\ref{sup-fig:cof3}a, and it preserves the entire space group. The magnetic space group is thus $R\bar{3}c~(167.103)$.

The magnetic atoms occupy the $6b$ Wyckoff position, and the spin-wave variables transform in the ${}^1E_g\oplus{}^{2}E_g$ representation of the site-symmetry group $\bar{3}$, i.e.\@ two inversion-even irreps with $C_3=e^{\pm 2\pi/3}$. The band representations induced from ${}^1E_g$ and ${}^2E_g$ are summarized in \tablename~\ref{sup-tab:cof3-bandrep}. Inspecting the irrep dimensions, we find that the two magnon bands are connected at four TRIM points (the $k$-stars of $\Gamma$, $L$ and $T$).  According to the subgroup tables (see the auxiliary supplementary materials), all magnetic subgroups with a non-trivial classification are centrosymmetric, and thus we need only consider magnetic field and strain perturbations to lift these degeneracies. Now we describe how Weyl magnons are induced by a combination of these perturbations.

First, the application of uniaxial strain in the $[110]$ direction breaks down $R\bar{3}c\rightarrow C2/c~(15.85)$ with conventional axes (in the $b$-unique setting)
\begin{align}
  \begin{pmatrix}
    \bm{a}\\
    \bm{b}\\
    \bm{c}
  \end{pmatrix}
  \rightarrow
  \begin{pmatrix}
 1/3& -1/3& 2/3\\
 -1& -1& 0\\
 1/3& -1/3& -1/3
  \end{pmatrix}
  \begin{pmatrix}
    \bm{a}\\
    \bm{b}\\
    \bm{c}
  \end{pmatrix}
\end{align}

As illustrated in \figurename~\ref{sup-fig:cof3}d, two (out of four) TRIM degeneracies are lifted (at $k=V,V'$). What is the contribution of the new states at $k=V,V'$ to the inversion $\mathbb{Z}_4$ of the higher band? Thanks to the glide symmetry of $C2/c$, the states at $V$ and $V'$ with the same energy have matching inversion eigenvalues, and thus their total contribution to $\mathbb{Z}_4$ must be an even integer, regardless of the details of strain-induced spin interactions.

Finally, if we additionally apply a small magnetic field along $[001]$ of $R\bar{3}c$, the remaining two TRIM degeneracies are lifted, with $+1$ contribution to the higher band (\figurename~\ref{sup-fig:cof3}e). In total, each band is assigned $1,3\in\mathbb{Z}_4$, signaling Weyl magnons.

Note that another choice of strain direction ($\perp[110]$) breaks down $R\bar{3}c$ into the same $C2/c$ group, and therefore also leads to Weyl magnons. A different $C2/c$ group can be obtained with strain $\parallel[100]$ or $\perp[100]$, and it turns out a similar argument also predicts Weyl magnons upon an additional $\bm{B}\parallel[001]$ perturbation.

\subsection{\texorpdfstring{FeF$_3$}{FeF3}}

\begin{figure*}[ht]
  \centering
  \includegraphics[width=0.8\linewidth]{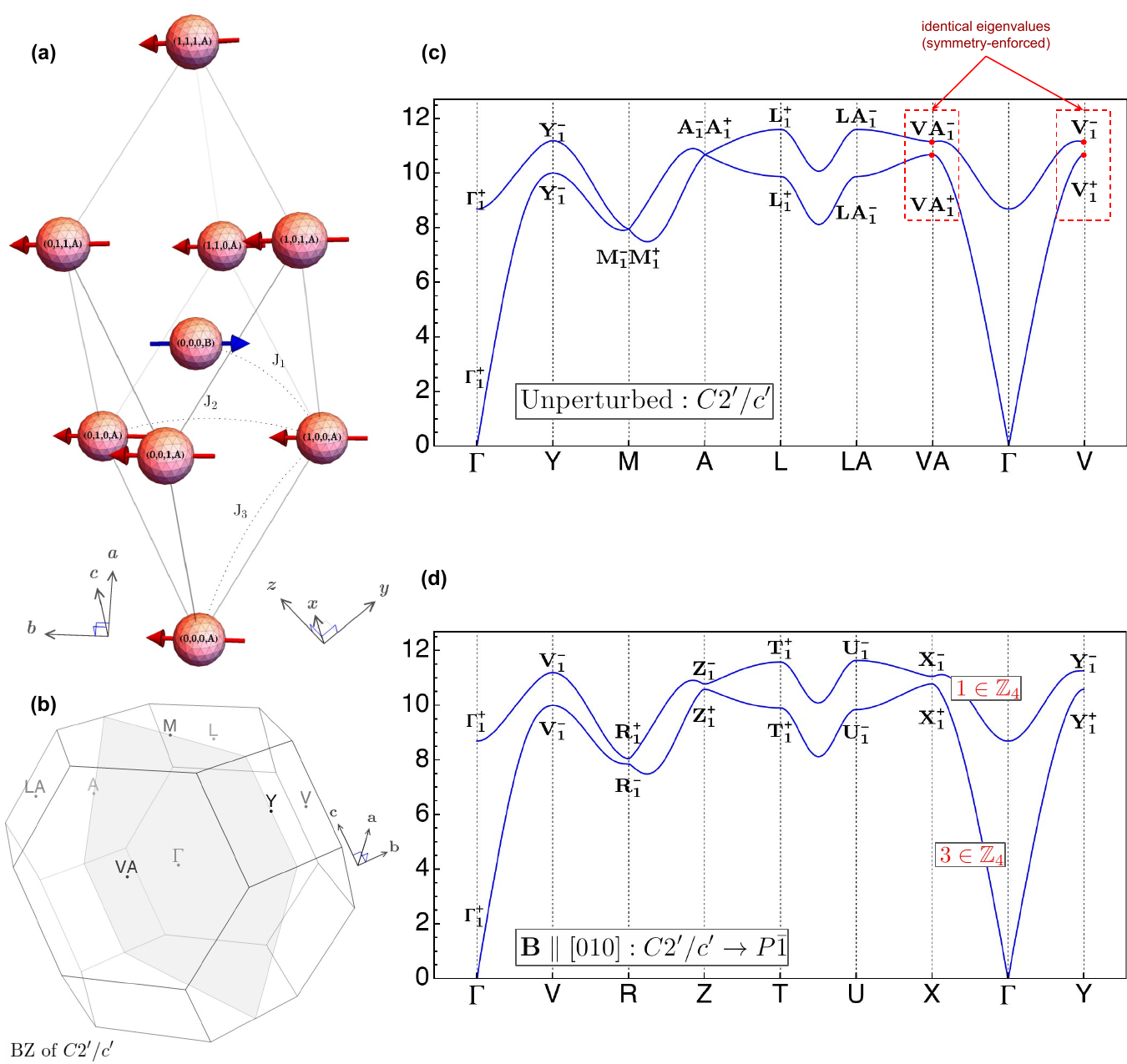}
  \caption{\textbf{Weyl magnons in FeF$_3$ induced by a magnetic field.} \textbf{a}-\textbf{b}, The magnetic structure of FeF$_3$ and the Brillouin zone. \textbf{c}, Unperturbed magnon spectrum with $J_1=1.00$, $J_2=-0.40$, $\Gamma_1^{xy}=-0.31$ and  $\Gamma_2^{xy}=0.25$. For any choice of spin model, the total contribution of $k=\Gamma,Y,L,\mathit{LA}$ to $\mathbb{Z}_4$ of the higher band is fixed, as only one parity of inversion arises at each of these points. In contrast, both parities appear at $k=V$ and $k=\mathit{VA}$. However, $V$ and $\mathit{VA}$ are symmetry-related, constraining their total contribution to $0$, or $2$. \textbf{d}, Magnon spectrum after a $\bm{B}\parallel[010]$ perturbation ($h=0.1$), leading to $1,3\in\mathbb{Z}_4$ inversion SI, signaling Weyl magnons.\label{sup-fig:fef3}}
\end{figure*}

The space group symmetry ($R\bar{3}c$) and the magnetic atom coordinates are identical to CoF$_3$ in Sec.~\ref{sup-subsec:cof3}. However the magnetic structure (\figurename~\ref{sup-fig:fef3}) is different, having monoclinic $C2'/c'~(15.89)$ symmetry. In contrast to CoF$_3$, note that the 3-fold axis is broken by the magnetic order, leading to a Goldstone mode in bilinear spin models.

The conventional axes of $C2'/c'$ (in the $b$-unique setting) can be obtained from the hexagonal axes of $R\bar{3}c$ by
\begin{align}
  \begin{pmatrix}
    \bm{a}\\
    \bm{b}\\
    \bm{c}
  \end{pmatrix}
  \rightarrow
  \begin{pmatrix}
 1/3& 2/3& 2/3\\
 -1& 0& 0\\
 -1/3& -2/3& 1/3
  \end{pmatrix}
  \begin{pmatrix}
    \bm{a}\\
    \bm{b}\\
    \bm{c}
  \end{pmatrix}
\end{align}

The magnetic atoms occupy the $4d$ Wyckoff position of $C2'/c'$, with site-symmetry group $\bar{1}$. The spin-wave variables consist of two copies of the even irrep (i.e.\@ $A_g\oplus A_g$) of $\bar{1}$, and the induced band representation is described in \tablename~\ref{sup-tab:fef3-bandrep}. We find two degenerate irreps at the TRIM points $k=M,A$.
\begin{table}[htb!]
  \centering
  \begin{tabular}{cc}
    \toprule
    $k$ point & $A_g\uparrow$ MSG $\downarrow G_{\bm{k}}$\\
    \midrule
    $\Gamma(0,0,0)$ & $2 \Gamma_1^+(1)$\\[2pt]
    $A(0,0,\frac{1}{2})$ & $A_1^-A_1^+(2)$\\[2pt]
    $L(\frac{1}{2},\frac{1}{2},\frac{1}{2})$ & $2L_1^+(1)$\\[2pt]
    $LA(\frac{1}{2},-\frac{1}{2},\frac{1}{2})$ & $2LA_1^-(1)$\\[2pt]
    $M(0,1,\frac{1}{2})$ & $M_1^-M_1^+(2)$\\[2pt]
    $V(\frac{1}{2},\frac{1}{2},0)$ & $V_1^-(1)\oplus V_1^+(1)$\\[2pt]
    $VA(\frac{1}{2},-\frac{1}{2},0)$ & $VA_1^-(1)\oplus VA_1^+(1)$\\[2pt]
    $Y(0,1,0)$ & $2 Y_1^+(1)$\\
    \bottomrule
  \end{tabular}
  \caption{Band representation induced from the $A_g$ irrep of $\bar{1}$, the site-symmetry group of the $4d$ Wyckoff position in $C2'/c'$. The magnon band representation in FeF$_3$ is induced from two copies of $A_g$ (see \tablename~\ref{sup-tab:irreps}.)\label{sup-tab:fef3-bandrep}}
\end{table}

Before the system is perturbed, the gapped TRIM points $k=\Gamma,L$ have no contribution to the inversion $\mathbb{Z}_4$ index of the higher band, while each of $k=Y,LA$ contributes $+1$, independently of the spin model  (as only one type of inversion parity arises at each of these points). In contrast, the points $k=V,\mathit{VA}$ individually contribute wither $0$ or $+1$, depending on the model details. However, thanks to the anti-unitary glide $c'$ of $C2'/c'$, the combined contributions of $V$ and $\mathit{VA}$ always add up to an even number. The magnon spectrum and irrep labels are illustrated in \figurename~\ref{sup-fig:fef3}c.

A magnetic field perturbation ($\bm{B}\parallel[010]$) along the moment directions breaks down $C2'/c'$ into $P\bar{1}$, lifting the two degeneracies with a $+1$ contribution to $\mathbb{Z}_4$. In total, the $\mathbf{Z}_4$ is odd for each band, diagnosing Weyl magnons and Fermi arcs. The band structure after the perturbation is illustrated in \figurename~\ref{sup-fig:fef3}d.
 
 Alternatively, we consider strain-induced Heisenberg interactions. It turns out that only the $A_1^-A_1^+$ degeneracy is lifted. This combined with a subsequent magnetic field perturbation allows to tune a Hamiltonian with $2\in\mathbb{Z}_4$, indicating a magnon axion insulator.

\subsection{\texorpdfstring{SrRu$_2$O$_6$}{SrRu2O6}}
\begin{figure*}[htb!]
  \centering
  \includegraphics[width=0.8\linewidth]{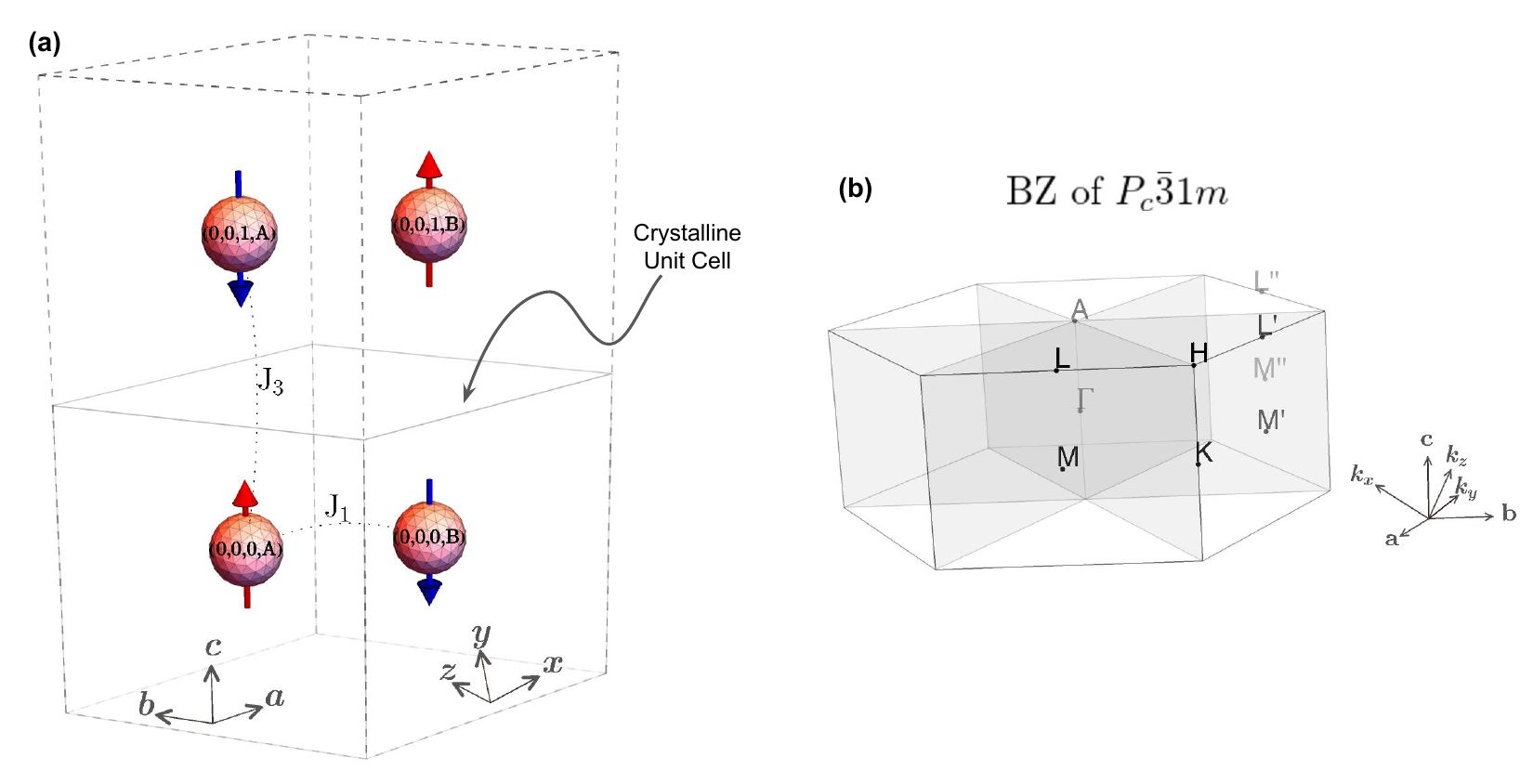}
  \caption{\textbf{The magnetic structure of SrRu$_2$O$_6$.} \textbf{a}, The magnetic unit cell which is twice as large as the crystalline unit cell. \textbf{b}, The magnetic Brillouin zone.\label{sup-fig:srru2o6-1}}
\end{figure*}

The structure of SrRu$_2$O$_6$ has trigonal $P\bar{3}1m~(162)$ symmetry. The structural and magnetic unit cells, hosting 2 and 4 Ru atoms respectively, are illustrated in \figurename~\ref{sup-fig:srru2o6-1}a.

In Cartesian coordinates, the primitive (hexagonal) translations of the crystal are
\begin{equation}
  \bm{a}_1=\frac{a}{\sqrt{6}}(1,1,-2),\qquad
  \bm{a}_2=\frac{a}{\sqrt{6}}(-2,1,1),\qquad
  \bm{a}_3=\frac{c}{\sqrt{3}}(1,1,1),
\end{equation}
and the two Ru atoms in the structural unit cell are positioned at
\begin{align}
  \bm{r}_{\mathrm{A}}&=\frac{1}{3}\bm{a}_1+\frac{2}{3}\bm{a}_2+\frac{1}{2}\bm{a}_3\\
  \bm{r}_{\mathrm{B}}&=\frac{2}{3}\bm{a}_1+\frac{1}{3}\bm{a}_2+\frac{1}{2}\bm{a}_3.
\end{align}

The two sublattices transform under $P\bar{3}1m$ as follows.
\begin{itemize}
  \item $i$, bond-centered inversion: $\mathrm{A}\leftrightarrow\mathrm{B}$ within the same cell.
  \item $3_{001}$, three-fold axis along $\bm{a}_3$:
    \begin{align*}
      (x_1,x_2,x_3,s)&\longrightarrow
      \begin{cases}
        (-1-x_2,-1+x_1-x_2,x_3,\mathrm{A}),&s=\mathrm{A}\\
        (-1-x_2,x_1-x_2,x_3,\mathrm{B}),&s=\mathrm{B}
      \end{cases}
    \end{align*}
  \item $M_{1\bar{1}0}$, mirror plane $u(\bm{a}_1+\bm{a}_2)+v\bm{a}_3$:
    \begin{align*}
      (x_1,x_2,x_3,s)&\longrightarrow
      \begin{cases}
        (x_2,x_1,x_3,\mathrm{B}),&s=\mathrm{A}\\
        (x_2,x_1,x_3,\mathrm{A}),&s=\mathrm{B}
      \end{cases}
    \end{align*}
\end{itemize}

The first NN exchange
\begin{equation}
  \mathsf{J}_{1}:(0,0,0,\mathrm{A})\leftrightarrow(0,0,0,\mathrm{B})
\end{equation}
is constrained by mirror and inversion symmetries to the form

\begin{equation}
  \mathsf{J}_1=\begin{pmatrix}
    J_1&\Gamma_1^{xy}&\Gamma_1^{xz}\\
    \Gamma_1^{xy} & J_1+J_1^y & \Gamma_1^{xy}\\
    \Gamma_1^{xz} & \Gamma_1^{xy} & J_1
  \end{pmatrix},
\end{equation}
while the third NN exchange
\begin{equation}
  \mathsf{J}_{3}:(0,0,0,\mathrm{A})\leftrightarrow(0,0,1,\mathrm{A})
\end{equation}
is constrained by the three-fold axis and a perpendicular two-fold axis to the form
\begin{equation}
  \mathsf{J}_3=\begin{pmatrix}
    J_3 & J_3^{xy} & J_3^{xz}\\
    J_3^{xz} & J_3 & J_3^{xy}\\
    J_3^{xy} & J_3^{xz} & J_3
  \end{pmatrix}=
  \begin{pmatrix}
    J_3 & \Gamma_3+D_3 & \Gamma_3-D_3\\
    \Gamma_3-D_3 & J_3 & \Gamma_3+D_3\\
    \Gamma_3+D_3 &  \Gamma_3-D_3 & J_3
  \end{pmatrix}.
\end{equation}

\subsubsection{Spin-wave spectrum and topology}
The magnetic structure (\figurename~\ref{sup-fig:srru2o6-1}a) preserves the generators of $P\bar{3}1m$ except $T_{\bm{a}_3}$, a translation by $\bm{a}_3$. However, it preserves magnetic translation $\tilde{T}_{\bm{a}_3}=\tilde{\mathcal{T}}\times T_{\bm{a}_3}$, a combination of time-reversal $\tilde{\mathcal{T}}$ and $T_{\bm{a}_3}$. The magnetic space group is therefore $P_c\bar{3}1m~(162.78)$, and the unit cell is doubled. The magnetic Bravais translations are given by
\begin{equation}
  \begin{pmatrix}
    \bm{a}\\
    \bm{b}\\
    \bm{c}
  \end{pmatrix}
  =\begin{pmatrix}
    1& & \\
    &1&\\
    &&2
  \end{pmatrix}
  \begin{pmatrix}
    \bm{a}_1\\
    \bm{a}_2\\
    \bm{a}_3
  \end{pmatrix}.
\end{equation}

The magnetic Fe atoms sit on the $4d$ Wyckoff position of $P_c\bar{3}1m$, with site-symmetry group $32'$. Thus the spin-wave variables furnish a direct sum of two irreps (labeled ${}^1E$ and ${}^2E$) with $C_{3}=e^{\pm i 2\pi/3}$ eigenvalues of the three-fold rotation. Shown in \tablename~\ref{sup-tab:srru2o6-bandrep} is the band representations induced from ${}^1E$ and ${}^2E$.

\begin{table}[htb!]
  \centering
  \begin{tabular}{ccc}
    \toprule
    $k$ point & ${}^1E\uparrow$ MSG $\downarrow G_{\bm{k}}$ & ${}^2E\uparrow$ MSG $\downarrow G_{\bm{k}}$\\
    \midrule
    $\Gamma(0,0,0)$ & $\Gamma_3^+(2)\oplus \Gamma_3^-(2)$ & $\Gamma_3^+(2)\oplus \Gamma_3^-(2)$\\[2pt]
    $A(0,0,\frac{1}{2})$ & $A_3^-A_3^+(4)$ & $A_3^-A_3^+(4)$\\[2pt]
    $H(\frac{1}{3},\frac{1}{3},\frac{1}{2})$ & $H_1(1)\oplus H_2(1)\oplus H_3(2)$ & $H_1(1)\oplus H_2(1)\oplus H_3(2)$\\[2pt]
    $K(\frac{1}{3},\frac{1}{3},0)$ & $K_1(1)\oplus K_2(1)\oplus K_3(2)$ & $K_1(1)\oplus K_2(1)\oplus K_3(2)$\\[2pt]
    $L(\frac{1}{2},0,\frac{1}{2})$ & $L_1^+L_2^-(2)\oplus L_1^-L_2^+(2)$ & $L_1^+L_2^-(2)\oplus L_1^-L_2^+(2)$\\[2pt]
    $M(\frac{1}{2},0,0)$ & $M_1^+(1)\oplus M_1^-(1)\oplus M_2^+(1)\oplus M_2^-(1)$ & $M_1^+(1)\oplus M_1^-(1)\oplus M_2^+(1)\oplus M_2^-(1)$\\
    \bottomrule
  \end{tabular}
  \caption{Band representation induced from the $E_1$ and $E_2$ irreps of $32'$, the site-symmetry group of the $4d$ Wyckoff position in $P_c\bar{3}1m$. The magnon band representation in SrRu$_2$O$_6$ is induced from ${}^1E\oplus {}^2E$ (see \tablename~\ref{sup-tab:irreps}.)\label{sup-tab:srru2o6-bandrep}}
\end{table}

\begin{figure*}[htb!]
  \centering
  \includegraphics[width=0.95\linewidth]{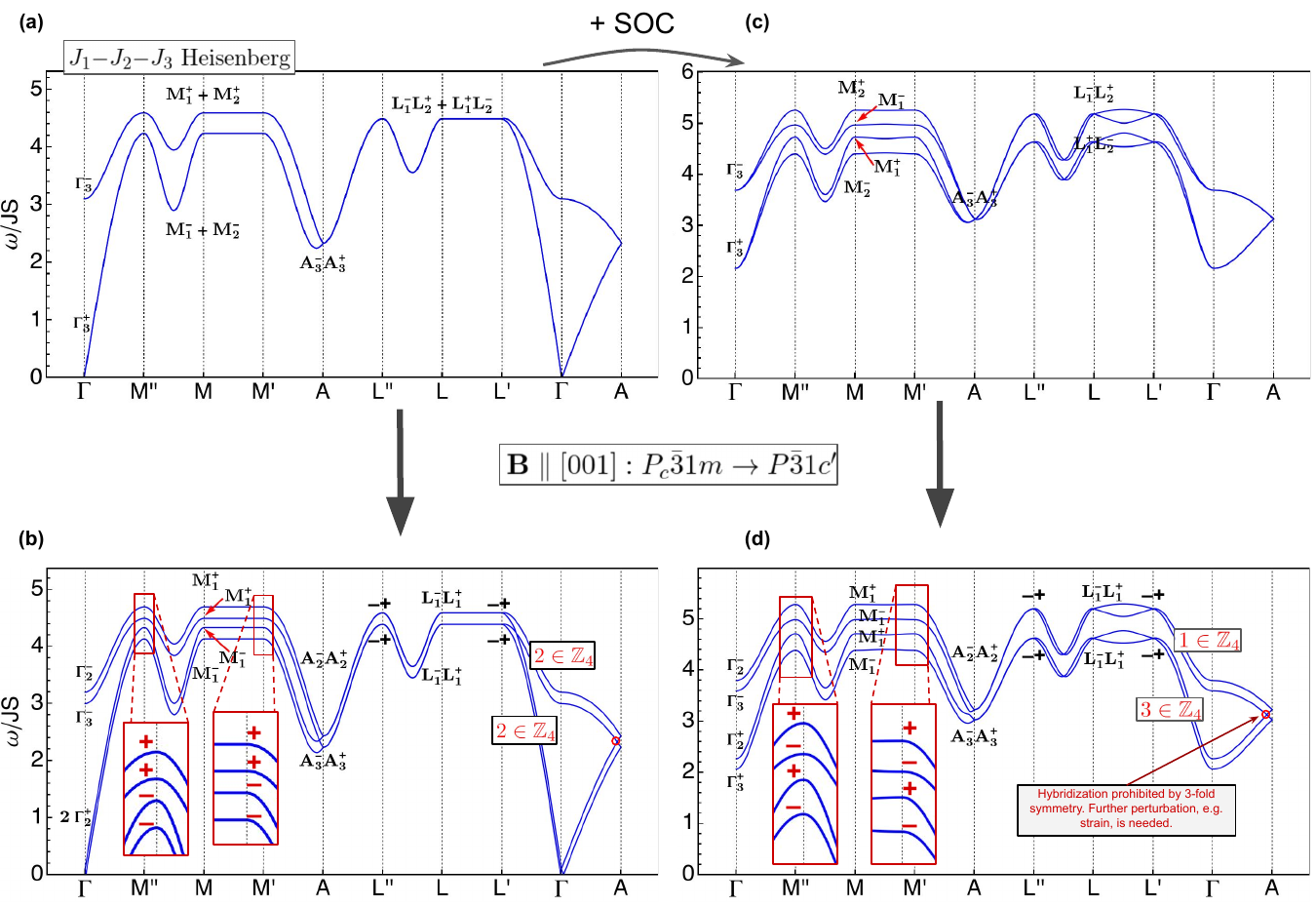}
  \caption{\textbf{B-field induced topological magnons in SrRu$_2$O$_6$.} Panels \textbf{a} and \textbf{c} (\textbf{b} and \textbf{d}) show the magnon spectrum without (with) a $\bm{B}$-field perturbation. In \textbf{a} and \textbf{c} we use a Heisenberg model with $J_1=1.00$ and $J_3=0.40$.\label{sup-fig:srru2o6-2}}
\end{figure*}

Now we proceed to show that a $\bm{B}\parallel[001]$ perturbation shall separate the four bands (previously connected by the 4D irrep $A_3^-A_3^+$) at all TRIM points, and furthermore assign a non-trivial SI to the higher pair of bands. The argument can be summarized in the following points.
\begin{itemize}
  \item The perturbation $\bm{B}\parallel[001]$ breaks down $P_c\bar{3}1m$ down into $P\bar{3}1c~(163.83)$, and as a result, the 4D irrep into two 2D irreps, each of which containing a single inversion-odd component:
    \begin{equation}
      A_3^-A_3^+(4)\longrightarrow A_2^-A_2^+(2)\oplus A_3^-A_3^+(2)
    \end{equation}
  \item At $k=L$ there exist two copies of the 2D irrep $L_1^-L_1^+$, each containing a single inversion-odd component. Furthermore, the orbit of $L$ (i.e.\@ its $k$-star) contains two other $k$-points related by the 3-fold symmetry. This is true both before and after the perturbation.

    Therefore, following the perturbation, the combined contribution of $A$ and $L$ to $\mathbb{Z}_4$ of the higher bands is always
    \begin{equation}
      1+3=0\mod4.
    \end{equation}
  \item The contribution of the remaining 4 TRIM points ($\Gamma$ and the orbit of $M$) cannot be fixed by symmetry alone, but requires a concrete spin model. Specifically, we show below that in any spin model with dominant $J_1$-$J_2$-$J_3$ Heisenberg interactions, the combined contribution of $\Gamma$ and $M$ is constrained to $1,3,2\in\mathbb{Z}_4$, with the latter (i.e.\@ $2\in\mathbb{Z}_4$) corresponding to the limit of vanishing SOC terms.
\end{itemize}
Therefore the contribution of the 8 TRIM points shall be $1,3,2\in\mathbb{Z}_4$. The perturbed magnon spectrum is illustrated in \figurename~\ref{sup-fig:srru2o6-2}.

  However, note that although the higher two bands are separated at all TRIM points following the $\bm{B}$-field perturbation, they do not satisfy the compatibility relations, which means it is impossible to consistently patch together the two bands while maintaining an energy gap at all high-symmetry lines and planes. In our particular case, this results from a mismatch in the $C_3$ eigenvalues of the representations at $\Gamma$ and $A$ within the higher two bands. It follows that the $C_3$-symmetric line $\Gamma-A$, connecting $\Gamma$ and $A$, must host a gapless point to correct for this mismatch.

  To eliminate this gapless point, we need to break the $C_3$ symmetry. This can be achieved by a generic strain perturbation.

  Finally, we show by a direct calculation the contribution of $\Gamma$ and $M$ to $\mathbb{Z}_4$ of the higher bands. The magnon energies at $k=M$ are given by
\begin{align}
  {(E_{M_1^+})}^2&=\frac{1}{3}\bigl(4J_1-8J_2+\alpha\bigr)\bigl(6J_1-24J_2+12J_3+\beta\bigr)\\
  {(E_{M_2^+})}^2&=\frac{1}{3}\bigl(12J_1-24J_2+\gamma\bigr)\bigl(2J_1-8J_2+4J_3+\delta\bigr)\\
  {(E_{M_1^-})}^2&=\frac{1}{3}\bigl(2J_1-8J_2+\delta\bigr)\bigl(12J_1-24J_2+12J_3+\gamma\bigr)\\
  {(E_{M_2^-})}^2&=\frac{1}{3}\bigl(6J_1-24J_2+\beta\bigr)\bigl(4J_1-8J_2+4J_3+\alpha\bigr)
\end{align}
where $\alpha,\beta,\gamma,\delta$ are SOC terms, given by (up to 2NN bonds)
\begin{align}
  \alpha&=-6A+2\Gamma_1^{xy}+3\Gamma_1^{xz}-6\Gamma_2^{xy}-4\Gamma_2^{xz}+2J_1^y\\
  \beta&=-18A+10\Gamma_1^{xy}+11\Gamma_1^{xz}-2\Gamma_2^{xy}-28\Gamma_2^{xz}+4J_1^y\\
  \gamma&=-18A+14\Gamma_1^{xy}+\Gamma_1^{xz}-2\Gamma_2^{xy}-28\Gamma_2^{xz}+2J_1^y\\
  \delta&=-6A+6\Gamma_1^{xy}+\Gamma_1^{xz}-6\Gamma_2^{xy}-4\Gamma_2^{xz}
\end{align}
It follows that
\begin{align}
  {(E_{M_1^+})}^2-{(E_{M_2^-})}^2&=\frac{4}{3}J_3 (6J_1+3\alpha-\beta)\label{sup-eq:deltaE1}\\
  &= \frac{8}{3} J_3 (3J_1+J_1^y-2\Gamma_1^{xy}-\Gamma_1^{xz}-8\Gamma_2^{xy}+8\Gamma_2^{xz}),
\end{align}
and
\begin{align}
  {(E_{M_2^+})}^2-{(E_{M_1^-})}^2&=\frac{4}{3}(6J_1-3\delta+\gamma)J_3\label{sup-eq:deltaE2}\\
  &=\frac{8}{3}J_3(3J_1+J_1^y-2\Gamma_1^{xy}-\Gamma_1^{xz}+8\Gamma_2^{xy}-8\Gamma_2^{xz}).
\end{align}
Furthermore, for the two 2D states at $k=\Gamma$ we find that
\begin{equation}\label{sup-eq:deltaE3}
  {(E_{\Gamma_3^-})}^2-{(E_{\Gamma_3^+})}^2=8J_3(3J_1+J_1^y-\Gamma_1^{xy}-\Gamma_1^{xz}),
\end{equation}

In the following, we assume the SOC terms are small compared to $J_1$.

For $J_1J_3>0$, Eqs.~\ref{sup-eq:deltaE1} and~\ref{sup-eq:deltaE2} rule out the possibility of having two inversion-odd eigenvalues in the higher two bands. On the other hand, Eq.~\ref{sup-eq:deltaE3} forces 2D irrep $\Gamma_3^-$ to be higher than 2D irrep $\Gamma_3^+$. Thus, the number of inversion-odd eigenvalues at $\Gamma$ and the orbit of $M$ is constrained to $2$ or $5$. Conversely, if $J_1J_3<0$, the total count is either $3$ or $6$.

\subsection{\texorpdfstring{LaSrFeO$_4$}{LaSrFeO4}}
\begin{figure*}[htb!]
  \centering
  \includegraphics[width=0.8\linewidth]{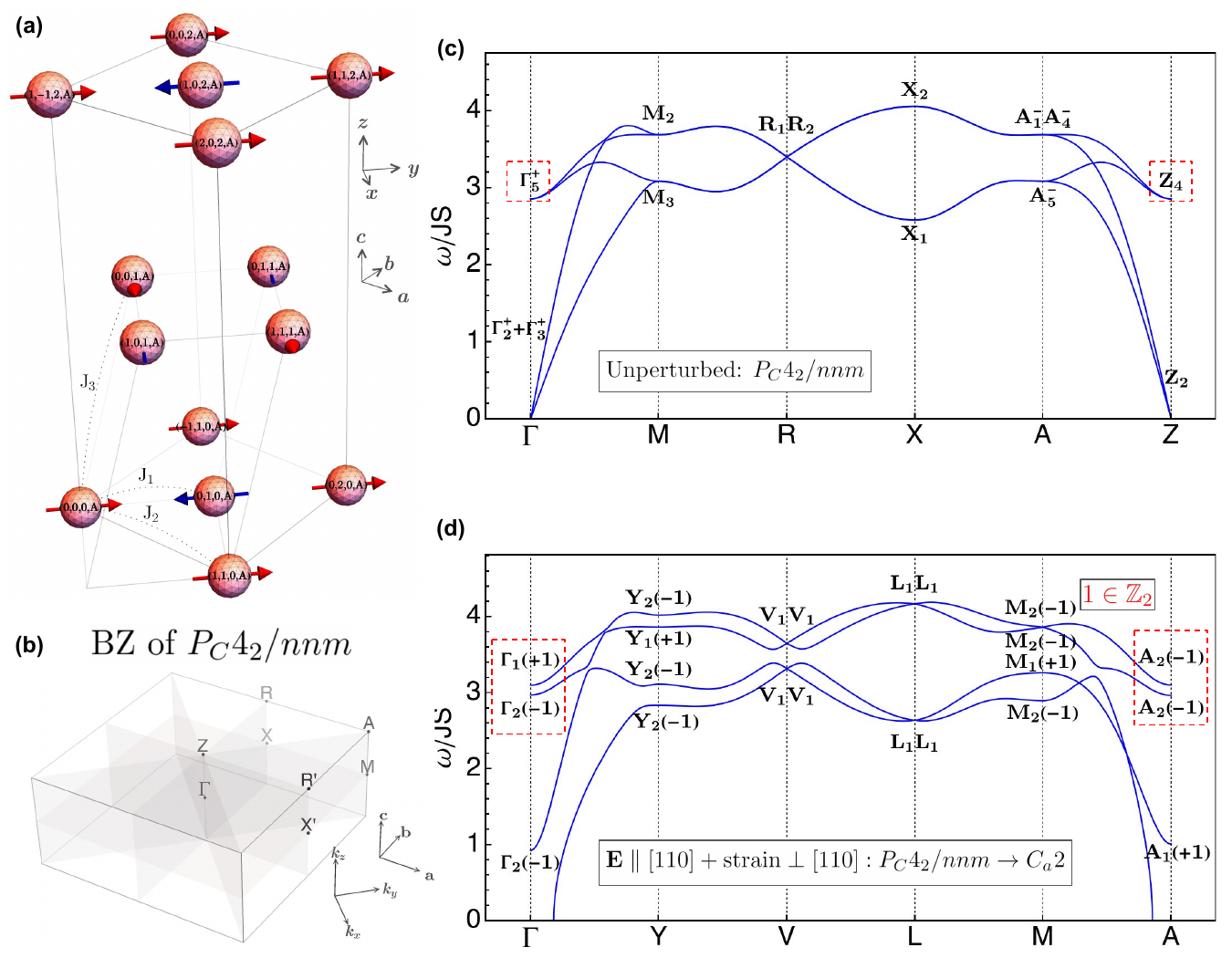}
  \caption{\textbf{Topological magnon bands in LaSrFeO$_4$.} \textbf{a}-\textbf{b}, The magnetic unit cell and the magnetic Brillouin zone. \textbf{c}, Magnon spectrum with $J_1=1.00$, $J_3=0.36$, $J_1^y=-0.30$ and $J_1^z=-0.45$. Without inter-planar interactions, the states $\Gamma_5^+$ and $Z_4$ are equal in energy. In \textbf{d}, $\Gamma_5^+$ splits into two states with $\pm1$ eigenvalues of the $2_{010}$ symmetry of the magnetic subgroup $C_a2$, while $Z_4$ splits into $-1$ states. These irreps in the subgroup cannot arise from atomic insulators, and they correspond to the nontrivial indicator $1\in\mathbb{Z}_2$ of $C_a2$.\label{sup-fig:lasrfeo4}}
\end{figure*}

The structure of LaSrFeO$_4$ is illustrated in \figurename~\ref{sup-fig:lasrfeo4}a. The magnetic unit cell has four times the size of the structural cell, with 4 Fe atoms in former and only one in the latter.

The primitive translations of the crystal are
\begin{align}
  \bm{a}_1=a(1,0,0),\qquad \bm{a}_2=a(0,1,0),\qquad \bm{a}_3=\frac{1}{2}(-a,a,c),
\end{align}
and the Fe atom sits at the unit cell corner,
\begin{align}
  \bm{r}_{\mathrm{A}}=(0,0,0).
\end{align}

The space group $I4/mmm~(139)$ transforms Fe lattice sites as follows.
\begin{itemize}
  \item $4_{z}$, four-fold axis $(0,0,z)$:
    \begin{equation}
      (x_1,x_2,x_3,A)\longrightarrow
      (-x_2,x_1-x_3,x_3,\mathrm{A})
    \end{equation}
  \item $M_{z}$, mirror plane $(x,y,0)$:
    \begin{equation}
      (x_1,x_2,x_3,A)\longrightarrow
      (x_1-x_3,x_2+x_3,-x_3,\mathrm{A})
    \end{equation}
  \item $M_{x}$, mirror plane $(0,y,z)$:
    \begin{equation}
      (x_1,x_2,x_3,A)\longrightarrow
      (-x_1+x_3,x_2,x_3,\mathrm{A})
    \end{equation}
  \item $M_{110}$, mirror plane $(x,-x,z)$:
    \begin{equation}
      (x_1,x_2,x_3,A)\longrightarrow
      (-x_2,-x_1,x_3,\mathrm{A})
    \end{equation}
\end{itemize}

The first 3 NN exchange paths are
\begin{align}
  \mathsf{J}_{1}&:(0,0,0,\mathrm{A})\leftrightarrow(0,1,0,\mathrm{A})\\
  \mathsf{J}_{2}&:(0,0,0,\mathrm{A})\leftrightarrow(1,1,0,\mathrm{A})\\
  \mathsf{J}_{3}&:(0,0,0,\mathrm{A})\leftrightarrow(0,0,1,\mathrm{A}),
\end{align}
with generic exchange matrices,
\begin{gather}
  \mathsf{J}_1=\begin{pmatrix}
    J_1&&\\
    &J_1+J_1^y&\\
    &&J_1+J_1^z
  \end{pmatrix},\qquad
  \mathsf{J}_2=\begin{pmatrix}
    J_2&\Gamma_2^{xy}&\\
    \Gamma_2^{xy}&J_2&\\
    &&J_2+J_2^z
  \end{pmatrix},\qquad\\
  \mathsf{J}_3=\begin{pmatrix}
    J_3& \Gamma_3^{xy}& \Gamma_3^{xz}\\
    \Gamma_3^{xy}& J_3& -\Gamma_3^{xz}\\
    \Gamma_3^{xz}& -\Gamma_3^{xz}& J_3+J_3^z
  \end{pmatrix}.
\end{gather}

\begin{table}[htb!]
  \centering
  \begin{tabular}{cc}
    \toprule
    $k$ point & $B_g\uparrow$ MSG $\downarrow G_{\bm{k}}$ \\
    \midrule
    $\Gamma(0,0,0)$ & $\Gamma_2^+(1)\oplus \Gamma_3^+(1)\oplus \Gamma_5^+(2)$\\[2pt]
    $A(\frac{1}{2},\frac{1}{2},\frac{1}{2})$ & $A_1^-A_4^-(2)\oplus A_5^-(2)$\\[2pt]
    $M(\frac{1}{2},\frac{1}{2},0)$ & $M_2(2)\oplus M_3(2)$\\[2pt]
    $R(0,\frac{1}{2},\frac{1}{2})$ & $R_1R_2(4)$\\[2pt]
    $X(0,\frac{1}{2},0)$ & $X_1(2)\oplus X_2(2)$\\[2pt]
    $Z(0,0,\frac{1}{2})$ & $Z_2(2)\oplus Z_4(2)$\\
    \bottomrule
  \end{tabular}
  \caption{Band representation induced from the $B_g$ irrep of $m'm'm$, the site-symmetry group of the $4d$ Wyckoff position in $P_C4_2/nnm$. The magnon band representation in LaSrFeO$_4$ is induced from two copies of $B_g$ (see \tablename~\ref{sup-tab:irreps}.)\label{sup-tab:lasrfeo4-bandrep}}
\end{table}

\subsubsection{Spin-wave spectrum and topology}
The antiferromagnetic structure of LaSrFeO$_4$ is described by tetragonal $P_C4_2/nnm~(134.481)$ symmetry, with a magnetic unit cell illustrated in \figurename~\ref{sup-fig:lasrfeo4}a. The magnetic Fe atoms occupy the $4d$ Wyckoff position, and the spin-wave variables transform as two copies of the $B_g$ irrep of the site-symmetry group $m'm'm$ ($B_g$ labels the 1D inversion-even mirror-odd irrep.) \tablename~\ref{sup-tab:lasrfeo4-bandrep} summarized the band representation induced from $B_g$, indicating a four-fold degeneracy at $k=R$ connecting all magnon bands together.

A combined perturbation of $\bm{E}\parallel[110]$ and strain $\perp[110]$ reduces $P_C4_2/nnm\longrightarrow C_a2~(5.17)$ with monoclinic axes (in the $b$-unique setting) obtained by
\begin{align}
  \begin{pmatrix}
    \bm{a}\\
    \bm{b}\\
    \bm{c}
  \end{pmatrix}
  \rightarrow
  \begin{pmatrix}
    1&-1&\\
    1&\hphantom{-}1&\\
    &&1
  \end{pmatrix}
  \begin{pmatrix}
    \bm{a}\\
    \bm{b}\\
    \bm{c}
  \end{pmatrix}
\end{align}
and a $\mathbb{Z}_2$ index counting the number of $C_2=-1$ eigenvalues of 2-fold rotation symmetry at both $k=\Gamma$ and $k=Z$. The nontrivial element $1\in\mathbb{Z}_2$ corresponds to Weyl points in a plane perpendicular to the 2-fold axis of $C_a2$.

For the unperturbed system, a spin model with $J_1$-$J_2$-$J_3$ Heisenberg exchange as well as all symmetry-allowed terms up to 2NN leads to gaps at $\Gamma$ and $Z$ with
\begin{align}
  {\bigl(E_{Z_4}\bigr)}^2-{\bigl(E_{Z_2}\bigr)}^2&= {\bigl(E_{\Gamma_5^+}\bigr)}^2-{\bigl(E_{\Gamma_2^+\oplus\Gamma_3^+}\bigr)}^2\\
   &= 8 (2 J_1 + J_1^y) (J_1^y - 2 J_1^z + 2 J_2^z).
\end{align}
Thus, the 2D irreps $\Gamma_5^+$ and $Z_4$ belong the higher two bands. Additionally, within $C_a2$ these irreps break down into
\begin{align}
  \Gamma_5^+&\longrightarrow \Gamma_1(C_2=+1)\oplus \Gamma_2(C_2=-1)\\
  Z_4&\longrightarrow A_2(C_2=-1)\oplus A_2(C_2=-1),
\end{align}
and thus the higher two bands must be nontrivial, following the perturbation.

The nontrivial $\mathbb{Z}_2$ symmetry indicator physically corresponds to Weyl magnons located in lines $k_a=k_b=\pm\pi/2$ in the BZ, for the following reason. First of all, the magnetic translation $\tilde{\mathcal{T}}\equiv T_1\cdot\mathcal{T}$ satisfies: 
\begin{align}
    &\tilde{\mathcal{T}}^2=(C_2\cdot\tilde{\mathcal{T}})^2=T_aT_b=e^{i(k_b+k_a)}
\end{align}
Since the 2-fold rotation $C_2$ is perpendicular to $\hat z$-axis, the combined $C_2\cdot\tilde{\mathcal{T}}$ symmetry plays the role of a PT symmetry in in any one-dimensional manifold $k_a=k_b=k_0,~\forall~k_0$. As a result, the polarization in any line $k_a=k_b=k_0,~\forall~k_0$ must be quantized with a $\mathbb{Z}_2$ topological invariant\cite{Ahn2019}. For a set of trivial bands separated from other bands, as we vary $k_0\in[-\pi,\pi)$ in the BZ, this $\mathbb{Z}_2$-valued invariant (i.e. polarization) should remain the same: otherwise, the gap must close and Weyl points between this set of bands and other bands must appear. 

In the presence of unitary $C_2$ symmetry for the case of $k_0=0,\pi$ with $(C_2\cdot\tilde{\mathcal{T}})^2=1$, the $\mathbb{Z}_2$-valued polarization is exactly given by the aforementioned $\mathbb{Z}_2$ symmetry indicator. Meanwhile, for $k_0=\pm\pi/2$ lines, the anti-unitary PT symmetry satisfies $(C_2\cdot\tilde{\mathcal{T}})^2=-1$ with a Kramers degeneracy, and its polarization must be trivial for any gapped bands, hence a contradiction with the nontrivial polarization at $k_0=0,\pi$. As a result, Weyl points must show up along the $k_a=k_b=\pm\pi/2$ lines.

\subsection{MnTe}
\subsubsection{Crystal symmetries and spin model}
The MnTe crystal has hexagonal $P6_3/mmc~(194)$ symmetry. In Cartesian coordinates, we work with primitive hexagonal translations given by
\begin{equation}\label{mnte-hex-axes}
  \bm{a}_1=\frac{a}{\sqrt{2}}(1,0,-1),\qquad\bm{a}_2=\frac{a}{\sqrt{2}}(-1,1,0),\qquad\bm{a}_3=\frac{c}{\sqrt{3}}(1,1,1).
\end{equation}

The magnetic Mn atoms in a crystalline unit cell are
\begin{align}
  \bm{r}_{\mathrm{A}}&=(0,0,0)\\
  \bm{r}_{\mathrm{B}} &= \frac{1}{2}\bm{a}_3.
\end{align}

\begin{figure*}[b!]
  \centering
  \hfill
  \includegraphics[width=0.95\linewidth]{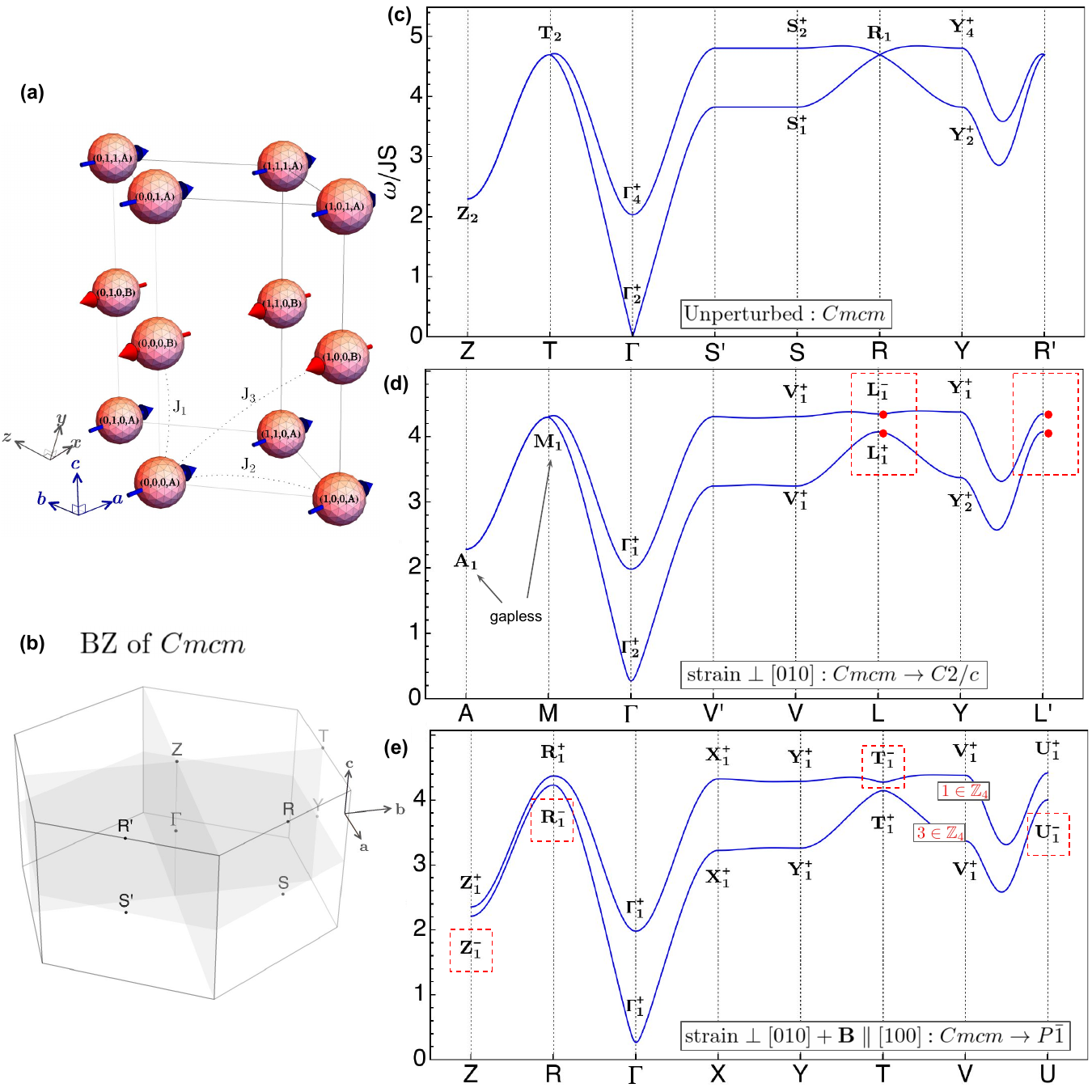}
  \hfill
  \caption{\textbf{Weyl magnons in MnTe induced by a combination of strain and magnetic field.} \textbf{a}-\textbf{b}, The magnetic structure of MnTe and the Brillouin zone. \textbf{c}, Unperturbed magnon spectrum with $J_1=1.00$, $\Gamma_1=-0.15$, $J_2=-0.30$, $J_3=0.22$ and $\Gamma_3^{xy}=-0.05$. \textbf{d}, Magnon spectrum perturbed by uniaxial strain $\parallel[101]$ with $\sigma=0.12$, reducing $Cmcm$ into $C2/c$. We use strain-induced terms $S_2^{xx,J}=.7$, $S_2^{xx,yz}=-0.13$. \textbf{e}, Further applying a magnetic field $h=0.07$ in the $[100]$ direction gaps out the remaining points, and separated bands acquire $1,3\in\mathbb{Z}_4$.\label{sup-fig:mnte-1}}
\end{figure*}

The generators of $P6_3/mmc$ and their action on Mn atoms are as follows.
\begin{itemize}
  \item $6_3$, six-fold screw axis along $w\bm{a}_3$:
    \begin{align*}
      (x_1,x_2,x_3,s)&\longrightarrow
      \begin{cases}
        (x_1-x_2,x_1,x_3,\mathrm{B}),&s=\mathrm{A}\\
        (x_1-x_2,x_1,1+x_3,\mathrm{A}),&s=\mathrm{B}
      \end{cases}
    \end{align*}
  \item $M_{001}$, mirror plane $u\bm{a}_1+v\bm{a}_2+\frac{1}{4}\bm{a}_3$:
    \begin{align*}
      (x_1,x_2,x_3,s)&\longrightarrow
      \begin{cases}
        (x_1,x_2,-x_3,\mathrm{B}),&s=\mathrm{A}\\
        (x_1,x_2,-x_3,\mathrm{A}),&s=\mathrm{B}
      \end{cases}
    \end{align*}
  \item $M_{100}$, mirror plane $u(\bm{a}_1+2\bm{a}_2)+w\bm{a}_3$:
    \begin{align*}
      (x_1,x_2,x_3,s)&\longrightarrow
      \begin{cases}
        (-x_1+x_2,x_2,x_3,\mathrm{A}),&s=\mathrm{A}\\
        (-x_1+x_2,x_2,x_3,\mathrm{B}),&s=\mathrm{B}
      \end{cases}
    \end{align*}
  \item $M_{120}$, glide plane $u\bm{a}_1+w\bm{a}_3$ with a fractional translation $\frac{1}{2}\bm{a}_3$:
    \begin{align*}
      (x_1,x_2,x_3,s)&\longrightarrow
      \begin{cases}
        (x_1-x_2,-x_2,x_3,\mathrm{B}),&s=\mathrm{A}\\
        (x_1-x_2,-x_2,1+x_3,\mathrm{A}),&s=\mathrm{B}
      \end{cases}
    \end{align*}
\end{itemize}

\begin{table}[htb!]
  \centering
  \begin{tabular}{cc}
    \toprule
    $k$ point & $B_g\uparrow$ MSG $\downarrow G_{\bm{k}}$ \\
    \midrule
    $\Gamma(0,0,0)$ & $\Gamma_2^+(1)\oplus\Gamma_4^+(1)$\\[2pt]
    $R(1/2,1/2,1/2)$ & $R_1(2)$\\[2pt]
    $S(1/2,1/2,0)$ & $S_1^+(1)\oplus S_2^+(1)$\\[2pt]
    $T(1,0,1/2)$ & $T_2(2)$\\[2pt]
    $Y(1,0,0)$ & $Y_2^+(1)\oplus Y_4^+(1)$\\[2pt]
    $Z(0,0,1/2)$ & $Z_2(2)$\\
    \bottomrule
  \end{tabular}
  \caption{Band representation induced from the $B_g$ irrep of $2/m$, the site-symmetry group of the $4a$ Wyckoff position in $Cmcm$. The magnon band representation in MnTe is induced from two copies of $B_g$ (see \tablename~\ref{sup-tab:irreps}.)\label{sup-tab:mnte-bandrep}}
\end{table}

The first three NN bonds are given as follows.
\begin{equation}
  \mathsf{J}_{1\mathrm{a}}:(0,0,0,\mathrm{A})\leftrightarrow(0,0,0,\mathrm{B})
\end{equation}
is constrained by a three-fold axis $3_{001}$, a perpendicular mirror $M_{001}$ and a parallel mirror $M_{100}$, leading to the generic form
\begin{equation}
  \mathsf{J}_1=\begin{pmatrix}
    J_{1} & \Gamma_{1} & \Gamma_{1}\\
    \Gamma_{1} & J_{1} & \Gamma_{1}\\
    \Gamma_{1} & \Gamma_{1} & J_{1}
  \end{pmatrix}
\end{equation}

\begin{equation}
  \mathsf{J}_2:(0,0,0,\mathrm{A})\leftrightarrow(1,0,0,\mathrm{A}),
\end{equation}
due to inversion symmetry and a perpendicular mirror, reads
\begin{equation}
  \mathsf{J}_2=\begin{pmatrix}
    J_{2} & \Gamma_{2}^{xy} & \Gamma_{2}^{xz}\\
    \Gamma_{2}^{xy} & J_{2}+J_{2}^{y} & \Gamma_{2}^{xy}\\
    \Gamma_{2}^{xz}& \Gamma_{2}^{xy} & J_{2}
  \end{pmatrix}
\end{equation}

\begin{figure*}[htb!]
  \centering
  \hfill
  \includegraphics[width=0.95\linewidth]{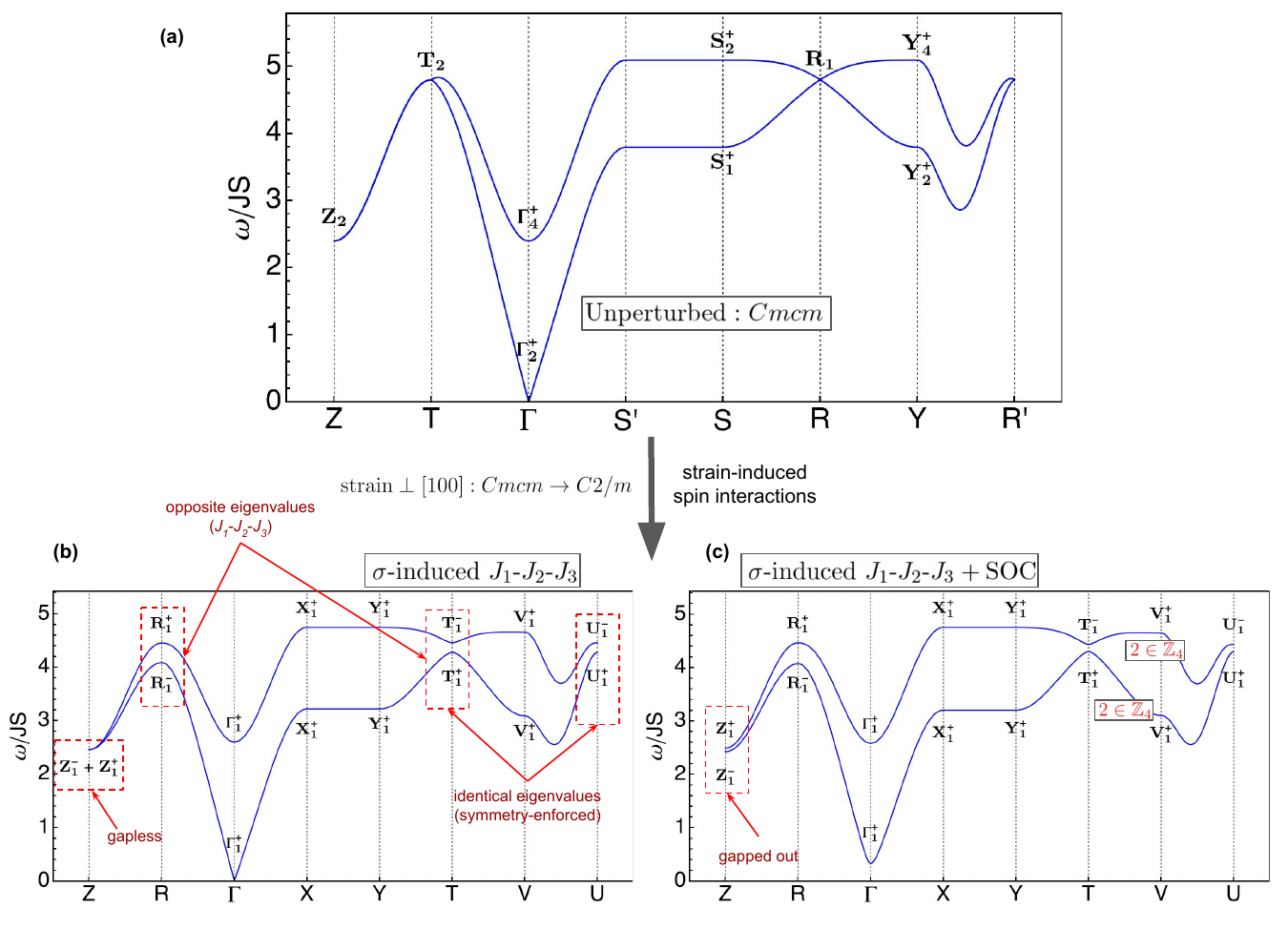}
  \hfill
  \caption{\textbf{Weyl/magnon axion insulator in MnTe induced by uniaxial strain in the $b$-$c$ plane of $Cmcm$.} The spin model parameters in \textbf{a} are $J_1=1.00$, $J_2=-0.30$, $J_3=0.20$, $\Gamma_1^{xy}=-0.23$ and $\Gamma_3^{xy}=-0.05$.  In \textbf{b}-\textbf{c}, we consider $[011]$ strain $\sigma=0.20$ with Heisenberg and SOC terms $S_2^{xx,J}=0.50$, $S_2^{xx,yz}=-0.15$.\label{sup-fig:mnte-2}}
\end{figure*}

\subsubsection{Spin-wave spectrum and topology}
The magnetic moments are aligned antiferromagnetically along the $[110]$ direction of $P6_3/mmc$ (\figurename~\ref{sup-fig:mnte-1}a), breaking the hexagonal symmetry down into $Cmcm~(63.457)$, with conventional monoclinic axes
\begin{equation}
  \begin{pmatrix}
    \bm{a}\\
    \bm{b}\\
    \bm{c}
  \end{pmatrix}
  =\begin{pmatrix}
    1&1&\\
    -1&1&\\
    &&1
  \end{pmatrix}
  \begin{pmatrix}
    \bm{a}_1\\
    \bm{a}_2\\
    \bm{a}_3
  \end{pmatrix},
\end{equation}
where $\{\bm{a}_i\}$ are the hexagonal axes defined in Eq.~\ref{mnte-hex-axes}.

The Mn atoms are positioned at the $4a$ Wyckoff position of $Cmcm$, and the magnon band representation is induced from two copies of the $B_g$ irrep of $2/m$, the site-symmetry group at $4a$ (see \tablename~\ref{sup-tab:mnte-bandrep}). The two magnon bands are connected at high-symmetry points $k=R,T,Z$, as shown in \figurename~\ref{sup-fig:mnte-1}c.

From the subgroup tables (see auxiliary supplementary materials), the only relevant subgroups are centrosymmetric, and thus we only consider magnetic field and strain perturbations. A uniaxial strain $\perp[010]$ breaks down the symmetry $Cmcm\rightarrow C2/c$, lifting two degeneracies at $k=L,L'$ in \figurename~\ref{sup-fig:mnte-1}d. These two $k$-points are related by glide symmetry, and their inversion eigenvalues are opposite, thus they always contribute $+1$ in total to $\mathbb{Z}_4$. The remaining two gapless TRIM points can be gapped out with $\bm{B}\parallel[100]$, contributing a multiple of $2$ to $\mathbb{Z}_4$ (\figurename~\ref{sup-fig:mnte-1}e), and thus the SI of either band is odd, signaling Weyl magnons.

Alternatively, a strain-only perturbation can induce topological magnons, as summarized in \figurename~\ref{sup-fig:mnte-2}. In particular, uniaxial strain $\perp[100]$ reduces $Cmcm\rightarrow C2/m$, gapping out all TRIM points. It turns out that within $J_1$-$J_2$-$J_3$ Heisenberg interactions induced by strain, only 3 (out of 4) points are gapped out, and their total contribution to $\mathbb{Z}_4$ is either $+1$ or $+2$ (\figurename~\ref{sup-fig:mnte-2}b). The remaining point is gapped out by strain-induced SOC terms, and contributes either $0$ or $+1$ to $\mathbb{Z}_4$, depending on microscopic details, and thus the result must be either Weyl magnons or magnon axion insulator.

\clearpage
\section{Holstein-Primakoff boson representation in site-symmetry groups}
As discussed in the methods section, the magnon band representation is induced from the representation of the spin-wave variables $S_{x}$ and $S_{y}$ (or equivalently, the Holstein-Primakoff boson $b=(S_x+\imth S_y)/\sqrt{2S}$ and $b^\dagger$) of the site-symmetry group of a given Wyckoff position. \tablename~\ref{sup-tab:irreps} lists the irreducible representation (irrep) content of the spin-wave variables in all point groups compatible with magnetism.

\begin{table}[htb!]
  \centering
  \begin{tabular}[t]{cc}
    \toprule
    $G$&$\rho_{(S_x,S_y)}\downarrow G$\\
    \midrule
    $1$ & $2A$\\
    $\bar{1}$ & $2A_g$\\
    $2$ & $2B$\\
    $2'$ & $2A$\\
    $m$ & $2A''$\\
    $m'$ & $2A$\\
    $2/m$ & $2B_g$\\
    $2'/m'$ & $2A_g$\\
    $2'2'2$ & $2B$\\
    $m'm'2$ & $2B$\\
    $m'm2'$ & $2A''$\\
    $m'm'm$ & $2B_g$\\
    \bottomrule
  \end{tabular}\hspace{.2in}%
  \begin{tabular}[t]{cc}
    \toprule
    $G$&$\rho_{(S_x,S_y)}\downarrow G$\\
    \midrule
    $4$ & $\prescript{1}{}{E}\oplus\prescript{2}{}{E}$\\
    $\bar{4}$ & $\prescript{1}{}{E}\oplus\prescript{2}{}{E}$\\
    $4/m$ & $\prescript{1}{}{E_g}\oplus\prescript{2}{}{E_g}$\\
    $42'2'$ & $\prescript{1}{}{E}\oplus\prescript{2}{}{E}$\\
    $4m'm'$ & $\prescript{1}{}{E}\oplus\prescript{2}{}{E}$\\
    $\bar{4}2'm'$ & $\prescript{1}{}{E}\oplus\prescript{2}{}{E}$\\
    $4/mm'm'$ & $\prescript{1}{}{E_g}\oplus\prescript{2}{}{E_g}$\\
    $3$ & $\prescript{1}{}{E}\oplus\prescript{2}{}{E}$\\
    $\bar{3}$ & $\prescript{1}{}{E_g}\oplus\prescript{2}{}{E_g}$\\
    $32'$ & $\prescript{1}{}{E}\oplus\prescript{2}{}{E}$\\
    $3m'$ & $\prescript{1}{}{E}\oplus\prescript{2}{}{E}$\\
    $\bar{3}m'$ & $\prescript{1}{}{E_g}\oplus\prescript{2}{}{E_g}$\\
    \bottomrule
  \end{tabular}\hspace{.2in}%
  \begin{tabular}[t]{cc}
    \toprule
    $G$&$\rho_{(S_x,S_y)}\downarrow G$\\
    \midrule
    $6$ & $\prescript{1}{}{E_2}\oplus\prescript{2}{}{E_2}$\\
    $\bar{6}$ & $\prescript{1}{}{E''}\oplus\prescript{2}{}{E''}$\\
    $6/m$ & $\prescript{1}{}{E_{2g}}\oplus\prescript{2}{}{E_{2g}}$\\
    $62'2'$ & $\prescript{1}{}{E_2}\oplus\prescript{2}{}{E_2}$\\
    $6m'm'$ & $\prescript{1}{}{E_2}\oplus\prescript{2}{}{E_2}$\\
    $\bar{6}m'2'$ & $\prescript{1}{}{E''}\oplus\prescript{2}{}{E''}$\\
    $6/mm'm'$ & $\prescript{1}{}{E_{2g}}\oplus\prescript{2}{}{E_{2g}}$\\
    \bottomrule
  \end{tabular}

  \caption{Representation of spin-wave variables $(S_x,S_y)$ in all magnetic point groups compatible with magnetism. The notation in Ref.~\cite{Bradley1972} is used for point-group irrep labels.\label{sup-tab:irreps}}
\end{table}

\newcommand*{\Scale}[2][4]{\scalebox{#1}{$#2$}}%

\section{Illustration of magnon surface states}

Below, we illustrate the topologically protected magnon surface states in magnetically ordered materials. We focus on three representative cases relevant to the high-$T_c$ topological magnon materials presented in the main text: (1) surface magnon arcs associated with bulk Weyl magnons; (2) surface chiral magnons in magnon axion insulators, where the bulk Chern number does not vanish; and (3) magnon hinge states in magnon axion insulators, where the bulk Chern number vanishes. We first illustrate the surface magnon arcs that arise due to Weyl points in the bulk magnon spectrum, using TbFeO$_3$ as an example. Next, we discuss two types of topological magnon surface states associated with a magnon axion insulator in the bulk, diagnosed by a $\mathbb{Z}_4$ inversion symmetry indicator $\nu=2\bmod4$. When the Chern number in one 2D plane of the 3D Brillouin zone is nonzero, the magnon axion insulator exhibits chiral surface magnons, which we illustrate using $\alpha$-Fe$_2$O$_3$ as an example. When the Chern number in any 2D plane of the 3D Brillouin zone vanishes, the magnon axion insulator exhibits chiral hinge magnons, which we illustrate using HoRh as an example.

\subsection{Surface magnon arcs}
\begin{figure}[htpb!]
  \centering
  \subfigure[]{\includegraphics[width=0.7\textwidth]{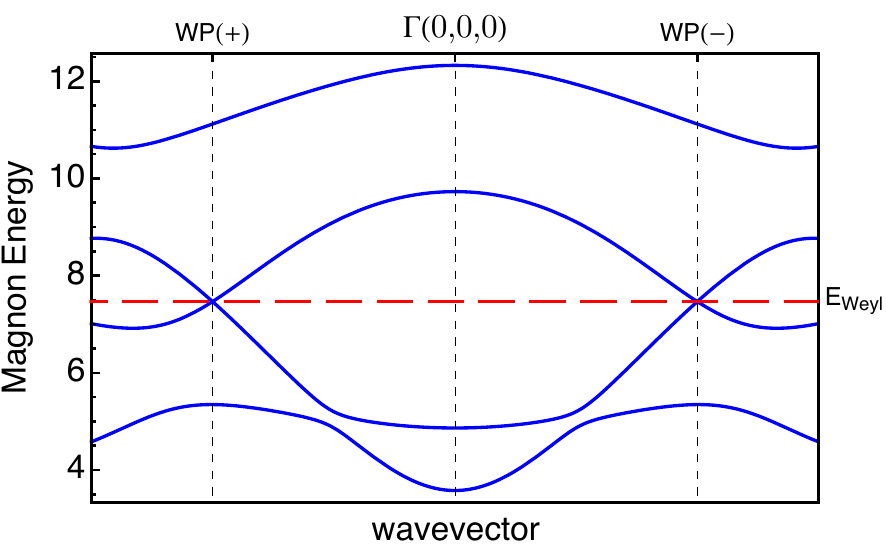}}
  \subfigure[]{\includegraphics[width=0.7\textwidth]{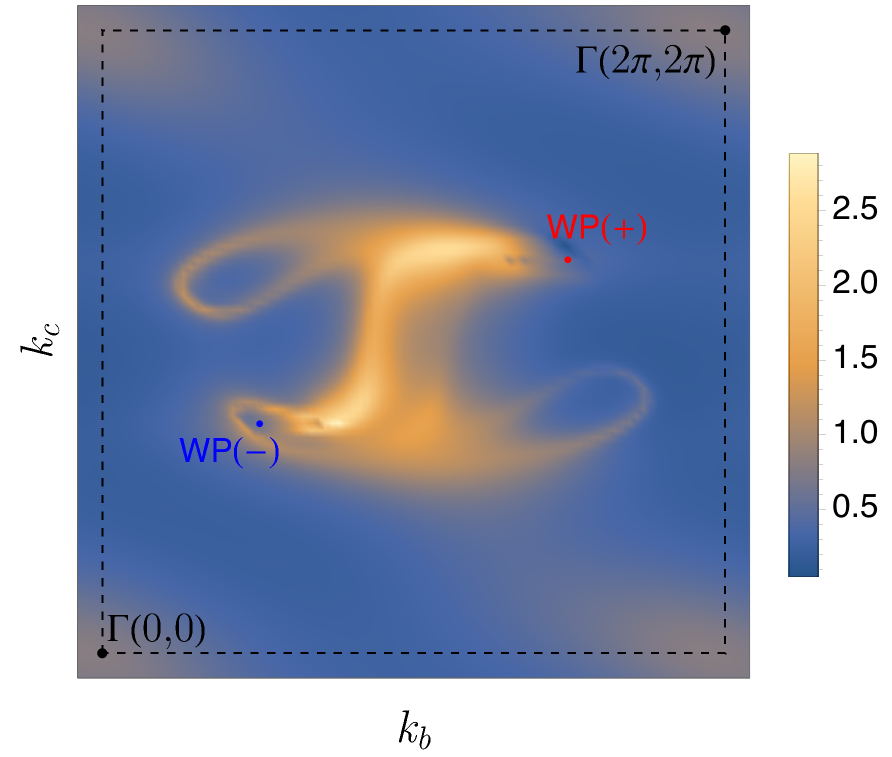}}
  \caption{\label{fig:magnon-arc}\textbf{Weyl points and surface magnon arc in TbFeO$_3$.} \textbf{(a)}, The bulk magnon spectrum along a straight line in the Brillouin zone connecting the two Weyl points. We use the following spin model parameters: $J_{1}^{y}=0.8$, $J_{1}^{z}=1.5$, $J_{2}^{y}=-0.5$, $\Gamma_{2}^{xy}=-0.6$, $D_{2}^{z}=0.8$, $J_{4}^{y}=-1.0$, $J_{4}^{z}=-0.5$, $\Gamma_{4}^{xy}=-0.7$, and a magnetic field $h_c=1.0$. The dashed, red line indicates the energy of the Weyl points, $E=E_{\textrm{Weyl}}$. \textbf{(b)}, The spectral function computed at a surface perpendicular to the $\bm{a}$-direction showing a magnon arc connecting the projections of the Weyl points on the surface.}
\end{figure}
To illustrate the surface magnon arcs that arise in TbFeO$_3$ upon the application of an external magnetic field, we consider a spin model with parameters $J_{1}^{y}=0.8$, $J_{1}^{z}=1.5$, $J_{2}^{y}=-0.5$, $\Gamma_{2}^{xy}=-0.6$, $D_{2}^{z}=0.8$, $J_{4}^{y}=-1.0$, $J_{4}^{z}=-0.5$, $\Gamma_{4}^{xy}=-0.7$, and a magnetic field $h_c=1.0$. Fig.~S\ref{fig:magnon-arc}(a) shows the magnon spectrum along a line across the Brillouin zone, which connects the two Weyl points in the Brillouin zone. Note that no other states exist in the bulk magnon spectrum at the energy $E=E_{\textrm{Weyl}}$, indicated with a red dashed line in the figure.

Next, we consider an open boundary in the $\bm{a}$-direction and compute the surface Green function using the iterative scheme described in Ref.~\cite{Sancho1985}. The surface spectral function, $\rho_{k_b,k_c}(E=E_{\textrm{Weyl}})=-\frac{1}{\pi}\operatorname{Im} G_{k_b,k_c}(E_{\textrm{Weyl}}+i0^+)$, is plotted in Fig.~S\ref{fig:magnon-arc}(b), showing the surface arc connecting the projections of the two Weyl points.

\subsection{Chiral surface magnons in magnon axion insulators}
In this section we illustrate the surface chiral magnons with the example of $\alpha$-$\text{Fe}_2\text{O}_3$. With the parameters $J_1=-0.1,J_2^z=-0.1,J_2=-1.1,\Gamma_2^{xy}=\Gamma_2^{xz}=\Gamma_2^{yz}=-1$, we find that there's an indirect magnon energy gap between the two upper bands and the two lower bands: there's no states in the energy range $1.41-2.06$. In the 2D plane with momentum $k_z=0$, we find that the total Chern number of the two upper bands is 2. The Chern number for a set of bands isolated from other bands in a 2d plane can be defined (Ref.\cite{thouless1982quantized}) using $H_f$, the Hermitian counterpart of the dynamical matrix $YR$.

Since there's a gap between the upper two magnon bands and the lower two bands, the total Chern number of the two upper bands is 2 for any 2D plane with a fixed momentum $k_z$.

\begin{figure}[htpb!]
  \centering
{\includegraphics[width=1.0\textwidth]{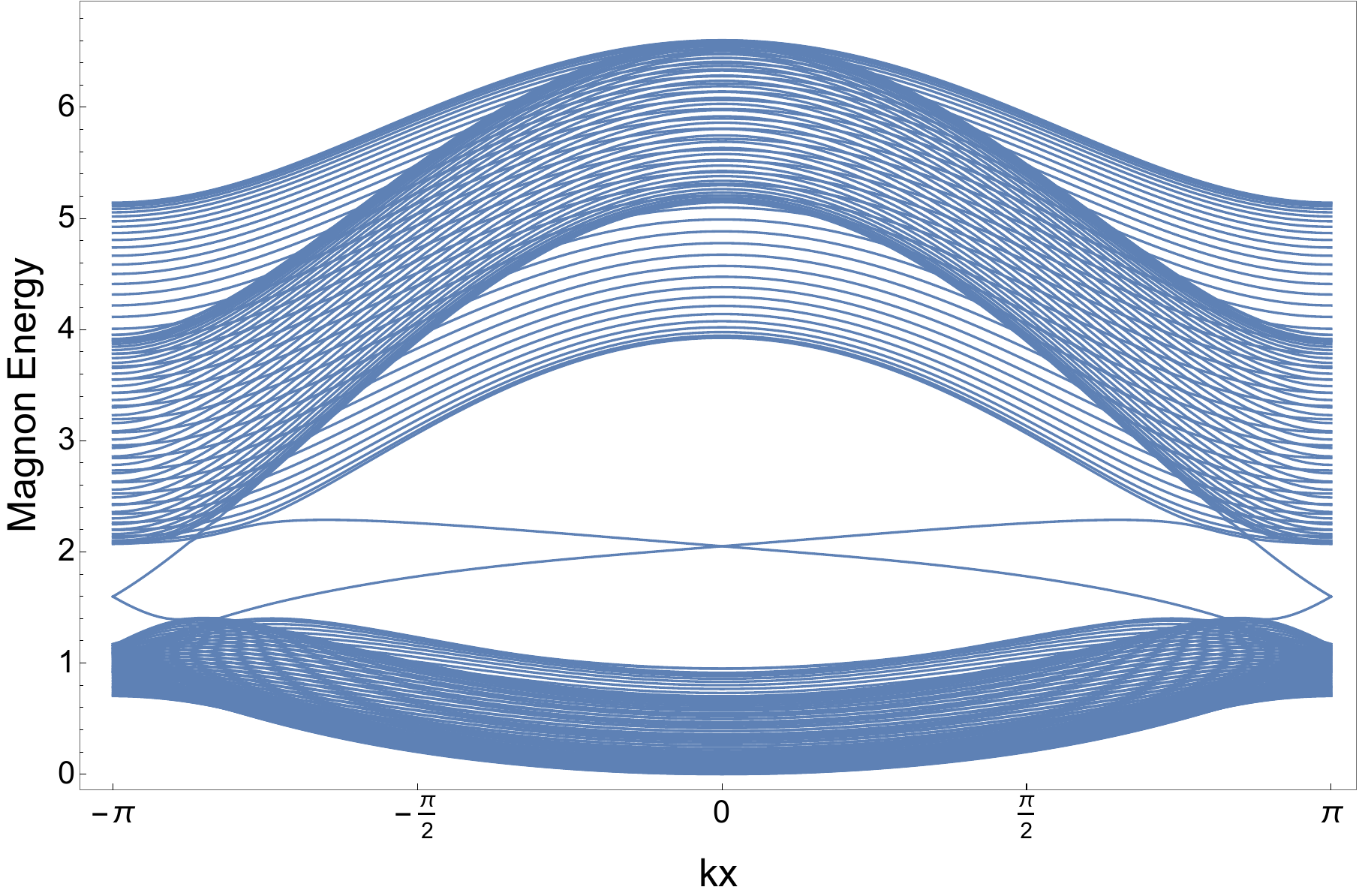}}
  \caption{\label{fig:chiraledge}\textbf{Surface chiral magnons in $\alpha-$Fe$_2$O$_3$.} The parameters we use are $J_1=-0.1,J_2^z=-0.1,J_2=-1.1,\Gamma_2^{xy}=\Gamma_2^{xz}=\Gamma_2^{yz}=-1$. We keep the system periodic along $\bm{a}_1,\bm{a}_3$ direction and open along $\bm{a}_2$ direction. We have included 40 unit-cells along $\bm{a}_2$ direction. In this figure we have plotted the magnon spectrum as a function of $k_x$ with $k_z=0$ fixed. The bulk spectrum has an indirect energy gap ($1.41-2.06$) between the two lower bands and the two upper bands. With the presence of an edge in the x-direction, two pairs of chiral edge states interpolating between the upper bands and the lower bands are clearly seen, consistent with the fact that Chern number is 2 for the upper bands.}
\end{figure}

Let's restrict ourselves to the effective 2D Hamiltonian with $k_z=0$ as an illustration. We keep the system periodic along $\bm{a}_1$ direction and make it open along $\bm{a}_2$ direction. We then plot the magnon spectrum as a function of $k_x$ in Fig. S\ref{fig:chiraledge}. Two pairs of chiral edge states interpolating between the two upper bands and the two lower bands are clearly seen.

\subsection{Hinge magnons in magnon axion insulators}

The magnon hinge states observed in this work are different from what was previously discussed in Ref.\cite{Mook2021} and \cite{Park2021}. Specifically, in Ref.\cite{Mook2021}, the chiral hinge magnons are protected by a mirror Chern number of the gapped bulk magnons, and therefore are robust only on a hinge that preserves the mirror symmetry. In Ref.\cite{Park2021}, the magnon hinge states coexist with magnon band crossings in the bulk, in analogy to topological semimetals in the context of electron bands. Moreover, the $\mathbb{Z}_2$ topological protection of the hinge magnons is provided by a 2-fold rotational symmetry $C_{2x}$ in Ref.\cite{Park2021}. In contrast, the hinge magnons of magnon axion insulators discussed in this work are protected by the inversion symmetry\cite{Yue2019}, different from both Ref.\cite{Mook2021} and \cite{Park2021}.

For an axion insulator with a $\mathbb{Z}_4$ inversion symmetry indicator $\nu=2\bmod4$, the reason for protected hinge magnons can be understood as follows\cite{Yue2019}. When its Chern number in any 2D plane in the momentum space vanishes, the wavefunction of an axion insulator can be adiabatically tuned into that of a 3D topological insulator, which exhibits Dirac surface states. Now consider a finite system with two open surfaces which join at a hinge. Each of the two surfaces preserve no symmetry, and hence the Dirac surface states should be gapped out by a mass term. Meanwhile, the inversion symmetry guarantees the two surfaces related by inversion have opposite mass terms\cite{Yue2019}. As a result, the hinge is a mass domain wall of the Dirac surface states, giving rise to a 1D chiral hinge mode in 3D axion insulators with a vanishing Hall conductance tensor. 

In the following, we illustrate the hinge magnon states using HoRh as an example. We start by first summarizing the results, and subsequently describing more details of the system.

\subsubsection{Hinge magnon modes in HoRh}
\begin{figure}[htpb!]
  \centering
{\includegraphics[width=.55\textwidth]{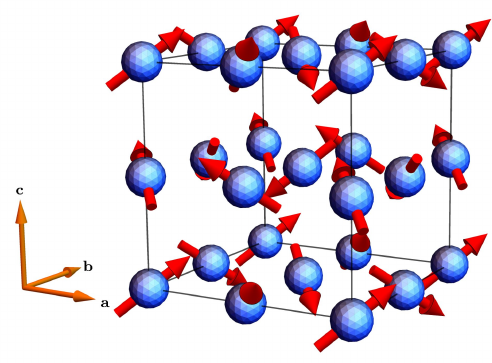}}
  \caption{\label{fig:HoRh-order}
  \textbf{The magnetic order of HoRh below $T_{\textrm{c}}=\SI{3.2}{\kelvin}$.} Only the magnetic atoms (Ho) are shown, and they occupy the $8b$ WP of $P_{I}a\bar{3}~(205.36)$ with a site symmetry group $\bar{3}$.}
\end{figure}

Below $T_{\mathrm{c}}=\SI{3.2}{\kelvin}$, the magnetic moments of the holmium atoms form a non-collinear anti-ferromagnetic order~\cite{HoRh-magnetic-order}, as illustrated in Fig.~S\ref{fig:HoRh-order}. More specifically, the magnetic structure is symmetric under the MSG $P_{I}a\bar{3}~(205.36)$ and the magnetic atoms are located at the $8b$ Wyckoff position. The coordinates and moments of the (magnetic) holmium atoms, belonging to the $8b$ Wyckoff position, are given by (in basis of primitive, cubic Bravais vectors)
\begin{align*}
  \Bigl\{\Bigl(0,0,0\Bigr),(m_{x},m_{x},m_{x})\Bigr\},&\qquad \Bigl\{\Bigl(\frac{1}{2},\frac{1}{2},0),(m_{x},-m_{x},-m_{x}\Bigr)\Bigr\},\\
  \Bigl\{\Bigl(0,\frac{1}{2},\frac{1}{2}\Bigr),(-m_{x},m_{x},-m_{x})\Bigr\},&\qquad \Bigl\{\Bigl(\frac{1}{2},0,\frac{1}{2}\Bigr),(-m_{x},-m_{x},m_{x})\Bigr\},\\
  \Bigl\{\Bigl(\frac{1}{2},\frac{1}{2},\frac{1}{2}\Bigr),(-m_{x},-m_{x},-m_{x})\Bigr\},&\qquad \Bigl\{\Bigl(0,0,\frac{1}{2}),(-m_{x},m_{x},m_{x}\Bigr)\Bigr\},\\
  \Bigl\{\Bigl(\frac{1}{2},0,0\Bigr),(m_{x},-m_{x},m_{x})\Bigr\},&\qquad \Bigl\{\Bigl(0,\frac{1}{2},0),(m_{x},m_{x},-m_{x}\Bigr)\Bigr\}
\end{align*} 

The 8 magnon bands have the following band representation:
\begin{align*}
{(^{1}E_{g})}_{8b}\uparrow P_{I}a\bar{3}~(205.36)&=
  \Gamma_{2}^{+}\Gamma_{3}^{+}(2)\oplus 2\Gamma_{4}^{+}(3)\\
  &\quad \oplus R_{1}^{+}R_{3}^{-}(4)\oplus R_{2}^{-}R_{2}^{+}(4)\\
&\quad\oplus 2X_{1}(2)\oplus 2X_{2}(2)\\
  &\quad\oplus 2M_{1}(2)\oplus 2M_{2}(2)
\end{align*}
where 
\begin{align*}
  \bm{k}_{\Gamma}&=(0,0,0),\\
  \bm{k}_R&=(\pi,\pi,\pi),\\
  \star\bm{k}_{X}&=\{(\pi,0,0),(0,\pi,0),(0,0,\pi)\},\\
  \star\bm{k}_{M}&=\{(0,\pi,\pi),(\pi,0,\pi),(\pi,\pi,0)\}.
\end{align*}

A crucial property of this band representation is that, at any given TRIM point, all the irreps have the same number of $I=-1$ inversion eigenvalues. For example, at $\bm{k}_M$, each of $M_1(2)$ and $M_2(2)$ contains one inversion-even and one inversion-odd modes. Additionally, by simple counting, the highest 4 magnon bands must always have an inversion SI $\nu=2\bmod 4$. Therefore, if we energetically separate the highest 4 bands by applying an inversion-preserving perturbation, the result is always a magnon axion insulator, independent of the details of the unperturbed spectrum and the perturbation specifics. Thus, HoRh is a type-I topological magnon candidate.

\begin{figure}[htpb!]
  \centering
{\includegraphics[width=.8\textwidth]{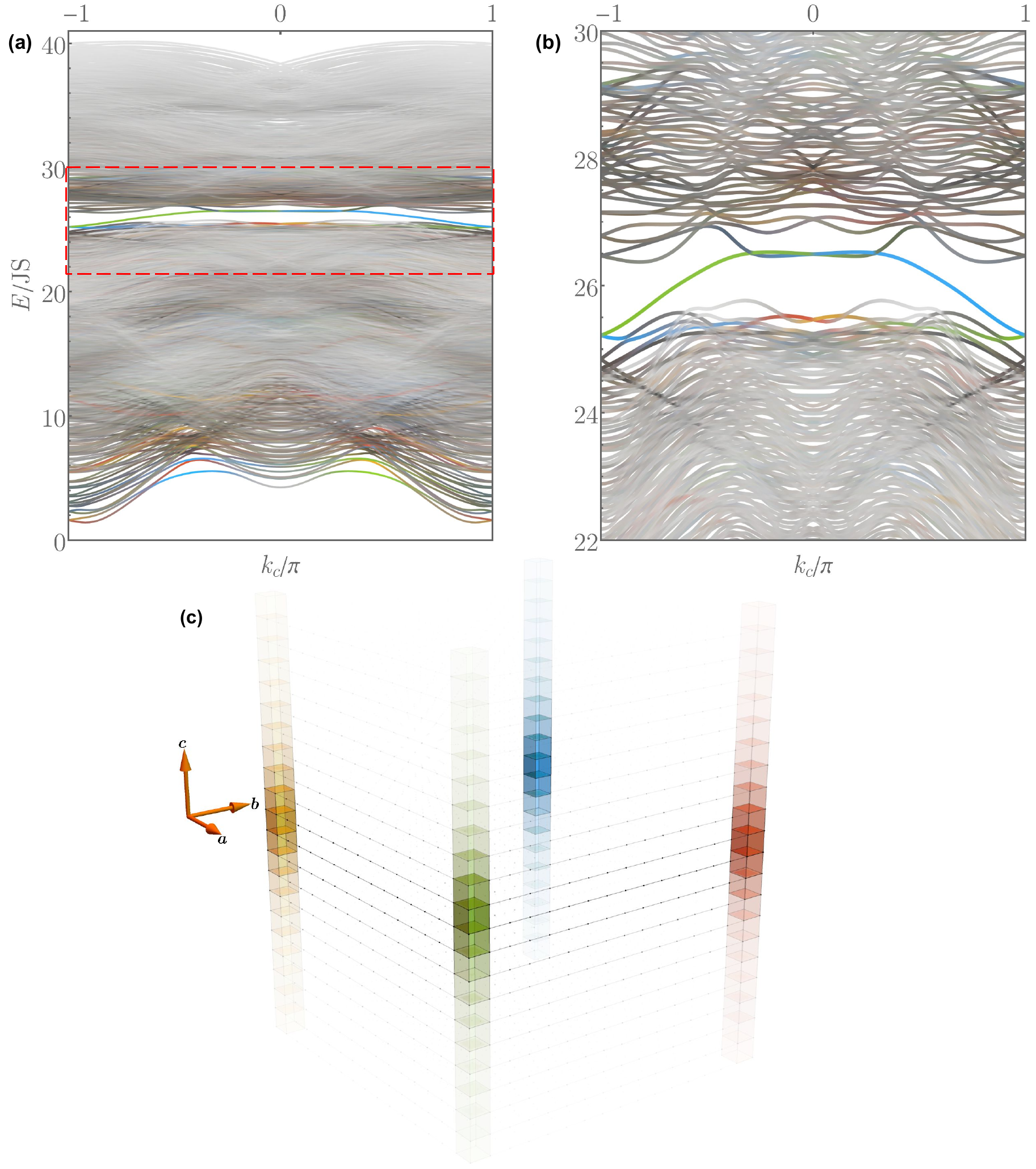}}
  \caption{\label{fig:hinge}
  \textbf{1D chiral hinge magnons in HoRh after the application of mechanical strain in a low-symmetry direction.} \textbf{(a)}, The magnon spectrum upon perturbing the system, using open boundaries in the $\bm{a}$- and $\bm{b}$-directions and a periodic $\bm{c}$-direction. We include 16 crystalline unit cells in each open direction. \textbf{(b)}, The same band structure but focusing on states near the gap (see the red, dashed rectangle in (a)). The colors in the spectrum are given by the weights of the wave function in the $\bm{a}$-$\bm{b}$ plane with different regions contributing different colors. Specifically, bulk and surface states tend to be gray and dark gray, respectively, and states localized at a hinge tends towards the corresponding color shown in (c). \textbf{(c)}, A schematic of the system with two open boundaries and one periodic, as well as a color code to label hinge modes. In the spectrum in (a) and (b), one can clearly see the existence of in-gap 1D chiral modes localized at opposite hinges.
  }
\end{figure}

Next, we consider a system with open boundaries in the $\bm{a}$- and $\bm{b}$-directions and we keep it periodic in the $\bm{c}$-direction. We construct a spin model (see details in the next subsection), as well as an inversion-preserving perturbation that separates the highest four bands and creates a gap both in the bulk and the 2D surfaces. We summarize the resulting spectrum in Fig.~S\ref{fig:hinge}. This figure illustrates the appearance of two chiral 1D modes localized at two opposite hinges and propagating in opposite directions along the $\bm{c}$-direction.

\subsubsection{Additional details: the magnon Hamiltonian}
Here, we provides details about the Hamiltonian we used to compute the spectrum in Fig.~S\ref{fig:hinge}. We start by identifying all symmetry-allowed, bilinear interaction terms up to the 3\textsuperscript{rd} nearest-neighbor bonds.

The space group of HoRh is $Pm\bar{3}m~(221)$, with a primitive cubic lattice. There is one Ho atom per crystalline unit cell and it is located at the origin, where the site symmetry group is $m\bar{3}m$. We use a triplet of integers, $(x_1x_2x_3)$, to label a general Ho site, with a location at $\bm{R}_{(x_1x_2x_3)}=x_1\bm{a}_1+x_2\bm{a}_2+x_3\bm{a}_3$. Here $\{\bm{a}_i\}$ are the primitive translations, given in Cartesian coordinates as $a(1,0,0)$,\; $a(0,1,0)$,\; and $a(0,0,1)$ respectively.

The 1NN interactions are specified by two free parameters. For example, the interaction matrix between sites $(000)$ and $(100)$ reads
\begin{equation}
  \mathsf{J}_{1}=\begin{pmatrix}
    J_1+J_1^x&&\\
    &J_1&\\
    &&J_1
  \end{pmatrix}.
\end{equation}
For the 2NN interaction between sites $(000)$ and $(110)$, the interaction matrix is
\begin{equation}
  \mathsf{J}_{2}=\begin{pmatrix}
    J_2&\Gamma_2&\\
    \Gamma_2&J_2&\\
    &&J_2+J_2^z
  \end{pmatrix}
\end{equation}
For the 3NN interaction between sites $(000)$ and $(111)$, we have
\begin{equation}
  \mathsf{J}_{3}=\begin{pmatrix}
  J_3& \Gamma_3& \Gamma_3\\
  \Gamma_3& J_3& \Gamma_3\\
  \Gamma_3& \Gamma_3& J_3\\
  \end{pmatrix}
\end{equation}
All other 1NN, 2NN and 3NN interactions are related to the bonds above by a symmetry in $Pm\bar{3}m$, and therefore they are all fixed by the same parameters.

Next, we compute the interaction matrices that couple to the strain tensor
\begin{equation}
  \sigma=\begin{pmatrix}
    \sigma^{xx}&\sigma^{xy}&\sigma^{xz}\\
    \sigma^{xy}&\sigma^{yy}&\sigma^{yz}\\
    \sigma^{xz}&\sigma^{xy}&\sigma^{zz}
  \end{pmatrix}.
\end{equation} 
For the 1NN bond $(000)-(100)$, the symmetry allowed exchange matrices are
\begin{equation}
  \Scale[0.8]{
    \mathsf{\Sigma}_1=
\left(
\begin{array}{ccc}
 \left(
\begin{array}{ccc}
 \Sigma _1^{{xx,xx}} & {} & {} \\
  {} & \Sigma _1^{{xx,yy}} & {} \\
 {} & {} & \Sigma _1^{{xx,yy}} \\
\end{array}
\right) & \left(
\begin{array}{ccc}
 {} & \Sigma _1^{{xy,xy}} & {} \\
 \Sigma _1^{{xy,xy}} & {} & {} \\
 {} & {} & {} \\
\end{array}
\right) & \left(
\begin{array}{ccc}
 {} & {} & \Sigma _1^{{xy,xy}} \\
 {} & {} & {} \\
 \Sigma _1^{{xy,xy}} & {} & {} \\
\end{array}
  \right) \\[20pt]
 \left(
\begin{array}{ccc}
 {} & \Sigma _1^{{xy,xy}} & {} \\
 \Sigma _1^{{xy,xy}} & {} & {} \\
 {} & {} & {} \\
\end{array}
\right) & \left(
\begin{array}{ccc}
 \Sigma _1^{{yy,xx}} & {} & {} \\
 {} & \Sigma _1^{{yy,yy}} & {} \\
 {} & {} & \Sigma _1^{{yy,zz}} \\
\end{array}
\right) & \left(
\begin{array}{ccc}
 {} & {} & {} \\
 {} & {} & \Sigma _1^{{yz,yz}} \\
 {} & \Sigma _1^{{yz,yz}} & {} \\
\end{array}
  \right) \\[20pt]
 \left(
\begin{array}{ccc}
 {} & {} & \Sigma _1^{{xy,xy}} \\
 {} & {} & {} \\
 \Sigma _1^{{xy,xy}} & {} & {} \\
\end{array}
\right) & \left(
\begin{array}{ccc}
 {} & {} & {} \\
 {} & {} & \Sigma _1^{{yz,yz}} \\
 {} & \Sigma _1^{{yz,yz}} & {} \\
\end{array}
\right) & \left(
\begin{array}{ccc}
 \Sigma _1^{{yy,xx}} & {} & {} \\
 {} & \Sigma _1^{{yy,zz}} & {} \\
 {} & {} & \Sigma _1^{{yy,yy}} \\
\end{array}
\right) \\
\end{array}
\right)
}
\end{equation} 
For the 2NN bond $(000)-(110)$, we obtain
\begin{equation}
  \Scale[0.8]{
    \mathsf{\Sigma}_2=
\left(
\begin{array}{ccc}
 \left(
\begin{array}{ccc}
 \Sigma _2^{{xx,xx}} & \Sigma _2^{{xx,xy}} & {} \\
 \Sigma _2^{{xx,xy}} & \Sigma _2^{{xx,yy}} & {} \\
 {} & {} & \Sigma _2^{{xx,zz}} \\
\end{array}
\right) & \left(
\begin{array}{ccc}
 \Sigma _2^{{xy,xx}} & \Sigma _2^{{xy,xy}} & {} \\
 \Sigma _2^{{xy,xy}} & \Sigma _2^{{xy,xx}} & {} \\
 {} & {} & \Sigma _2^{{xy,zz}} \\
\end{array}
\right) & \left(
\begin{array}{ccc}
 {} & {} & \Sigma _2^{{xz,xz}} \\
 {} & {} & \Sigma _2^{{xz,yz}} \\
 \Sigma _2^{{xz,xz}} & \Sigma _2^{{xz,yz}} & {} \\
\end{array}
\right) \\[20pt]
 \left(
\begin{array}{ccc}
 \Sigma _2^{{xy,xx}} & \Sigma _2^{{xy,xy}} & {} \\
 \Sigma _2^{{xy,xy}} & \Sigma _2^{{xy,xx}} & {} \\
 {} & {} & \Sigma _2^{{xy,zz}} \\
\end{array}
\right) & \left(
\begin{array}{ccc}
 \Sigma _2^{{xx,yy}} & \Sigma _2^{{xx,xy}} & {} \\
 \Sigma _2^{{xx,xy}} & \Sigma _2^{{xx,xx}} & {} \\
 {} & {} & \Sigma _2^{{xx,zz}} \\
\end{array}
\right) & \left(
\begin{array}{ccc}
 {} & {} & \Sigma _2^{{xz,yz}} \\
 {} & {} & \Sigma _2^{{xz,xz}} \\
 \Sigma _2^{{xz,yz}} & \Sigma _2^{{xz,xz}} & {} \\
\end{array}
\right) \\[20pt]
 \left(
\begin{array}{ccc}
 {} & {} & \Sigma _2^{{xz,xz}} \\
 {} & {} & \Sigma _2^{{xz,yz}} \\
 \Sigma _2^{{xz,xz}} & \Sigma _2^{{xz,yz}} & {} \\
\end{array}
\right) & \left(
\begin{array}{ccc}
 {} & {} & \Sigma _2^{{xz,yz}} \\
 {} & {} & \Sigma _2^{{xz,xz}} \\
 \Sigma _2^{{xz,yz}} & \Sigma _2^{{xz,xz}} & {} \\
\end{array}
\right) & \left(
\begin{array}{ccc}
 \Sigma _2^{{zz,xx}} & \Sigma _2^{{zz,xy}} & {} \\
 \Sigma _2^{{zz,xy}} & \Sigma _2^{{zz,xx}} & {} \\
 {} & {} & \Sigma _2^{{zz,zz}} \\
\end{array}
\right) \\
\end{array}
\right)
}
\end{equation} 
Finally, for the 3NN bond $(000)-(111)$, we have
\begin{equation}
  \Scale[0.8]{
    \mathsf{\Sigma}_3=
\left(
\begin{array}{ccc}
 \left(
\begin{array}{ccc}
 \Sigma _3^{{xx,xx}} & \Sigma _3^{{xx,xy}} & \Sigma _3^{{xx,xy}} \\
 \Sigma _3^{{xx,xy}} & \Sigma _3^{{xx,yy}} & \Sigma _3^{{xx,yz}} \\
 \Sigma _3^{{xx,xy}} & \Sigma _3^{{xx,yz}} & \Sigma _3^{{xx,yy}} \\
\end{array}
\right) & \left(
\begin{array}{ccc}
 \Sigma _3^{{xy,xx}} & \Sigma _3^{{xy,xy}} & \Sigma _3^{{xy,xz}} \\
 \Sigma _3^{{xy,xy}} & \Sigma _3^{{xy,xx}} & \Sigma _3^{{xy,xz}} \\
 \Sigma _3^{{xy,xz}} & \Sigma _3^{{xy,xz}} & \Sigma _3^{{xy,zz}} \\
\end{array}
\right) & \left(
\begin{array}{ccc}
 \Sigma _3^{{xy,xx}} & \Sigma _3^{{xy,xz}} & \Sigma _3^{{xy,xy}} \\
 \Sigma _3^{{xy,xz}} & \Sigma _3^{{xy,zz}} & \Sigma _3^{{xy,xz}} \\
 \Sigma _3^{{xy,xy}} & \Sigma _3^{{xy,xz}} & \Sigma _3^{{xy,xx}} \\
\end{array}
\right) \\[20pt]
 \left(
\begin{array}{ccc}
 \Sigma _3^{{xy,xx}} & \Sigma _3^{{xy,xy}} & \Sigma _3^{{xy,xz}} \\
 \Sigma _3^{{xy,xy}} & \Sigma _3^{{xy,xx}} & \Sigma _3^{{xy,xz}} \\
 \Sigma _3^{{xy,xz}} & \Sigma _3^{{xy,xz}} & \Sigma _3^{{xy,zz}} \\
\end{array}
\right) & \left(
\begin{array}{ccc}
 \Sigma _3^{{xx,yy}} & \Sigma _3^{{xx,xy}} & \Sigma _3^{{xx,yz}} \\
 \Sigma _3^{{xx,xy}} & \Sigma _3^{{xx,xx}} & \Sigma _3^{{xx,xy}} \\
 \Sigma _3^{{xx,yz}} & \Sigma _3^{{xx,xy}} & \Sigma _3^{{xx,yy}} \\
\end{array}
\right) & \left(
\begin{array}{ccc}
 \Sigma _3^{{xy,zz}} & \Sigma _3^{{xy,xz}} & \Sigma _3^{{xy,xz}} \\
 \Sigma _3^{{xy,xz}} & \Sigma _3^{{xy,xx}} & \Sigma _3^{{xy,xy}} \\
 \Sigma _3^{{xy,xz}} & \Sigma _3^{{xy,xy}} & \Sigma _3^{{xy,xx}} \\
\end{array}
  \right) \\[20pt]
 \left(
\begin{array}{ccc}
 \Sigma _3^{{xy,xx}} & \Sigma _3^{{xy,xz}} & \Sigma _3^{{xy,xy}} \\
 \Sigma _3^{{xy,xz}} & \Sigma _3^{{xy,zz}} & \Sigma _3^{{xy,xz}} \\
 \Sigma _3^{{xy,xy}} & \Sigma _3^{{xy,xz}} & \Sigma _3^{{xy,xx}} \\
\end{array}
\right) & \left(
\begin{array}{ccc}
 \Sigma _3^{{xy,zz}} & \Sigma _3^{{xy,xz}} & \Sigma _3^{{xy,xz}} \\
 \Sigma _3^{{xy,xz}} & \Sigma _3^{{xy,xx}} & \Sigma _3^{{xy,xy}} \\
 \Sigma _3^{{xy,xz}} & \Sigma _3^{{xy,xy}} & \Sigma _3^{{xy,xx}} \\
\end{array}
\right) & \left(
\begin{array}{ccc}
 \Sigma _3^{{xx,yy}} & \Sigma _3^{{xx,yz}} & \Sigma _3^{{xx,xy}} \\
 \Sigma _3^{{xx,yz}} & \Sigma _3^{{xx,yy}} & \Sigma _3^{{xx,xy}} \\
 \Sigma _3^{{xx,xy}} & \Sigma _3^{{xx,xy}} & \Sigma _3^{{xx,xx}} \\
\end{array}
\right) \\
\end{array}
\right)
}
\end{equation} 
Similar to the unperturbed terms, all other 1NN, 2NN and 3NN interactions induced by strain are related to the bonds above by a symmetry in $Pm\bar{3}m$, and therefore they are also all fixed by the same parameters shown above.

\begin{figure}[ht!]
  \centering
\includegraphics[width=1\textwidth]{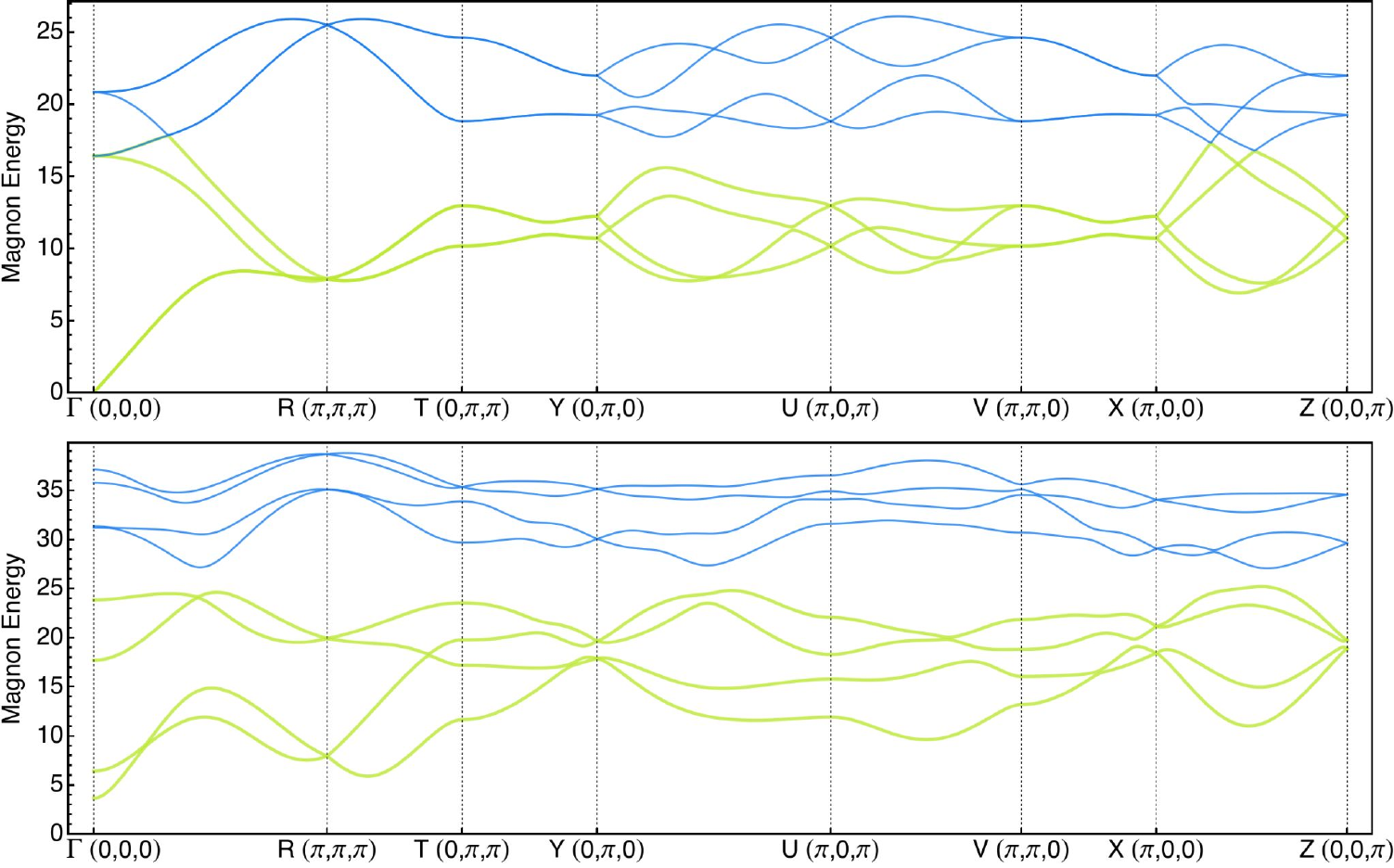}
  \caption{\label{fig:HoRh-bulk} \textbf{The bulk magnon spectrum of HoRh.} The top (bottom) panel shows the magnon spectrum before (after) perturbing the system with a mechanical strain (see the supplementary text for a complete description of the perturbation specifics).
  }  
\end{figure}
  We included the following parameters in our calculation in this section (both bulk and hinge calculations): 
  \begin{gather*}
  J_1=0.40,J_1^x=-0.15,J_2^z=3.30,\Gamma_2=2.20,J3=0.50,\Gamma_3=0.40,\\
\Sigma_1^{xyxy} = 0.5, \Sigma_1^{yyxx} = -0.5, \Sigma_2^{xxxx} = 0.34, \Sigma_2^{xxxy} = -0.94, \\
    \Sigma_2^{xxyy} = 0.26, \Sigma_2^{xxzz} = 0.27, \Sigma_2^{xyxx} = -0.50, \Sigma_2^{xyxy} = 1.34,\\
    \Sigma_2^{xyzz} = -0.86, \Sigma_2^{xzxz} = -0.2^0, \Sigma_2^{xzyz} = -0.3^2, \Sigma_2^{zzxx} =  0.06,\\
    \Sigma_2^{zzxy} = -0.81, \Sigma_3^{xxxx} = -0.30, \Sigma_3^{xxxy} = 0.50, \Sigma_3^{xxyy} = -0.75,\\
    \Sigma_3^{xyxx} =  -0.60, \Sigma_3^{xyxy} = 0.05, \Sigma_3^{xyxz} = -0.25, \Sigma_3^{xyzz} = 0.50,\\
    \sigma^{xx} = -0.10,
    \sigma^{xy} =  -0.64,
    \sigma^{xz} = 0.52,
    \sigma^{yy} = -0.33,
    \sigma^{yz} = 0.79,
    \sigma^{zz} = -0.12,
  \end{gather*} 
  in addition to a staggered magnetic field ($h=0.7$) to stabilize the magnetic structure.

  The bulk magnetic structure before and after the perturbation is illustrated in Fig.~S\ref{fig:HoRh-bulk}.

\input{supplement-section-subgroups.tex}
\section{BCS database materials passing our group-theoretical filters}
 Listed in \tablename~\ref{sup-tab:materials} are the magnetic materials in the Bilbao database which passed our filters for magnetic space group and Wyckoff position, as explained in the main text. Entries are listed in a decreasing order of their transition temperature. A green-colored Wyckoff position label indicates that the induced magnon band representation contains one or more magnon band degeneracies, whereas a black label indicates that the site of the corresponding magnetic atom is incompatible with magnetism (and thus the magnetic moment is $|\bf{m}|=0$.)
\begin{center}
\setlength\LTleft{-.5in}
\begin{longtable}{ccccccc}
    \caption{Bilbao database materials that passed our filters for magnetic space group and Wyckoff position.\label{sup-tab:materials}}\\
\toprule
    & Compound & SG \& MSG & $T_{\rm{c}}$ & (WP, $|\bm{m}|$) & Band Gap ($\si{\eV}$)\\
  \midrule
\endfirsthead
\multicolumn{5}{@{}l}{\ldots continued}\\
\toprule
  & Compound & SG \& MSG & $T_{\rm{c}}$ & (WP, $|\bm{m}|$)\\
  \midrule
  \endhead 
\multirow{2}{*}{1} & \multirow{2}{*}{\href{http://webbdcrista1.ehu.es/magndata/index.php?this_label=0.458}{CsFeO\textsubscript{2}}} & \multirow{1}{*}{$Fd\bar{3}m~(227)$ } & \multirow{2}{*}{1055} & \multirow{2}{*}{({\color{green}{\textbf{4a}}}, 3.7)} & \\
 &  & \multirow{1}{*}{ \hyperref[subsec:141.556]{$I4_{1}'/a'm'd~(141.556)$}} &  &  & \\\midrule\multirow{2}{*}{2} & \multirow{2}{*}{\href{http://webbdcrista1.ehu.es/magndata/index.php?this_label=0.456}{RbFeO\textsubscript{2}}} & \multirow{1}{*}{$Fd\bar{3}m~(227)$ } & \multirow{2}{*}{1027} & \multirow{2}{*}{({\color{green}{\textbf{4a}}}, 2.75)} & \multirow{2}{*}{$1.70$~\cite{bandgap-rbfeo2}}\\
    &  & \multirow{1}{*}{ \hyperref[subsec:141.556]{$I4_{1}'/a'm'd~(141.556)$}} &  &  & \\\midrule\multirow{2}{*}{3} & \multirow{2}{*}{\href{http://webbdcrista1.ehu.es/magndata/index.php?this_label=0.65}{$\alpha$-Fe\textsubscript{2}O\textsubscript{3}}} & \multirow{1}{*}{$R\bar{3}c~(167)$ } & \multirow{2}{*}{955} & \multirow{2}{*}{({\color{green}{\textbf{8f}}}, 4.11)} & \multirow{2}{*}{$2.2$~\cite{bandgap-fe2o3}}\\
 &  & \multirow{1}{*}{ \hyperref[subsec:15.89]{$C2'/c'~(15.89)$}} &  &  & \\\midrule\multirow{2}{*}{4} & \multirow{2}{*}{\href{http://webbdcrista1.ehu.es/magndata/index.php?this_label=1.466}{MnPt\textsubscript{0.5}Pd\textsubscript{0.5}}} & \multirow{1}{*}{$P4/mmm~(123)$ } & \multirow{2}{*}{866} & \multirow{2}{*}{({\color{green}{\textbf{4g}}}, 4.09)} & \\
 &  & \multirow{1}{*}{ \hyperref[subsec:67.509]{$C_{a}mma~(67.509)$}} &  &  & \\
\bottomrule
\end{longtable}
\end{center}


%